\documentclass[prd,preprint,superscriptaddress,preprintnumbers,eqsecnum,showpacs,nofootinbib,nobibnotes]{revtex4}
\usepackage{amsfonts,bm}
\usepackage{amsfonts,amssymb,amsmath}
\usepackage{graphicx}
\usepackage{paralist}
\usepackage[english]{babel}
\usepackage{slashed}
\usepackage[usenames]{xcolor}
\usepackage{float}
\usepackage{pstricks}

\newcommand{\be}{\begin{equation}}
\newcommand{\bea}{\begin{eqnarray}}
\newcommand{\ee}{\end{equation}}
\newcommand{\eea}{\end{eqnarray}}

\def\s#1{{\scriptscriptstyle #1}}
\def\ie{{\it i.e.}, }
\def\eg{{\it e.g.}, }

\def\1eq#1{Eq.~(\ref{#1})}

\def\2eqs#1#2{Eqs.~(\ref{#1}) and~(\ref{#2})}
\def\3eqs#1#2#3{Eqs.~(\ref{#1}),~(\ref{#2}) and~(\ref{#3})}

\def\n#1{({\it #1}\,)}

\newcommand{\JBC}{J_{\textnormal{\tiny \textsc{BC}}}}  

\def\cd{\!\cdot\!}

\newcommand{\fatg}{{\rm{I}}\!\Gamma}             
\newcommand{\Gnp}{\Gamma}        
\newcommand{\Gp}{V}          
\newcommand{\GL}{{\Gamma}_{\!\!{\s{\mathbf{L}}}}} 
\newcommand{\GT}{{\Gamma}_{\!\!{\s{\mathbf{T}}}}}              
 
\newcommand{\Glatt}{L^{\rm{sym}}}       

\begin{document}
\title{Nonperturbative Ball-Chiu construction of the three-gluon vertex}

\author{A.~C. Aguilar}
\affiliation{\mbox{University of Campinas - UNICAMP, Institute of Physics ``Gleb Wataghin,''} \\
13083-859 Campinas, S\~{a}o Paulo, Brazil}

\author{M.~N. Ferreira}
\affiliation{\mbox{University of Campinas - UNICAMP, Institute of Physics ``Gleb Wataghin,''} \\
13083-859 Campinas, S\~{a}o Paulo, Brazil}
\affiliation{\mbox{Department of Theoretical Physics and IFIC, 
University of Valencia and CSIC},
E-46100, Valencia, Spain}

\author{C.~T. Figueiredo}
\affiliation{\mbox{University of Campinas - UNICAMP, Institute of Physics ``Gleb Wataghin,''} \\
13083-859 Campinas, S\~{a}o Paulo, Brazil}
\affiliation{\mbox{Department of Theoretical Physics and IFIC, 
University of Valencia and CSIC},
E-46100, Valencia, Spain}

\author{J. Papavassiliou}
\affiliation{\mbox{Department of Theoretical Physics and IFIC, 
University of Valencia and CSIC},
E-46100, Valencia, Spain}

\begin{abstract}

We present  the detailed  derivation of the  longitudinal part  of the
three-gluon  vertex  from   the  Slavnov-Taylor  identities  that  it
satisfies,  by  means  of  a  nonperturbative  implementation  of  the
Ball-Chiu construction; the latter, in its original form, involves the
inverse  gluon propagator,  the ghost  dressing function,  and certain
form factors of the ghost-gluon  kernel.  The main conceptual subtlety
that renders  this endeavor nontrivial  is the infrared  finiteness of
the gluon  propagator, and the  resulting need to separate  the vertex
into two pieces,  one that is intimately connected  with the emergence
of a  gluonic mass scale, and  one that satisfies the  original set of
Slavnov-Taylor  identities,  but  with the  inverse  gluon  propagator
replaced  by  its ``kinetic''  term.   The  longitudinal form  factors
obtained by  this construction  are presented for  arbitrary Euclidean
momenta, as well as  special kinematic configurations, parametrized by
a single momentum. A particularly preeminent feature of the components
comprising the tree-level vertex  is their considerable suppression for
momenta  below  1  GeV,  and  the  appearance  of  the  characteristic
``zero-crossing'' in  the vicinity of  \mbox{$100-200$ MeV}. 
Special combinations of the form factors derived with this method 
are compared  with the results of 
recent large-volume  lattice simulations, and  
are found  to  capture faithfully  the  rather complicated  curves
formed by the  data. A similar comparison with results obtained from
Schwinger-Dyson equations 
reveals a fair overall agreement, but with appreciable differences 
at intermediate energies.
A variety of issues related  to the distribution
of the pole terms responsible for
the gluon mass generation are discussed in detail, and their impact on
the structure of  the transverse parts is elucidated.   In addition, a
brief  account  of several theoretical  and  phenomenological
possibilities involving these newly acquired results is presented.

\end{abstract}

\pacs{
12.38.Aw,  
12.38.Lg, 
14.70.Dj 
}

\maketitle

\section{\label{sec:intro} Introduction}

The three-gluon vertex of QCD, to be denoted by $\fatg_{\alpha\mu\nu}$, is inseparably linked with the
non-Abelian nature of the theory~\cite{Marciano:1977su}, and is crucial for its most celebrated perturbative property,
namely asymptotic freedom~\cite{Gross:1973id,Politzer:1973fx}.
In addition, in recent years, the paramount importance of $\fatg_{\alpha\mu\nu}$ for a plethora of
nonperturbative phenomena has become increasingly evident among practitioners,
leading to a vigorous activity for unraveling its infrared properties~\cite{Cornwall:1989gv,Alkofer:2004it,Binger:2006sj,Fischer:2006vf,Fischer:2006ub,Binosi:2011wi, Huber:2012kd,Huber:2012zj,Binosi:2013rba,Pelaez:2013cpa,Blum:2014gna,Eichmann:2014xya,Williams:2015cvx,Aguilar:2013xqa,Aguilar:2013vaa,Blum:2015lsa,Blum:2016fib,Cyrol:2016tym,Corell:2018yil,Cucchieri:2006tf,Cucchieri:2008qm,Athenodorou:2016oyh,Sternbeck:2017ntv,Boucaud:2017obn,Vujinovic:2018nqc}.
In particular, distinct but equally remarkable aspects 
of the three-gluon vertex are intimately associated with the emergence of a gluonic mass scale~\cite{Cornwall:1981zr,Aguilar:2004sw,Aguilar:2002tc, Aguilar:2008xm,Binosi:2009qm,Cornwall:2010upa, Binosi:2012sj,Ibanez:2012zk,Aguilar:2015bud,Aguilar:2016vin,Binosi:2017rwj,Aguilar:2017dco,Gao:2017uox},
the masses and properties of glueballs~\cite{Meyers:2012ka,Strauss:2012dg,Sanchis-Alepuz:2015hma,Fukamachi:2016wxf}, and the potential formation 
of hybrids and exotics states~\cite{Xu:2018cor}.

Perhaps the most intriguing nonperturbative aspect of the three-gluon vertex in the Landau gauge is its so-called  
``infrared suppression''. Specifically, the predominant form factors of $\fatg_{\alpha\mu\nu}$,
which at tree level are equal to unity, 
decrease gradually as the Euclidean momenta become comparable to the fundamental QCD scale,   
and eventually reverse their sign, displaying the characteristic ``zero-crossing''~\cite{Aguilar:2013vaa,Eichmann:2014xya,Blum:2014gna,Vujinovic:2014fza,Athenodorou:2016oyh,Boucaud:2017obn},  
finally diverging logarithmically at the origin. These exceptional features
have far-reaching theoretical and  phenomenological consequences. From the theoretical
point of view, the aforementioned behavior of the vertex 
hinges on the subtle interplay between dynamical effects 
originating from the two-point sector of the theory~\cite{Roberts:1994dr,Alkofer:2000wg,Fischer:2006ub,Cloet:2013jya,Aguilar:2015bud,Binosi:2014aea}. In particular,
while the gluon acquires dynamically an effective mass,
the ghost remains massless even nonperturbatively;
thus, loops containing gluons give rise to ``protected'' logarithms, 
whilst loops containing ghosts to divergent ones~\cite{Aguilar:2013vaa}.
From the phenomenological perspective, the infrared suppression of $\fatg_{\alpha\mu\nu}$, and the overall attenuation of the
interaction strength that this causes to the Bethe-Salpeter kernels~\cite{Meyers:2012ka,Fukamachi:2016wxf}, 
appears to be instrumental for the formation of glueball states with masses compatible with those obtained
from lattice simulations~\cite{Morningstar:1999rf}. Moreover, the necessity of a considerable suppression has become evident also  
in a recent study of the hybrid states, in the framework of the Faddeev equations~\cite{Xu:2018cor}.

At the technical level, the nonperturbative study of the three-gluon vertex is particularly challenging,
mainly because it is composed by 14 form factors, which are complicated functions of three independent momenta ($q$, $r$, and $p$)~\cite{Ball:1980ax}.
The knowledge of the full momentum dependence of the form factors, in turn, may be crucial for the phenomenological applications
mentioned above, essentially because $\fatg_{\alpha\mu\nu}(q,r,p)$ appears usually inside ``loops'', and the evaluation 
of its contribution to the effective strength requires the integration over some of its momenta in the entire range of values. 
In order to acquire this type of detailed information, one has to turn to continuous approaches, such as the 
Schwinger-Dyson equations (SDEs)~\cite{Schleifenbaum:2004id,Huber:2012kd,Aguilar:2013xqa,Huber:2012zj,Blum:2014gna,Eichmann:2014xya,Williams:2015cvx, Binosi:2016wcx,Hawes:1998cw,Chang:2009zb,Qin:2011dd}
or the functional renormalization group~\cite{Corell:2018yil,Cyrol:2017ewj,Cyrol:2016tym}. 
Within these latter formalisms, the dynamical equations governing the momentum evolution of the form factors of $\fatg_{\alpha\mu\nu}$
(or selected subsets thereof) are projected out and solved, usually resorting to certain physically motivated assumptions and judiciously constructed 
Ans\"atze, in order to reduce, to some extent, the vast complexity of such undertakings.  

In the present work we employ an alternative procedure, which exploits the Slavnov-Taylor identities (STIs), and
amounts essentially to a contemporary application of the time-honored method known as
``gauge technique''~\cite{Salam:1963sa,Salam:1964zk,Delbourgo:1977jc,Delbourgo:1977hq}.
The central idea underlying this approach is to reconstruct 
the nontransverse part\footnote{In the original work by Ball and Chiu~\cite{Ball:1980ax}, this part is
referred to as ``longitudinal'', whereas, in some of the more recent literature, 
the alternative terms ``gauge''~\cite{Eichmann:2018ytt}, or ``STI saturating''\cite{Skullerud:2002ge} have been put forth as more accurate; throughout this work we 
 adhere to the initial term ``longitudinal''.} of the vertex from 
the quantities that enter in the STIs that $\fatg_{\alpha\mu\nu}(q,r,p)$ satisfies [see~\1eq{eq:sti_delta}].
In particular, Ball and Chiu (BC)~\cite{Ball:1980ax} cast the gluon propagator in the form  $\Delta^{-1}(q) = q^2 \JBC(q)$, 
and express the 10 longitudinal form factors
in terms of $\JBC(q)$, the ghost dressing function, $F(q)$, and a subset of the factors comprising the
so-called ``ghost-gluon scattering kernel'', $H_{\nu\mu}$. The nonperturbative
structure of all these quantities is in principle known:
both $\Delta(q)$ and $F(q)$ have been the focal point of intense investigations in a multitude of studies~\cite{Cucchieri:2007md,Cucchieri:2010xr,Cucchieri:2007rg,Bogolubsky:2007ud,Bogolubsky:2009dc,Oliveira:2009eh,Oliveira:2018lln,Oliveira:2010xc, Oliveira:2012eh, Bowman:2005vx, Ayala:2012pb,Boucaud:2008ky,Aguilar:2008xm,Dudal:2008sp,Fischer:2008uz,Mitter:2014wpa,Siringo:2015wtx,Huber:2018ned}, while the form factors of $H_{\nu\mu}$ have been computed in a recent work~\cite{Aguilar:2018csq},
using the one-loop dressed approximation of the SDEs they satisfy. 

It turns out, however, that the exercise at hand is considerably more subtle then the simple substitution
of  the aforementioned ingredients into the BC solution.
The first observation suggesting the need for a nonperturbative ``reinterpretation'' of the BC construction  
stems from the fact that the gluon propagator
is infrared finite; then, if the BC parametrization is taken at face value,
one realizes immediately that $\JBC(q)$ diverges at the origin as $\Delta^{-1}(0)/q^2$.
Thus, the ``naive'' use of the BC solution~\cite{Ball:1980ax} in the case of an infrared finite gluon propagator
would give rise to a longitudinal $\fatg_{\alpha\mu\nu}$ plagued with poles, which would diverge 
in the corresponding kinematic limits. 

At first sight, this observation alone may not be considered as sufficient cause for readjusting 
the BC construction; after all, as has been explained in a series of works,
the presence of massless poles in $\fatg_{\alpha\mu\nu}(q,r,p)$ is needed precisely for
obtaining an infrared finite solution out of the gluon SDE~\cite{Aguilar:2017dco,Binosi:2017rwj,Aguilar:2016vin,Aguilar:2015bud,Ibanez:2012zk,Binosi:2012sj,Binosi:2009qm,Aguilar:2008xm}.
The reader must note, however, an important caveat, spelled out in all works cited above: 
the massless poles contained in $\fatg_{\alpha\mu\nu}$, comprising a term to be denoted by 
$V_{\alpha\mu\nu}$, must be of a very special type.
In particular, they must be ``longitudinally coupled'', \ie 
appear {\it exclusively} in the form $q_{\alpha}/q^2$, $r_{\mu}/r^2$, or $p_{\nu}/p^2$,
or products thereof [see~\1eq{longcoupl}]~\cite{Aguilar:2017dco,Binosi:2017rwj,Aguilar:2016vin,Aguilar:2015bud,Ibanez:2012zk,Binosi:2012sj}; and this is clearly not the case
for the poles induced from the naive use of $\JBC(q)$. In fact, while the former
decouple from physical amplitudes and lattice observables, the latter would, in general, persist.

Instead, the self-consistent way to proceed may be briefly described as follows. 
\n{i} One starts by casting the gluon propagator  in the form~\cite{Aguilar:2011ux,Binosi:2012sj}\footnote{Note that, in contradistinction  to the
more familiar case of the quark propagator, 
this particular decomposition into a ``kinetic'' and a ``mass'' term is not mathematically unique~\cite{Aguilar:2014tka}.}
\mbox{$\Delta^{-1}(q) = q^2 J(q) + m^2(q)$},
where \mbox{$\Delta^{-1}(0) = m^2(0)$} (Euclidean space). Evidently, \mbox{$\JBC(q)\neq J(q)$}; in fact, while $\JBC(q)$
diverges as $1/q^2$ at the origin,  $J(q)$ diverges only logarithmically, precisely due to the
presence of massless ghost loops in its diagrammatic representation.

\n{ii} The STIs of $\fatg_{\alpha\mu\nu}(q,r,p)$ [see Eq.~\eqref{eq:sti_delta}] will be realized in a very particular way.
First, the above form of $\Delta^{-1}(q)$ is substituted on their r.h.s.
Then, on the l.h.s, $\fatg_{\alpha\mu\nu}(q,r,p)$ is written as the sum of the pole part, $V_{\alpha\mu\nu}$,
and a remainder, denoted by $\Gamma_{\alpha\mu\nu}$. At this point, given that the
origin of the terms $m^2(q)$ is inextricably connected to the existence of  $V_{\alpha\mu\nu}$,
it is natural to state that the divergence of  $V_{\alpha\mu\nu}$ on the l.h.s. is responsible
for the appearance of the mass terms $m^2(q)$ on the r.h.s., while the divergence of
$\Gamma_{\alpha\mu\nu}$ accounts for the ``kinetic'' terms $J(q)$. Thus, each original STI
is decomposed into two ``partial'' ones, one satisfied by $\Gamma_{\alpha\mu\nu}$ and one by $V_{\alpha\mu\nu}$
[see \2eqs{stig}{stiv}, respectively]~\cite{Aguilar:2017dco,Binosi:2017rwj,Aguilar:2016vin,Aguilar:2015bud,Ibanez:2012zk,Binosi:2012sj,Aguilar:2011xe}.

\n{iii}
This particular decomposition of the STIs converts the original exercise into the following equivalent task.
The partial STIs satisfied by $\Gamma_{\alpha\mu\nu}$ are precisely of the type appearing
in the original BC construction~\cite{Ball:1980ax}; indeed, now, on their r.h.s. one has only terms of the type $q^2 J(q)$,
which, up to the aforementioned logarithms, are well-behaved in the infrared (have no poles).
Thus, the BC construction may be applied {\it mutatis mutandis} for the determination of the
longitudinal part of $\Gamma_{\alpha\mu\nu}$. 

\n{iv} As for  $V_{\alpha\mu\nu}$, its form is {\it completely} determined from the corresponding partial STIs 
that it satisfies, together with the crucial requirement that it be ``longitudinally coupled''.
Its detailed construction and closed form have been worked out in~\cite{Ibanez:2012zk};  
see also \2eqs{longcoupl}{ABC} of the present article, and the related discussion.

In this work we carry out in the construction described in \n{i}-\n{iii}, whose
careful implementation furnishes the 10 longitudinal form factors of $\Gamma_{\alpha\mu\nu}$, 
for general values of their Euclidean momenta. 
The results obtained, in addition to displaying the special features
of general infrared suppression, zero crossing, and logarithmic divergence at the origin, 
compare rather favorably with the lattice data of~\cite{Athenodorou:2016oyh}.

The article is organized as follows. In Sec.~\ref{sec:theory}
we introduce the notation and set up the theoretical framework.
We pay particular attention to the connection between 
$\fatg^{abc}_{\alpha\mu\nu}(q,r,p)$ and the mechanism that
endows the gluons with a dynamical mass, and introduce the two basic components, 
$\Gamma_{\alpha\mu\nu}$ and $V_{\alpha\mu\nu}$, together with the ``partial'' STIs that they satisfy.
In Sec.~\ref{sec:BCS} we present the BC solution for the
longitudinal form factors $X_i$ of 
$\Gamma_{\alpha\mu\nu}$, derived from the aforementioned STIs, and comment
on the constraints imposed by Bose symmetry.
Our main results are presented in Sec.~\ref{sec:num}, where we explain the theoretical origin of    
the inputs used in our analysis, present and discuss several
three-dimensional (3D)  and two-dimensional (2D) plots for the $X_i$,
and compare them with the results of one-loop calculations. 
Next, in Sec.~\ref{sec:comp} we compare our findings  
with those of previous works based on SDEs, 
as well as with the two kinematic configurations obtained from recent lattice simulations.
In Sec.~\ref{furcons} we discuss a series of subtleties related with
the construction developed, paying particular attention to
the distribution and interpretation of the massless poles.
In Sec.~\ref{naiveBC} we elaborate on the complications associated with 
the ``naive'' implementation of the 
BC construction, discussing the necessary adjustments required for its applicability.
Sec.~\ref{sec:conc} is dedicated to a summary of our results, and the discussion of some possible future applications.
Finally, in  Appendixes~\ref{app:pert} and~\ref{changebasis} we present the   one-loop results for the form factors
in the ``totally symmetric''
and ``asymmetric'' configurations, and the transformation rules connecting the BC and the naive bases.

\section{\label{sec:theory} General framework and theoretical foundations}

In this section we introduce  the necessary notation and definitions,
review certain important relations,  
and elaborate on the main conceptual issues associated with the nonperturbative structure of the three-gluon vertex.

Throughout this article we work in the {\it Landau gauge}, where the 
gluon propagator $\Delta^{ab}_{\mu\nu}(q)=\delta^{ab}\Delta_{\mu\nu}(q)$ assumes the completely transverse form, 
\begin{align}
\Delta_{\mu\nu}(q) = -i\Delta(q)P_{\mu\nu}(q)\,, \qquad P_{\mu\nu}(q) = g_{\mu\nu} - \frac{q_\mu q_\nu}{q^2}\,.
\end{align}
In addition, we introduce the ghost propagator, \mbox{ $D^{ab}(q)= \deltaδ^{ab}D(q)$}, whose dressing 
function,  $F(q)$, is given by 
\be\label{eq:ghost_dressing}
D(q) =\frac{iF(q)}{q^2}\,.
\ee

\begin{figure}[ht]
\begin{center}
 \includegraphics[scale=0.7]{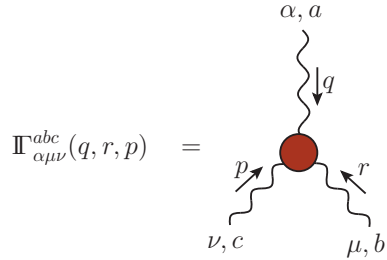}
\end{center}
\caption{The full three-gluon vertex with 
all momenta entering.}\label{fig:3g_Vertex}
\end{figure}

The focal point of the present work is the three-gluon vertex, to be denoted by  
$\fatg^{abc}_{\alpha\mu\nu}(q,r,p) = gf^{abc}\fatg_{\alpha\mu\nu}(q,r,p)$, which 
is diagrammatically represented  in Fig.~\ref{fig:3g_Vertex};
note that all momenta are considered to be incoming, so that $q + p + r = 0$. At tree level, $\fatg_{\alpha\mu\nu}(q,r,p) := \Gamma^{(0)}_{\alpha\mu\nu}(q,r,p)$, 
where 
\be
\Gamma^{(0)}_{\alpha\mu\nu}(q,r,p) = (q-r)_{\nu}g_{\alpha\mu} + (r-p)_{\alpha}g_{\mu\nu} + (p-q)_{\mu}g_{\alpha\nu}\,.
\label{treelevel}
\ee

The vertex $\fatg^{abc}_{\alpha\mu\nu}(q,r,p)$ displays full Bose symmetry,
\ie it remains invariant under the exchange of all ``indices'' associated with any two of its legs, such as, for example, 
$(a,\alpha, q) \leftrightarrow (b,\mu, r)$. This fundamental property, in turn, imposes
nontrivial constraints on the form factors comprising the three-gluon vertex [see Eqs.~\eqref{eq:Xi_more_bose},~\eqref{eq:Xi_bose}, and 
\eqref{eq:Yi_more_bose}]. 

The most important relations for the ensuing analysis 
are the three STIs  that $\fatg_{\alpha\mu\nu}$ satisfies when contracted
by $r^{\mu}$, $q^{\alpha}$, or $p^{\nu}$, given by~\cite{Ball:1980ax} 
\begin{align}
q^\alpha\fatg_{\alpha\mu\nu}(q,r,p) =& F(q)[\Delta^{-1}(p) P^{\alpha}_\nu(p)H_{\alpha\mu}(p,q,r) - \Delta^{-1}(r)P^{\alpha}_\mu(r)H_{\alpha\nu}(r,q,p)] \,, \nonumber\\
r^\mu\fatg_{\alpha\mu\nu}(q,r,p) =& F(r)[\Delta^{-1}(q) P^{\mu}_\alpha(q)H_{\mu\nu}(q,r,p) - \Delta^{-1}(p)P^{\mu}_\nu(p)H_{\mu\alpha}(p,r,q)] \,, \nonumber\\
p^\nu\fatg_{\alpha\mu\nu}(q,r,p) =& F(p)[\Delta^{-1}(r) P^{\nu}_\mu(r)H_{\nu\alpha}(r,p,q) - \Delta^{-1}(q)P^{\nu}_\alpha(q)H_{\nu\mu}(q,p,r)] \,.
\label{eq:sti_delta}
\end{align}

The $H_{\nu\mu}(q,p,r)$ appearing in the above STIs stands for 
the ghost-gluon scattering kernel, whose general Lorentz decomposition is given by~\cite{Ball:1980ax,Davydychev:1996pb}  
\be\label{eq:H}
H_{\nu\mu}(q,p,r) = g_{\mu\nu}A_1 + q_\mu q_\nu A_2 + r_\mu r_\nu A_3 + q_\mu r_\nu A_4 + r_\mu q_\nu A_5\,,
\ee
where the momentum dependence of the form factors, 
$A_i \equiv A_i(q,p,r)$, has been suppressed for compactness.
Notice that, at tree level, $H^{(0)}_{\nu\mu}(q,p,r) = g_{\mu\nu}$, so that $A^{(0)}_1 = 1$ and $A^{(0)}_i = 0$, for $i=2,\ldots,5$. The nonperturbative structure of the form factors $A_i$ is essential for the construction at hand, and 
has been studied in detail in~\cite{Aguilar:2018csq}.

Turning now to the relevant dynamical issues, let us first consider the gluon propagator.
As has been firmly established through a multitude of studies on the lattice~\cite{Cucchieri:2007md,Cucchieri:2010xr,Cucchieri:2007rg,Bogolubsky:2007ud,Bogolubsky:2009dc,Oliveira:2009eh,Oliveira:2018lln,Oliveira:2010xc, Oliveira:2012eh, Bowman:2005vx, Ayala:2012pb} and in the continuum~\cite{Boucaud:2008ky,Aguilar:2008xm,Dudal:2008sp,Fischer:2008uz,Mitter:2014wpa,Siringo:2015wtx,Huber:2018ned}, 
$\Delta(q)$ saturates in the deep infrared at a finite non-vanishing value, \ie \mbox{$\Delta^{-1}(0) =c \neq 0$}, both in the Landau gauge as well as away from it~\cite{Kondo:2001nq,Aguilar:2016ock,Siringo:2018uho,Mintz:2018hhx,Cucchieri:2009kk,Bicudo:2015rma, Cucchieri:2018rex}. 
This characteristic property, in turn, has been interpreted to signal the emergence of a mass scale in the gauge sector of QCD.
Motivated by this interpretation, it is natural to cast $\Delta(q)$ in the form  (Euclidean space)~\cite{Aguilar:2011ux,Binosi:2012sj} 
\be\label{eq:gluon_m_J}
\Delta^{-1}(q) = q^2J(q) + m^2(q)\,,
\ee
where $q^2J(q)$ corresponds to the so-called ``kinetic term'', while $m^2(q)$ to an effective (momentum-dependent)
gluon mass, with the property $m^2(0)=\Delta^{-1}(0)$.

\begin{figure}[ht]
\begin{center}
 	\includegraphics[scale=0.6]{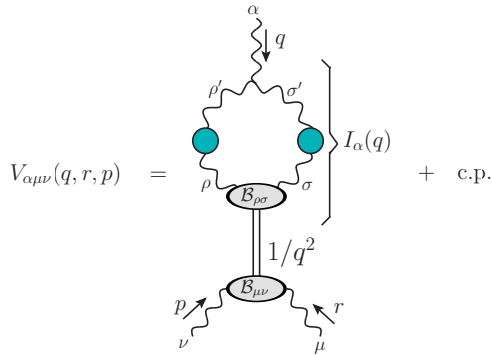}
\end{center}
\caption{ The pole vertex $\Gp_{\alpha\mu\nu}(q,r,p)$ is composed of three main ingredients: the transition amplitude, $I_{\alpha}(q)$, which mixes the gluon with a massless excitation, the propagator of the massless excitation $1/q^2$, while 
  ${\cal B}_{\mu\nu}$ (and ${\cal B}_{\rho\sigma}$)  denotes the proper vertex that couples the massless excitation to a pair of gluons, and 
``c.p.'' stands for ``cyclic permutations''.}
\label{fig:pole}
\end{figure}

The formalism
obtained from the fusion of the Pinch Technique~~\cite{Cornwall:1981zr,Cornwall:1989gv,Pilaftsis:1996fh,Binosi:2002ft,Binosi:2003rr,Binosi:2009qm} with the Background Field Method (PT-BFM)~\cite{DeWitt:1967ub,Honerkamp:1972fd,Kallosh:1974yh,KlubergStern:1974xv,Arefeva:1974jv,Hooft:1975vy,Abbott:1980hw,Abbott:1981ke} is particularly suited for addressing this fundamental question, by means of the special SDE governing the
dynamical evolution of $\Delta(q)$.
Within this latter framework, 
the emergence of a nontrivial $m^2(q)$ (\ie the existence of ``massive'' solutions)
proceeds through a non-Abelian realization of the {\it Schwinger mechanism}~\cite{Schwinger:1962tn,Schwinger:1962tp}, which, 
in the absence of fundamental scalar fields, endows gauge bosons with masses.
The implementation of this mechanism, in turn, hinges crucially  
on the presence of {\it ``longitudinally coupled''} massless poles in the vertex $\fatg_{\alpha\mu\nu}(q,r,p)$, 
which constitutes a key ingredient of the aforementioned gluon SDE~\cite{Jackiw:1973tr,Jackiw:1973ha,Cornwall:1973ts,Eichten:1974et,Poggio:1974qs,Smit:1974je}. 
In particular, $\fatg_{\alpha\mu\nu}(q,r,p)$ is composed by two distinct terms, namely   
\be\label{eq:Gnp}
\fatg_{\alpha\mu\nu}(q,r,p) =\Gnp_{\alpha\mu\nu}(q,r,p) +\Gp_{\alpha\mu\nu}(q,r,p)\,,
\ee
where $\Gp_{\alpha\mu\nu}(q,r,p)$ denotes the part associated with the massless poles, while $\Gnp_{\alpha\mu\nu}(q,r,p)$
captures all remaining contributions.
In what follows we briefly summarize some basic properties of $\Gp_{\alpha\mu\nu}(q,r,p)$, which 
is diagrammatically represented in the Fig.~\ref{fig:pole}; for further details, the reader is referred to the related literature~\cite{Aguilar:2011xe,Binosi:2012sj,Ibanez:2012zk,Binosi:2017rwj,Aguilar:2017dco}.

\n{i} The origin of the massless poles is dynamical rather than kinematic, 
in the sense that, for sufficiently strong binding, 
the mass of certain {\it colored} bound states may be
reduced to zero~\cite{Jackiw:1973tr,Jackiw:1973ha,Cornwall:1973ts,Eichten:1974et,Poggio:1974qs,Smit:1974je}.  
The actual nonperturbative realization of this possibility within a contemporary QCD framework  
has been demonstrated in~\cite{Aguilar:2017dco,Binosi:2017rwj,Aguilar:2016vin,Aguilar:2015bud,Ibanez:2012zk,Binosi:2012sj,Aguilar:2011xe},
where the homogeneous Bethe-Salpeter equation (BSE) that controls the formation of these bound states was investigated.

\n{ii} The term {\it ``longitudinally coupled''} means that $\Gp_{\alpha\mu\nu}(q,r,p)$ assumes the very special form  
\be
\Gp_{\alpha\mu\nu}(q,r,p) = \left(\frac{q_{\alpha}}{q^2}\right)A_{\mu\nu}(q,r,p) +
\left(\frac{r_{\mu}}{r^2}\right)B_{\alpha\nu}(q,r,p) + \left(\frac{p_{\nu}}{p^2}\right)C_{\alpha\mu}(q,r,p)\,,
\label{longcoupl}
\ee
and therefore,  $\Gp_{\alpha\mu\nu}(q,r,p)$ satisfies the crucial relation    
\be
\label{eq:transvp}
P_{\alpha\alpha^{\prime}}(q)P_{\mu\mu^{\prime}}(r)P_{\nu\nu^{\prime}}(p)\Gp^{\alpha\mu\nu}(q,r,p) = 0 \,.
\ee

\n{iii} We emphasize that the form of $\Gp_{\alpha\mu\nu}(q,r,p)$ given in \1eq{longcoupl} emerges automatically in the
dynamical framework put forth in the classic works of~\cite{Jackiw:1973tr,Jackiw:1973ha,Cornwall:1973ts,Eichten:1974et,Poggio:1974qs,Smit:1974je}, and its contemporary variations, studied in~\cite{Aguilar:2017dco,Binosi:2017rwj,Aguilar:2016vin,Aguilar:2015bud,Ibanez:2012zk,Binosi:2012sj}.
In particular, the pole in the $q$-channel is due to the propagation of a massless bound state excitation~\cite{Aguilar:2011xe}, as shown in the diagram of the Fig.~\ref{fig:pole}; the poles in the other two channels
are obtained through the cyclic permutations imposed by the Bose-symmetry of $\Gp_{\alpha\mu\nu}(q,r,p)$.
We emphasize that, within this scenario, 
Lorentz invariance alone forces the saturation of
$q$ by its own Lorentz index $\alpha$; in other words, the special form of \1eq{longcoupl}
is {\it not} assumed, but, instead, emerges automatically.

\n{iv} Note, in addition, that the pivotal property of \1eq{eq:transvp}
guarantees the {\it decoupling} of the massless excitations 
from physical ``on-shell'' amplitudes, as well as its vanishing 
from the {\it transversely projected}  
version of $\fatg_{\alpha\mu\nu}(q,r,p)$,
which constitutes the natural ingredient 
of lattice observables, such as those considered in Sec.~\ref{sec:lattice_comp}
[see \2eqs{projvert}{drop}].

\n{v} A crucial one-to-one correspondence between $m^2(q)$ and $\Gp_{\alpha\mu\nu}(q,r,p)$ is imposed by the 
need to evade, in a STI preserving way, the so-called ``seagull-identity''~\cite{Aguilar:2009ke,Aguilar:2015bud,Aguilar:2016vin}. Specifically,
if one substitutes \1eq{eq:Gnp} into the l.h.s and 
$\Delta^{-1}(q) = q^2J(q) - m^2(q)$ (Minkowski space) into the r.h.s. of \1eq{eq:sti_delta},
the full STI must be realized as the sum of two specific pieces, namely
\be
q^\alpha\Gnp_{\alpha\mu\nu}(q,r,p) = F(q)[p^2J(p) P^{\alpha}_\nu(p)H_{\alpha\mu}(p,q,r) - r^2J(r)P^{\alpha}_\mu(r)H_{\alpha\nu}(r,q,p)] \,,
\label{stig}
\ee
and
\be
q^\alpha\Gp_{\alpha\mu\nu}(q,r,p) = F(q)[m^2(r)P^{\alpha}_\mu(r)H_{\alpha\nu}(r,q,p) - m^2(p) P^{\alpha}_\nu(p)H_{\alpha\mu}(p,q,r)] \,.
\label{stiv}
\ee
Evidently, two additional pairs of similar (cyclically permuted) relations are obtained  
from the other two STIs of \1eq{eq:sti_delta}.

\n{vi}
It turns out that \1eq{eq:transvp}, together with \1eq{stiv} and the other two cyclic relations, determine completely the
form of $\Gp_{\alpha\mu\nu}(q,r,p)$~\cite{Ibanez:2012zk}; in particular,   
\begin{align}
A_{\mu\nu}(q,r,p)&=\frac{F(q^2)}{2}\left\{m^2(r^2)P^\rho_\mu(r)\left[g^\sigma_\nu+P^\sigma_\nu(p)\right]H_{\rho\sigma}(r,q,p)\right.&\nonumber\\
&\left.-m^2(p^2)P^\rho_\nu(p)\left[g^\sigma_\mu+P^\sigma_\mu(r)\right]H_{\rho\sigma}(p,q,r)\right\}\,,&\nonumber\\
B_{\alpha\nu}(q,r,p)&=\frac{F(r^2)}{2}\left\{m^2(p^2) P_\nu^\rho(p) \left[g^\sigma_\alpha+P^\sigma_\alpha(q)\right]H_{\rho\sigma}(p,r,q)\right.&\nonumber\\
&\left.-m^2(q^2) P_\alpha^\rho(q)\left[g^\sigma_\nu+P^\sigma_\nu(p)\right]H_{\rho\sigma}(q,r,p) \right\}\,,&\nonumber\\
C_{\alpha\mu}(q,r,p)&=\frac{F(p^2)}{2} \left\{m^2(q^2) P_\alpha^\rho(q) \left[g_\mu ^{\sigma }+P_\mu^\sigma(r)\right]H_{\rho\sigma}(q,p,r) \right.&\nonumber\\
&\left.-m^2(r^2) P_\mu^\rho(r)\left[g_{\alpha}^\sigma+P_\alpha^\sigma(q)\right]H_{\rho\sigma}(r,p,q)\right\} \,.&
\label{ABC}
\end{align}
Clearly, the substitution of the above terms into \1eq{longcoupl}
gives rise to a $\Gp_{\alpha\mu\nu}(q,r,p)$ that is manifestly Bose-symmetric.
Note that $\Gp_{\alpha\mu\nu}(q,r,p)$ contains single-, double-, and triple-pole terms, such as, for example,
$\frac{q_{\alpha}g_{\mu\nu}}{q^2}$, $\frac{q_{\alpha}r_{\mu}r_{\nu}}{q^2r^2}$, and $\frac{q_{\alpha}r_{\mu}p_{\nu}}{q^2r^2p^2}$~\cite{Ibanez:2012zk}.
On the other hand, terms having their momenta and Lorentz indices ``mismatched'', \eg $\frac{p_{\alpha}g_{\mu\nu}}{q^2}$ or
$\frac{r_{\alpha}p_{\mu}q_{\nu}}{q^2r^2p^2}$, are absent.

After this brief review, we return to the central objective of this work, namely  
the implementation of the BC solution in order to reconstruct the
longitudinal part of $\Gnp_{\alpha\mu\nu}(q,r,p)$ from \1eq{stig} and the other two similar
STIs obtained from the contraction by $r^\mu$ and $p^\nu$.
However, before embarking into the technical details of this construction, it is important to clarify 
an essential point regarding the central ingredient of the BC solution for $\Gnp_{\alpha\mu\nu}(q,r,p)$,
namely the function $J(q)$, and, in particular, the way in which it may be actually computed.

To that end, note that the special decomposition of the $\fatg_{\alpha\mu\nu}(q,r,p)$
given in \1eq{eq:Gnp} leads to the separation of the SDE for $\Delta(q)$ into two individual 
but coupled integral equations, governing the evolution of $J(q)$ and $m^2(q)$~\cite{Binosi:2012sj,Aguilar:2014tka}.
It turns out that the components of the equation that determines $m^2(q)$ are considerably better known than
those entering in the equation for $J(q)$; in particular, the four-gluon vertex drops practically out from the
former, but is present in the latter. 
Therefore, given these practical limitations, one proceeds as follows.
First, the equation for $m^2(q)$ is solved in isolation, using as input the lattice data for $\Delta(q)$,
together with certain simplifying assumptions related to multiplicative renormalization.
Then, one employs \1eq{eq:gluon_m_J} once again, and obtains 
$q^2 J (q)$ by {\it subracting} the solution for $m^2(q)$ from the lattice data for $\Delta^{-1}(q)$~\cite{Bogolubsky:2007ud}.
A different, but theoretically equivalent procedure, involves the derivation of a special
BSE, whose solution is identified with the first derivative of $m^2(q)$~\cite{Aguilar:2011xe,Binosi:2017rwj};
then, numerical integration furnishes $m^2(q)$, and its subtraction from $\Delta^{-1}(q)$, exactly as before, furnishes
$q^2 J (q)$. In Sec.~\ref{sec:num}  we will further elaborate on the structure of $J(q)$, its characteristic properties, and the
uncertainties in its determination.


\section{\label{sec:BCS}The Ball-Chiu solution for $\Gnp_{\alpha\mu\nu}(q,r,p)$}

For the actual construction of the vertex  $\Gnp_{\alpha\mu\nu}(q,r,p)$, let us cast it in the form
\be
\Gnp^{\alpha\mu\nu}(q,r,p) = \GL^{\alpha\mu\nu}(q,r,p) + \GT^{\alpha\mu\nu}(q,r,p)\,,
\ee
where the ``longitudinal'' part, $\GL^{\alpha\mu\nu}(q,r,p)$, saturates the relevant STIs [Eq.~\eqref{stig} and c.p. thereof], 
while the totally ``transverse'' part, $\GT^{\alpha\mu\nu}(q,r,p)$, satisfies
\be  
q_{\alpha}\GT^{\alpha\mu\nu}(q,r,p) = r_{\mu}\GT^{\alpha\mu\nu}(q,r,p) = p_{\nu} \GT^{\alpha\mu\nu}(q,r,p) =0 \,.
\label{stransv}
\ee
For the explicit tensorial decomposition of $\GL^{\alpha\mu\nu}(q,r,p)$ and $\GT^{\alpha\mu\nu}(q,r,p)$
we will employ the Bose symmetric basis introduced in~\cite{Ball:1980ax}.
Specifically, 
\be
\GL^{\alpha\mu\nu}(q,r,p) = \sum_{i=1}^{10} X_i(q,r,p) \ell_i^{\alpha\mu\nu} \,,
\label{eq:3g_sti_structure}
\ee
where the tensors $\ell_i^{\alpha\mu\nu}$ are given by
\be
\begin{tabular}{lll}
$\ell_1^{\alpha\mu\nu} =  (q-r)^{\nu} g^{\alpha\mu}\,,$
& 
$\ell_2^{\alpha\mu\nu} =  - p^{\nu} g^{\alpha\mu}\,,$\hspace{.75cm}
&
$\ell_3^{\alpha\mu\nu} =  (q-r)^{\nu}[q^{\mu} r^{\alpha} -  (q\cdot r) g^{\alpha\mu}]\,, $\\
$\ell_4^{\alpha\mu\nu} = (r-p)^{\alpha} g^{\mu\nu}\,,$
&
$\ell_5^{\alpha\mu\nu} =  - q^{\alpha} g^{\mu\nu}\,,$
&
$\ell_6^{\alpha\mu\nu} =  (r-p)^{\alpha}[r^{\nu} p^{\mu} -  (r\cdot p) g^{\mu\nu}]\,,$
\\
$\ell_7^{\alpha\mu\nu} =  (p-q)^{\mu} g^{\alpha\nu}\,,$
&
$\ell_8^{\alpha\mu\nu} = - r^{\mu} g^{\alpha\nu}\,,$
&
$\ell_9^{\alpha\mu\nu} = (p-q)^{\mu}[p^{\alpha} q^{\nu} -  (p\cdot q) g^{\alpha\nu}]\,,$
\\
$\ell_{10}^{\alpha\mu\nu} = q^{\nu}r^{\alpha}p^{\mu} + q^{\mu}r^{\nu}p^{\alpha}$\,, & &
\end{tabular}
\label{li}
\ee
and 
\be
\GT^{\alpha\mu\nu}(q,r,p) = \sum_{i=1}^{4}Y_i(q,r,p)t_i^{\alpha\mu\nu} \,,
\label{eq:3g_tr_structure}
\ee
with the $t_i^{\alpha\mu\nu}$ given by
\begin{align}
t_1^{\alpha\mu\nu} =& [(q\cdot r)g^{\alpha\mu} - q^{\mu}r^\alpha][(r\cdot p)q^\nu - (q\cdot p)r^\nu]\,,
\nonumber\\
t_2^{\alpha\mu\nu} =& [(r\cdot p)g^{\mu\nu} - r^{\nu}p^\mu][(p\cdot q)r^\alpha - (r\cdot q)p^\alpha]\,,
\nonumber\\
t_3^{\alpha\mu\nu} =& [(p\cdot q)g^{\nu\alpha} - p^{\alpha}q^\nu][(q\cdot r)p^\mu - (p\cdot r)q^\mu]\,,
\nonumber\\
t_4^{\alpha\mu\nu} =& g^{\mu\nu}[ (r\cdot q)p^\alpha - (p\cdot q)r^\alpha ] + g^{\nu\alpha}[ (p\cdot r)q^\mu - (q\cdot r)p^\mu ] + g^{\alpha\mu}[ (q\cdot p)r^\nu - (r\cdot p)q^\nu ]
\nonumber\\
& + r^\alpha p^\mu q^\nu - p^\alpha q^\mu r^\nu \,.
\label{ti}
\end{align}

At tree level, the only nonvanishing form factors are
\begin{align}
X^{(0)}_1(q,r,p) =  X^{(0)}_4(q,r,p) = X^{(0)}_7(q,r,p) = 1 \,. \nonumber
\end{align}
%

Bose symmetry with respect to the three legs requires that $\GL$ reverses sign under the interchange of the corresponding Lorentz indices and momenta
(remember that the color factor $f^{abc}$ has been factored out); this, in turn, imposes the following relations under the exchange of arguments~\cite{Ball:1980ax}
\begin{align}
X_1(q,r,p) &= X_1(r,q,p)\,, \quad    X_2(q,r,p) = - X_2(r,q,p)\,, \quad   X_3(q,r,p) = X_3(r,q,p)\,,   \nonumber    \\
X_4(q,r,p) &= X_4(q,p,r)\,, \quad   X_5(q,r,p) = - X_5(q,p,r)\,, \quad     X_6(q,r,p) = X_6(q,p,r)\,,       \nonumber\\
X_7(q,r,p) &= X_7(p,r,q)\,, \quad X_8(q,r,p) = - X_8(p,r,q)\,, \quad  X_9(q,r,p) = X_9(p,r,q)\,, \label{eq:Xi_more_bose} \\
X_{10}(q,r,p) &= - X_{10}(r,q,p), \; X_{10}(q,r,p) = - X_{10}(q,p,r),  \; X_{10}(q,r,p) = - X_{10}(p,r,q). \nonumber
\end{align}

In addition, Bose symmetry furnishes the following relations between different form factors~\cite{Ball:1980ax}
\begin{align}
X_4(q,r,p) &= X_1(r,p,q)\,, \qquad  X_5(q,r,p) = X_2(r,p,q)\,,  \qquad  X_6(q,r,p) = X_3(r,p,q)\,, \nonumber\\
X_7(q,r,p) &= X_1(p,q,r)\,, \qquad X_8(q,r,p) = X_2(p,q,r)\,,  \qquad X_9(q,r,p) = X_3(p,q,r) \,,
\label{eq:Xi_bose}
\end{align}
which reduce the number of independent form factors from the original ten to only four, namely  $X_1$, $X_2$, $X_3$, and $X_{10}$.
In particular, if the dependence of $X_1$, $X_2$, $X_3$ on $(q,r,p)$ could be cast in a closed functional form,
then all other $X_i$ would be obtained from them through a simple
interchange of the appropriate momenta, according to \1eq{eq:Xi_bose}.
However, in practice, 
$X_1$, $X_2$, $X_3$, and $X_{10}$ are computed numerically, and the reconstruction of the remaining $X_i$
requires a modest amount of additional numerical
effort; a concrete example of how to obtain $X_4$ from $X_1$ will be given in Sec.~\ref{otherxi}.

For the transverse part, Bose symmetry implies that~\cite{Ball:1980ax}
\begin{align}
Y_1(q,r,p) &= Y_1(r,q,p)\,, \qquad Y_2(q,r,p) = Y_2(q,p,r)\,, \qquad Y_3(q,r,p) = Y_3(p,r,q)\,,\label{eq:Yi_more_bose}
\end{align}
and 
\begin{align}
Y_2(q,r,p) &= Y_1(r,p,q)\,, \qquad Y_3(q,r,p) = Y_1(p,q,r) \,.
\label{eq:Yi_bose}
\end{align}
Therefore, there are only two independent transverse form factors, $Y_1$ and $Y_4$.

The form factors $X_i$ are fully determined in terms of the $A_j$, $F(q)$ and $J(q)$ by solving the system of linear equations generated by the identity given in Eq.~\eqref{stig} and its  cyclic permutations.  
Specifically, the solutions for $X_1$, $X_2$, $X_3$ and $X_{10}$, first obtained in Ref.~\cite{Ball:1980ax}, read
\begin{align}
X_1(q,r,p) &= \frac{1}{4}[2(a_{pqr} + a_{prq}) + p^2(b_{qrp} + b_{rqp}) + 2(\,q\cdot p\,d_{prq} + \,r\cdot p\,d_{pqr})
\nonumber\\
&\qquad + (q^2-r^2)(b_{rpq} + b_{pqr} - b_{qpr} - b_{prq})]\,, \nonumber
\\
X_2(q,r,p) &= \frac{1}{4}[2(a_{prq} - a_{pqr}) - (q^2 - r^2)(b_{qrp} + b_{rqp}) + 2(\,q\cdot p\,d_{prq} - \,r\cdot p\,d_{pqr})
\nonumber\\
&\qquad + p^2(b_{prq} - b_{pqr} + b_{qpr} - b_{rpq})]\,,
\nonumber
\\
X_3(q,r,p) &= \frac{1}{q^2 - r^2}[a_{rpq} - a_{qpr} + r\cdot p \, d_{qpr} - q\cdot p \, d_{rpq}]\,, \nonumber
\\
X_{10}(q,r,p) &= - \frac{1}{2}[b_{qrp} + b_{rpq} + b_{pqr} - b_{qpr} - b_{rqp} - b_{prq}]\,,
\label{eq:X_sol}
\end{align} 
where we introduce the following compact notation 
\begin{align}
a_{qrp} \equiv & F(r)J(p)A_1(p,r,q)\,,
\nonumber\\
b_{qrp} \equiv & F(r)J(p)A_3(p,r,q)\,,
\nonumber\\
d_{qrp} \equiv & F(r)J(p)[A_4(p,r,q)-A_3(p,r,q)]\,. 
\label{eq:ball_chiu_functions}
\end{align}

Clearly, these expressions satisfy the exchange symmetries of Eq.~\eqref{eq:Xi_more_bose}. The remaining six $X_i$ may be computed by permuting the arguments, according to Eq.~\eqref{eq:Xi_bose}. 

Notice that if the contributions from the ghost sector are turned off, 
by setting  \mbox{$H_{\nu\mu}(q,p,r)= g_{\nu\mu}$} and \mbox{$F(q)=1$} into Eqs.~\eqref{eq:X_sol}, 
we obtain the ``abelianized'' form factors, \mbox{${\widehat X}_i(q,r,p)$} (in Minkowski space), 
\begin{align}
{\widehat X}_1(q,r,p) &= \frac{1}{2}[ J(r) + J(q) ] \,,  \qquad   
{\widehat X}_3(q,r,p) = \frac{ [ J(q) - J(r) ] }{q^2-r^2}\,, 
\nonumber \\ 
{\widehat X}_2(q,r,p) &= \frac{1}{2}[ J(q) - J(r) ] \,. \qquad
{\widehat X}_{10}(q,r,p) = 0 \,,
\label{eq:X10_mBC}
\end{align} 
Evidently, the above expressions display the correct Bose symmetry properties required by Eqs.~\eqref{eq:Xi_more_bose} and \eqref{eq:Xi_bose}. 
In fact, this is exactly the result of the BC construction for 
a vertex with three ``background'' gluons, which satisfies ``abelian'' Ward identities~\cite{Cornwall:1989gv,Binosi:2011wi,Binosi:2013rba,Binger:2006sj};
of course, in that case, the additional replacement $J(q)\to J (q) [1+G(q)]^{-2}$ must be carried out,
where the function $1+G(q)$ has been studied extensively in the literature~\cite{Grassi:1999tp,Binosi:2009qm,Aguilar:2008xm,Binosi:2014aea,Binosi:2016nme}.

Let us finally emphasize that, as long as the quantities appearing on the  r.h.s. of \1eq{eq:X_sol} are properly renormalized,
the resulting $X_i(q,r,p)$ will be free of ultraviolet divergences. This is indeed the case, given that
$F(q)$, $J(q)$, and the form factors $A_i(q,p,r)$ have been duly renormalized~\cite{Aguilar:2018csq}, using a particular version of the
general momentum subtraction (MOM) scheme, known as ``Taylor scheme'' \cite{Boucaud:2008gn}.

In particular, let the renormalization constants of the gluon and ghost propagators and the ghost-gluon and three-gluon vertices be defined as~\cite{Aguilar:2013xqa} 
\begin{align}
\Delta_{\s R}(q) =& Z_A^{-1}\Delta(q) \,, \nonumber \\
F_{\!\s R}(q) =& Z_c^{-1}F(q) \,, \nonumber\\
\Gamma^{\mu}_{\!\s R}(q,p,r) =& Z_1 \Gamma^{\mu}(q,p,r) \,, \nonumber\\
\Gamma_{\!\s R}^{\alpha\mu\nu}(q,r,p) =& Z_3 \Gamma^{\alpha\mu\nu}(q,r,p) \,,
\end{align}
where the subscript ``$R$'' denotes renormalized quantities.

In the Taylor scheme, the renormalization constants $Z_A$ and $Z_c$ are defined by imposing
tree-level values for the propagators at $\mu$,  \ie 
$F(\mu)=1$ and $J(\mu)=1$,\footnote{Strictly speaking, given the form of $\Delta^{-1}(q)$ in \1eq{eq:gluon_m_J}, at $\mu$ one must impose the condition 
  \mbox{$\Delta^{-1}(\mu) = \mu^2 + m^2(\mu)$}, which yields $J(\mu)=1$; however, 
  in practice,  the same result emerges by imposing simply $\Delta^{-1}(\mu) = \mu^2$, given that, at $\mu = 4.3$ GeV,
  $m^2(\mu)$ is negligible.} while $A_1$ assumes its tree-level value in the ``soft-ghost'' kinematics,
where Taylor's theorem is valid (Landau gauge), and therefore one sets $Z_1 = 1$~\cite{Taylor:1971ff}.
Then, the remaining renormalization constant, $Z_3$,
is completely determined by appealing to the STI of Eq.~\eqref{eq:sti_delta}, which implies that
\be 
Z_3 = Z_A Z_1 Z_c^{-1} \,.
\label{eq:renorm_sti}
\ee

As as consequence of this particular choice, 
the results for the three-gluon vertex will not match exactly those obtained by renormalizing 
$X_1$ in the ``totally symmetric'' configuration, $q^2=p^2=r^2 =\mu^2$, often used in the literature.
Note, however, that, as has been explicitly shown in~\cite{Aguilar:2018csq}, when $X_1$ is renormalized in
the Taylor scheme and subsequently evaluated at the symmetric point, it departs from unity only by about 3\% (for $\mu =4.3$ GeV).

\section{\label{sec:num} Numerical analysis and main results}

In  this  section  we  present  the  numerical  analysis  and  main  results  for 
the form factors $X_1,$ $X_2$, $X_3$, and $X_{10}$, defined by Eq.~\eqref{eq:X_sol}.

\subsection{\label{subsec:ingredients} Inputs}

As can be observed from Eq.~\eqref{eq:X_sol}, the numerical evaluation of various $X_i$
requires  the  knowledge  of  the following additional  quantities:  
 ({\it i}) the  ghost  dressing  function, $F(q)$, ({\it ii}) the kinetic part of 
the gluon propagator,  $J(q)$, and ({\it iii})  the  form factors $A_1$, $A_3$, and $A_4$
of $H^{\nu\mu}(q,p,r)$. 
It is important to emphasize that  throughout this work, the renormalization point
will be fixed at \mbox{$\mu = 4.3 \text{ GeV}$}, and we will use \mbox{$\alpha_s(\mu) \equiv g^2/4\pi = 0.22$}.
In what follows, we will specify the main characteristics of $F(q)$, $J(q)$, and the $A_i$, which will be treated  as external inputs.
\begin{figure}[!ht]
\begin{minipage}[b]{0.45\linewidth}
\centering
\includegraphics[scale=0.32]{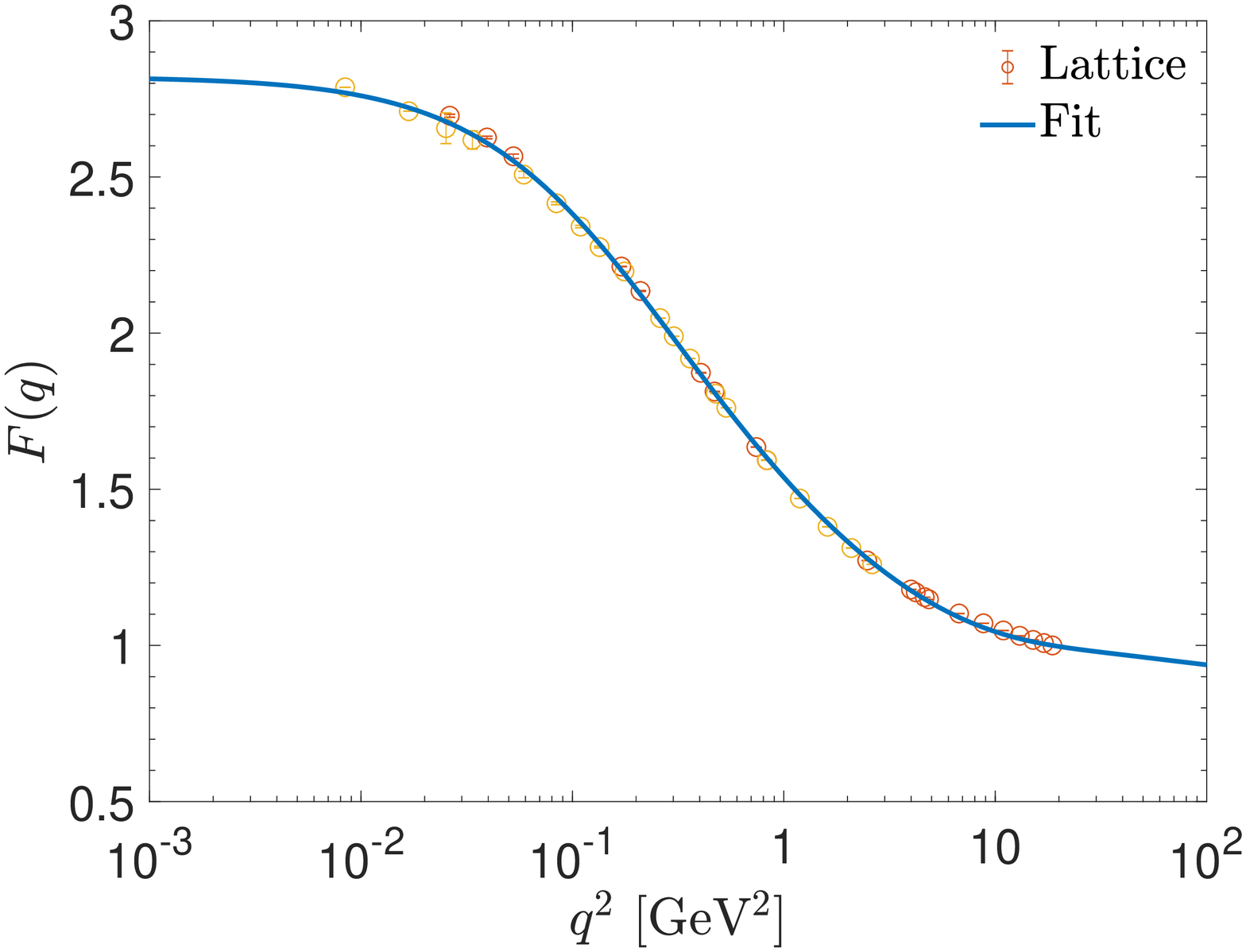}
\end{minipage}
\hspace{0.25cm}
\begin{minipage}[b]{0.45\linewidth}
\includegraphics[scale=0.32]{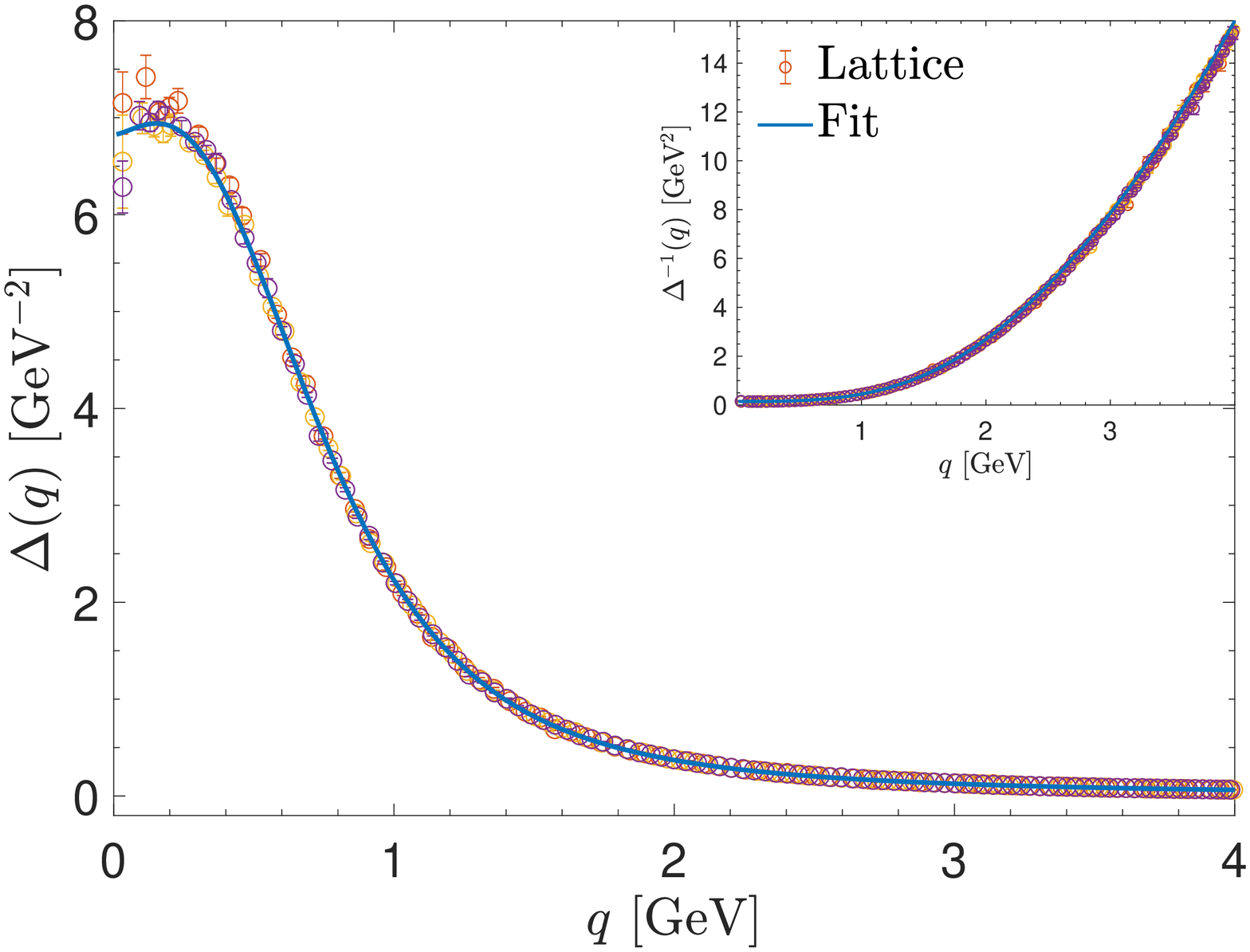}
\end{minipage}
\caption{The fits for  $F(q)$ (left panel) and $\Delta(q)$ (right panel) given by Eq.~\eqref{eq:ghost_logs} and the combination 
of  Eqs.~\eqref{eq:gmass} and~\eqref{eq:J_logs}, respectively (blue continuous curves). In the inset we show the inverse of the gluon propagator, $\Delta^{-1}(q)$. Note that, on theoretical grounds, $\Delta(q)$ must display a 
small local maximum in the infrared, which is discernible in the lattice data (around $q_{*} \approx 250$ MeV), and is duly captured by the fit; however, the corresponding minimum of $\Delta^{-1}(q)$ is not visible, given that \mbox{$d\Delta^{-1}(q)/dq^2 = - \Delta^{-2}(q)[d\Delta^{-1}(q)/dq^2]$}, and $\Delta^{-2}(q_{*}) \approx 1/50$. 
The lattice data are from~\cite{Bogolubsky:2007ud}.}
\label{fig:propagators}
\end{figure}

({\it i})
For the ghost dressing function $F(q)$ we employ a physically motivated fit of the solution obtained 
from the ghost SDE~\cite{Aguilar:2013xqa}, which is in excellent agreement with the lattice data of~\cite{Bogolubsky:2007ud}.
Specifically, the fit for $F(q)$ (in Euclidean space) is given by~\cite{Aguilar:2018epe,Aguilar:2018csq}
\be
F^{-1}(q) = 1 + \frac{9 C_\mathrm{A}\alpha_s}{48\pi}\left[ 1 + D\exp( - \rho_4 q^2 ) \right]\ln\left( \frac{q^2 + \rho_3 M^2(q)}{\mu^2} \right) \,, 
\label{eq:ghost_logs}
\ee
where
\be\label{eq:dyn_mass}
M^2(q) = \frac{m^2}{ 1 + q^2/{\rho_2^2}} \,,
\ee
and the fitting parameters are given by
\mbox{$m^2 = 0.16\,\text{GeV}^2$}, \mbox{$\rho_2^2 = 0.69\,\text{GeV}^2$}, \mbox{$\rho_3 = 0.89$}, \mbox{$\rho_4 = 0.12 \text{ GeV}^{-2}$} and \mbox{$D = 2.36$}. In  the left panel of Fig.~\ref{fig:propagators}, we show 
the  lattice data for $F(q)$  together
with its corresponding fit given by Eq.~\eqref{eq:ghost_logs}. Clearly, we see that  Eq.~\eqref{eq:ghost_logs} recovers the one-loop result for $F(q)$ for large values of $q^2$. 

({\it ii}) The way how the quantity $J(q)$ is obtained is considerably more subtle. In particular, as already mentioned at the end of Sec.~\ref{sec:theory},
the derivation of $J(q)$ is indirect, in the sense that one first obtains $m^2(q^2)$ 
and then subtracts it from the lattice data for $\Delta^{-1}(q)$~\cite{Bogolubsky:2007ud}, shown in the inset of Fig.~\ref{fig:propagators}.

In fact, two different but theoretically equivalent procedures~\cite{Ibanez:2012zk}
for obtaining the form of $m^2(q^2)$ have been developed in the literature:
({\it a}) one begins from the general SDE governing $\Delta(q)$, and, through a judicious separation of terms, derives 
an integral equation for $m^2(q^2)$~\cite{Binosi:2012sj,Aguilar:2014tka}, and ({\it b}) one solves the 
BSE responsible for the formation of the bound-state  poles~\cite{Aguilar:2011xe,Binosi:2017rwj}; the corresponding wave-function is known to coincide with the first derivative of $m^2(q^2)$~\cite{Aguilar:2011xe}, from which
$m^2(q^2)$ may be computed through simple numerical integration.
In practice, due to the approximations implemented~\cite{Ibanez:2012zk,Binosi:2017rwj},
these two methods yield very similar, but not identical results for $m^2(q^2)$; the common feature of all solutions
is that they are positive definite and monotonically decreasing, displaying the characteristic power-law running in the ultraviolet. 

In particular, the functional form of $m^2(q^2)$ can be accurately represented as~\cite{Aguilar:2017dco} 
\be 
m^2(q^2) =  \frac{m^2_0}{1+(q^2/\rho_m^2)^{1 + \gamma}} \,, 
\label{eq:gmass}
\ee
where  \mbox{$m^2_0=0.147\,\mbox{GeV}^2$} and the values of $\rho_m^2$ and $\gamma$
vary depending on the truncations employed.
For the purposes of the present article, we will consider that  $\rho_m^2 =1.18 \,\mbox{GeV}^2$ and we will vary $\gamma$  in the range
$[0,0.3]$. 

\begin{figure}[!h]
\begin{minipage}[b]{0.45\linewidth}
\centering
\includegraphics[scale=0.32]{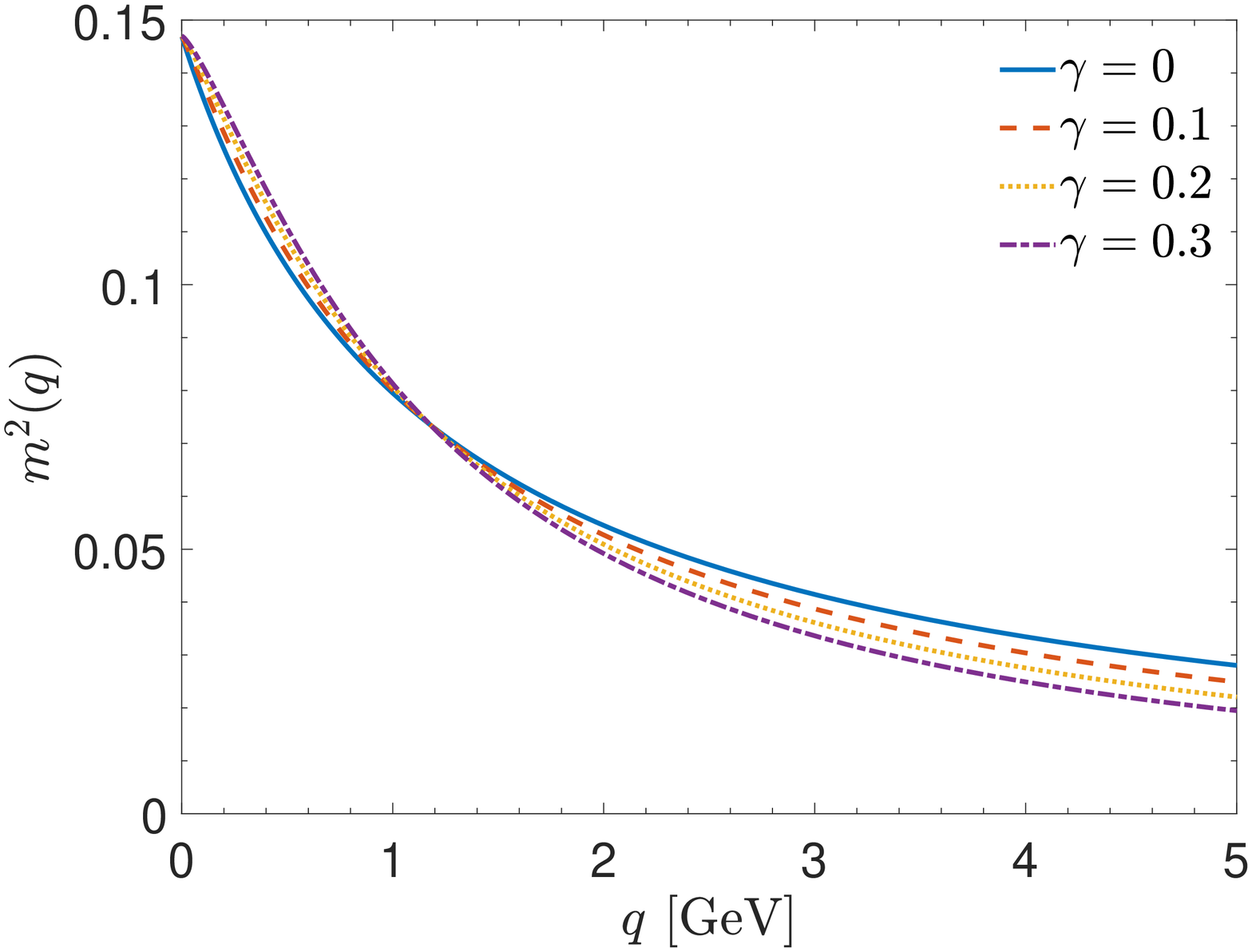}
\end{minipage}
\hspace{0.25cm}
\begin{minipage}[b]{0.45\linewidth}
\includegraphics[scale=0.32]{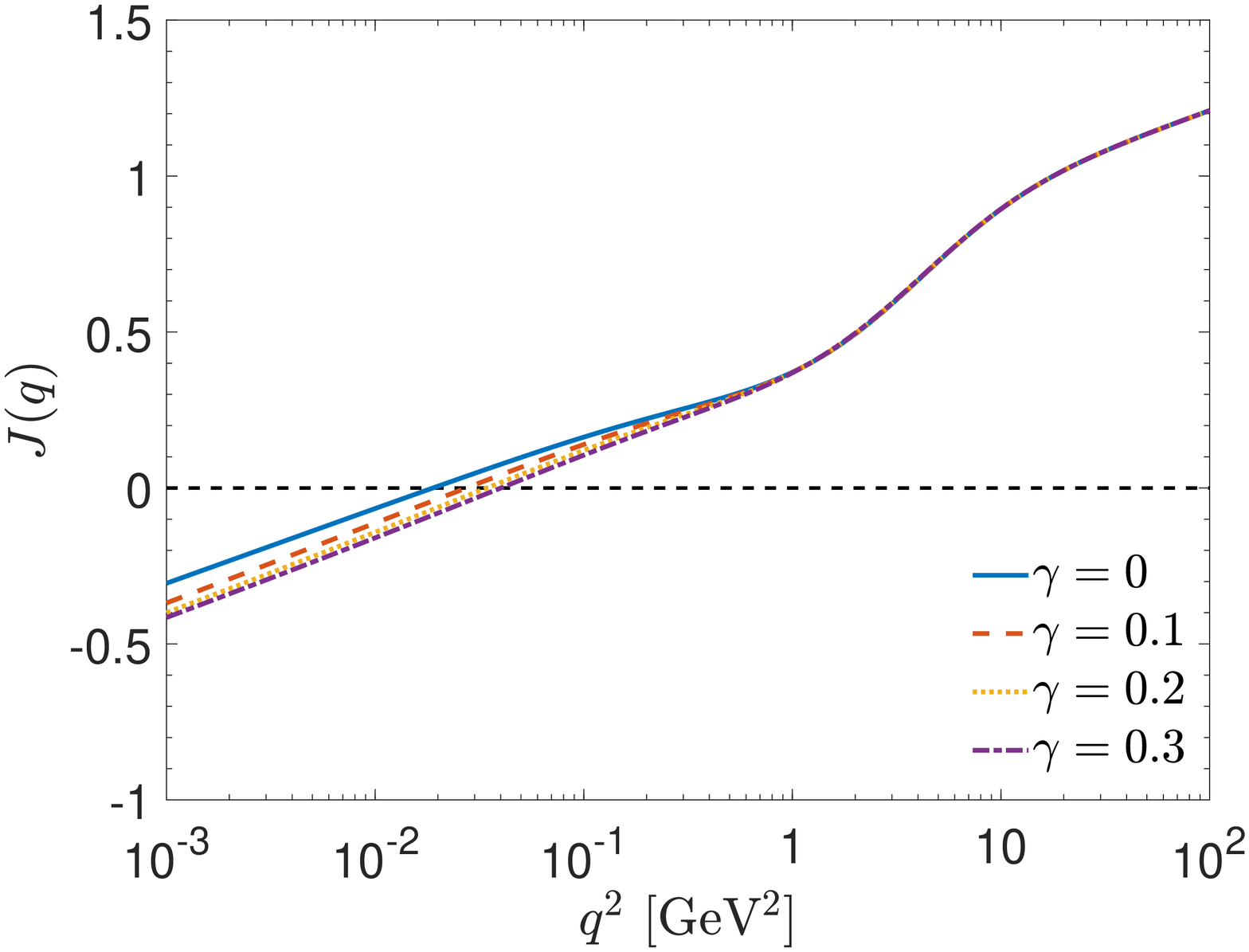}
\end{minipage}
\caption{The dynamical gluon mass, $m^2(q)$ given by Eq.~\eqref{eq:gmass} (left panel),  
and the corresponding kinetic term $J(q)$ given by Eq.~\eqref{eq:J_logs} (right panel). The curves were obtained 
for values of $\gamma$ in the range $[0,0.3]$.}
\label{fig:jm}
\end{figure}

In the left panel of Fig.~\ref{fig:jm}, we plot Eq.~\eqref{eq:gmass} for
different values of $\gamma$ in the range $[0,0.3]$, while in the 
right panel, we show the corresponding  $J(q)$, obtained after subtracting 
the $m^2(q)$ given by Eq.~\eqref{eq:gmass} from the lattice data $\Delta^{-1}(q)$, according to Eq.~\eqref{eq:gluon_m_J}.
All curves for $J(q)$ may be 
parametrized by the same functional form 
\be\label{eq:J_logs}
J(q) = 1 + \frac{C_\mathrm{A}\alpha_s}{4\pi}\left( 1 + \frac{\tau_1}{q^2 + \tau_2} \right)\left[ 2\ln\left(  \frac{q^2 + \rho m^2(q)}{\mu^2} \right) + \frac{1}{6}\ln\left( \frac{q^2}{\mu^2} \right) \right] \,,
\ee
where the $\gamma$ dependence of \mbox{$\rho \equiv \rho(\gamma)$}, \mbox{$\tau_1\equiv \tau_1(\gamma)$} and \mbox{$\tau_2\equiv \tau_2(\gamma)$}  has been suppressed for compactness. For values of $\gamma$ in the range $[0,0.3]$, these functions can be represented by 
\bea
\rho(\gamma) &=&  100.8 -82.21\gamma^{1.28} \,, \nonumber \\
\tau_1(\gamma) &=& 9.87 -6.96\gamma\,, \nonumber \\
\tau_2(\gamma) &=& 0.80 +0.11\exp(-10 \gamma)\,.
\label{pfit}
\eea

\begin{figure}[!htb]
\minipage{0.32\textwidth}
  \includegraphics[width=\linewidth]{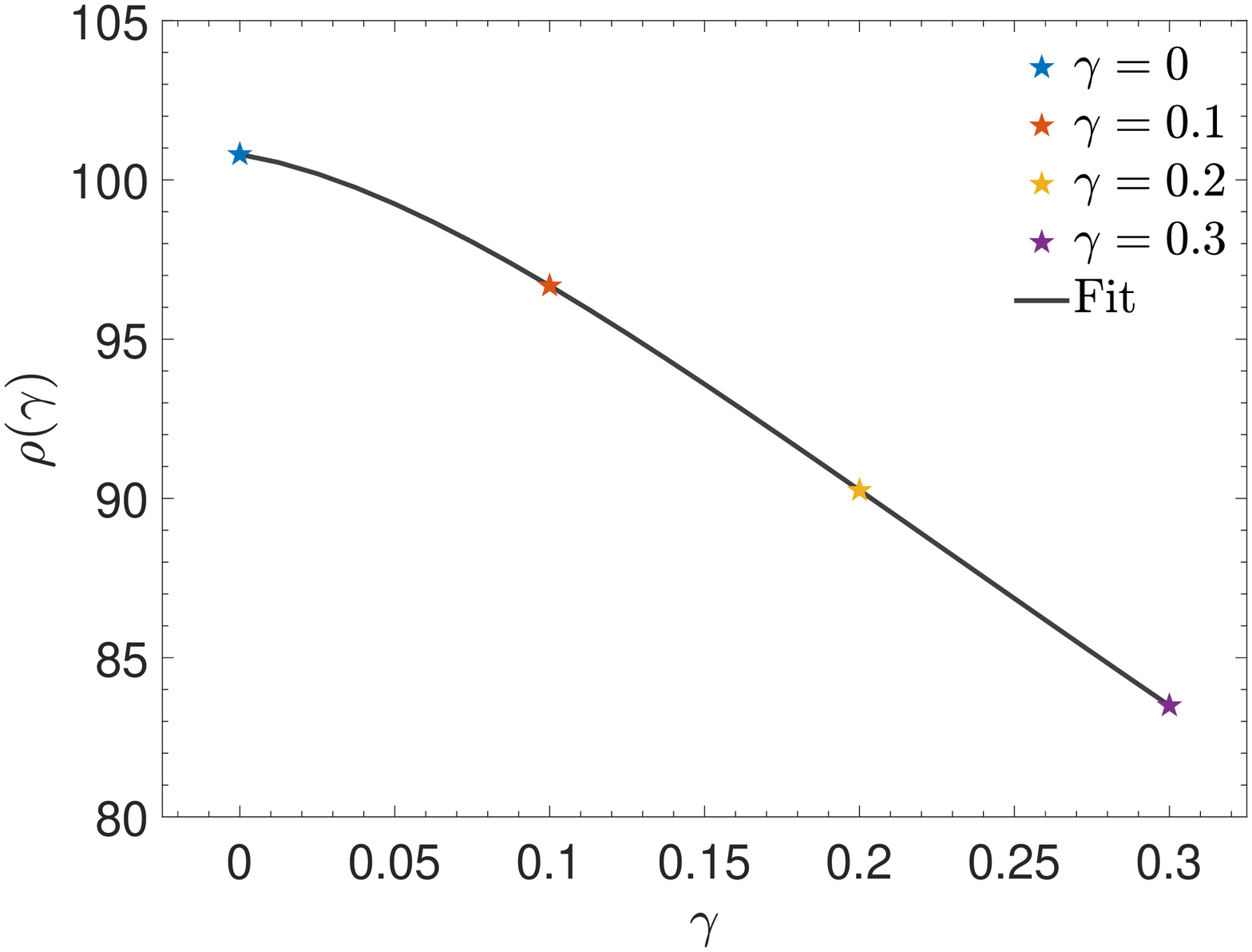}
\endminipage\hfill
\minipage{0.32\textwidth}
  \includegraphics[width=\linewidth]{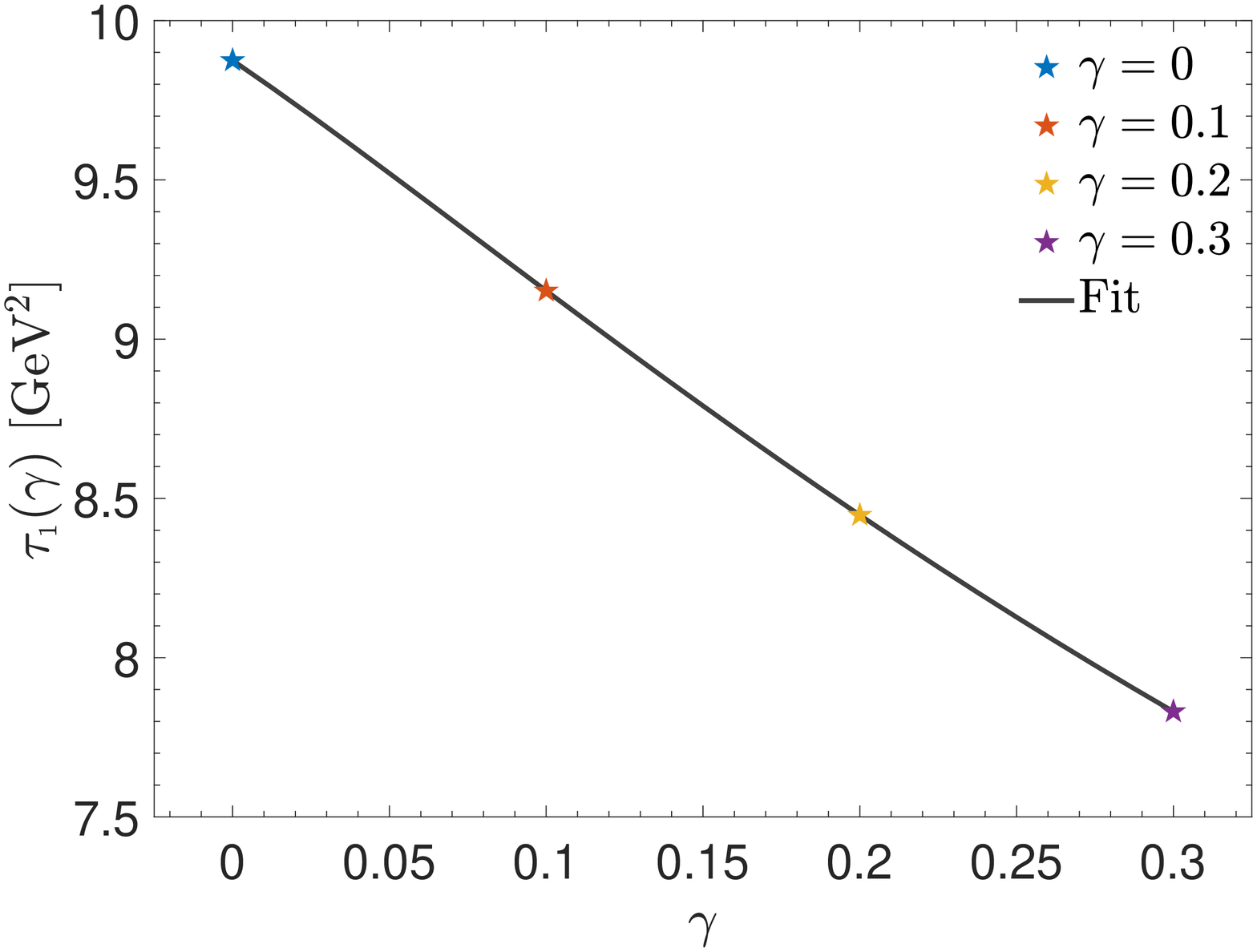}
\endminipage\hfill
\minipage{0.32\textwidth}%
  \includegraphics[width=\linewidth]{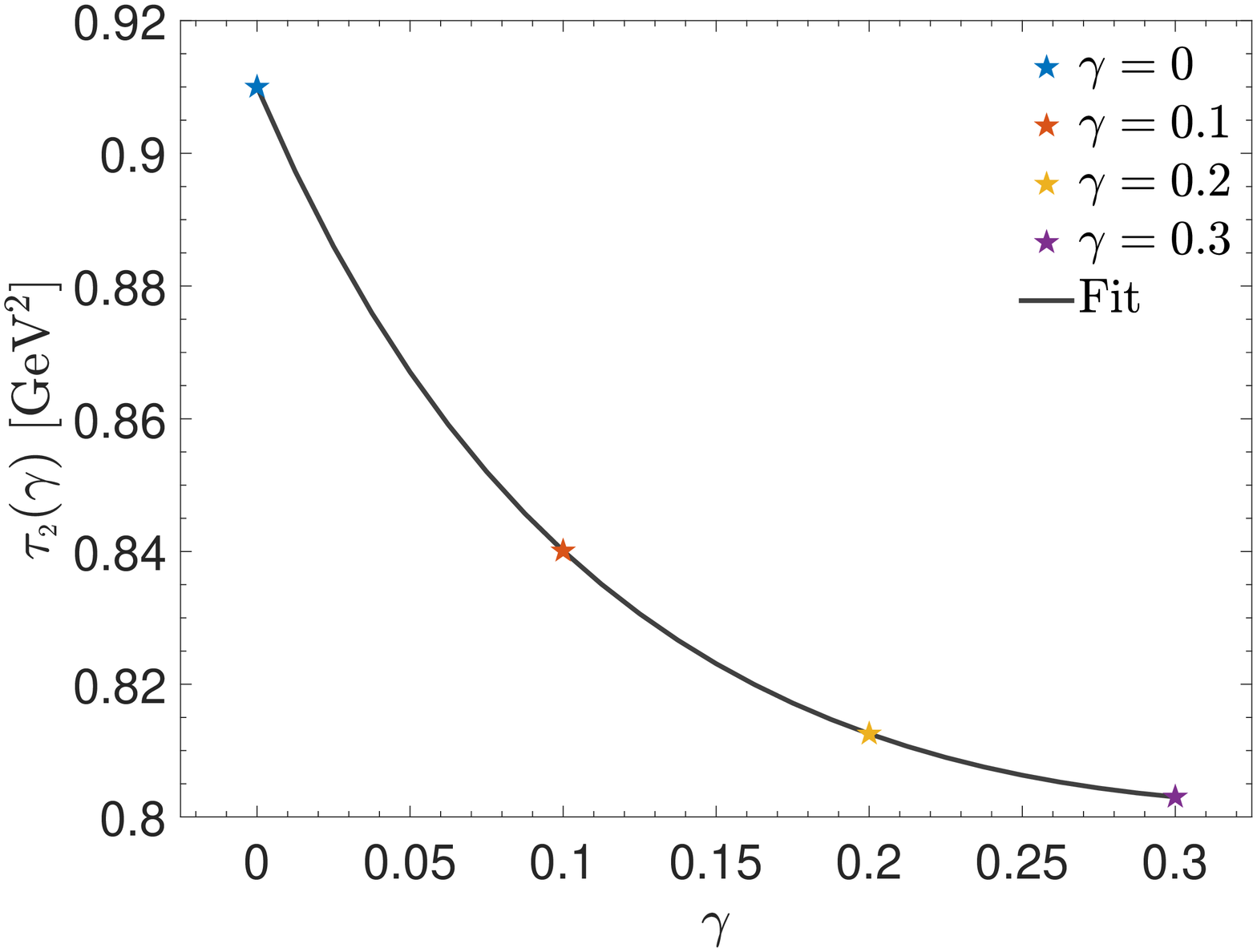}
\endminipage
\caption{The functions $\rho(\gamma)$ (left), $\tau_1(\gamma)$ (center), and  $\tau_2(\gamma)$ (right)  appearing in the Eq.~\eqref{eq:J_logs}; their
functional forms are given in Eq.~\eqref{pfit}. The stars
represent the following values for the set \mbox{$[\gamma,\rho,\tau_1(\mbox{in GeV}^2),\tau_2(\mbox{in GeV}^2)]$}:  \mbox{$[0,100.8,9.87,0.91]$} (blue stars),  \mbox{$[0.1,96.7,9.15,0.84]$} (red stars), \mbox{$[0.2,90.3,8.45,0.81]$} (yellow stars), and  \mbox{$[0.3,83.5,7.84,0.80]$} (purple stars).}
\label{fig:pfit}
\end{figure}

In Figs.~\ref{fig:pfit}, we show fits for the functions $\rho(\gamma)$, $\tau_1(\gamma)$, and  $\tau_2(\gamma)$; the values employed for obtaining
these curves are marked with stars. 
%

 In addition, notice that Eqs.~\eqref{eq:gmass} and~\eqref{eq:J_logs} not 
 only reproduce, by construction, the curve for $\Delta(q)$ shown in the right panel of  Fig.~\ref{fig:propagators},
 but also incorporate the following crucial features~\cite{Aguilar:2013vaa}:
 ({\it a}) the infrared finiteness of the gluon propagator;
 ({\it b}) the presence of  ``protected'' and  ``unprotected'' logarithms, {\it i.e.} $\ln(q^2 + m^2)$ and $\ln(q^2)$,
  originating, respectively from the gluon and ghost loops of the SDE for $\Delta(q)$; 
 ({\it c})  the massless ghost logarithms force $J(q)$ to
 reverse sign and diverge logarithmically in the infrared, with a \emph{zero-crossing} around the region of a few hundred MeV. 

 In the right panel of Fig.~\ref{fig:jm} one may appreciate that the precise location of the zero-crossing, $q_0^{\rm J}$, depends on the power $\gamma$,
 which controls the functional form of $m^2(q)$ in Eq.~\eqref{eq:gmass}.  More specifically, we have the following   values for the pair $[\gamma,q_0^{\rm J}\,\mbox(\textnormal{in MeV})]$: [$0$, $140$] (blue
  continuous), [$0.1$, $166$] (red dashed), 
  [$0.2$,$187$] (yellow dotted), and [$0.3$, $202$] (purple dash-dotted).
 
\begin{figure}[t]
\minipage{0.32\textwidth}
  \includegraphics[width=\linewidth]{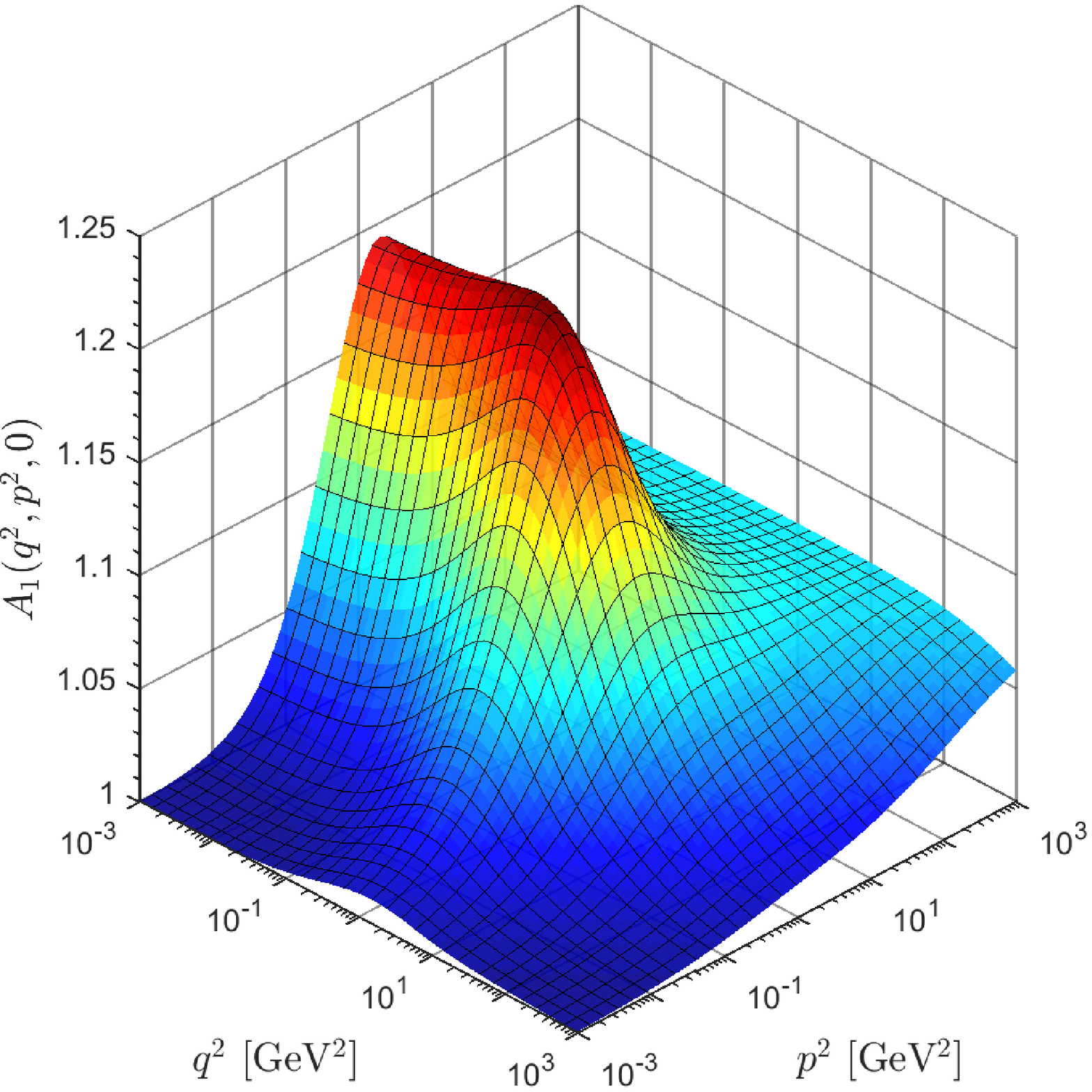}
\endminipage\hfill
\minipage{0.32\textwidth}
  \includegraphics[width=\linewidth]{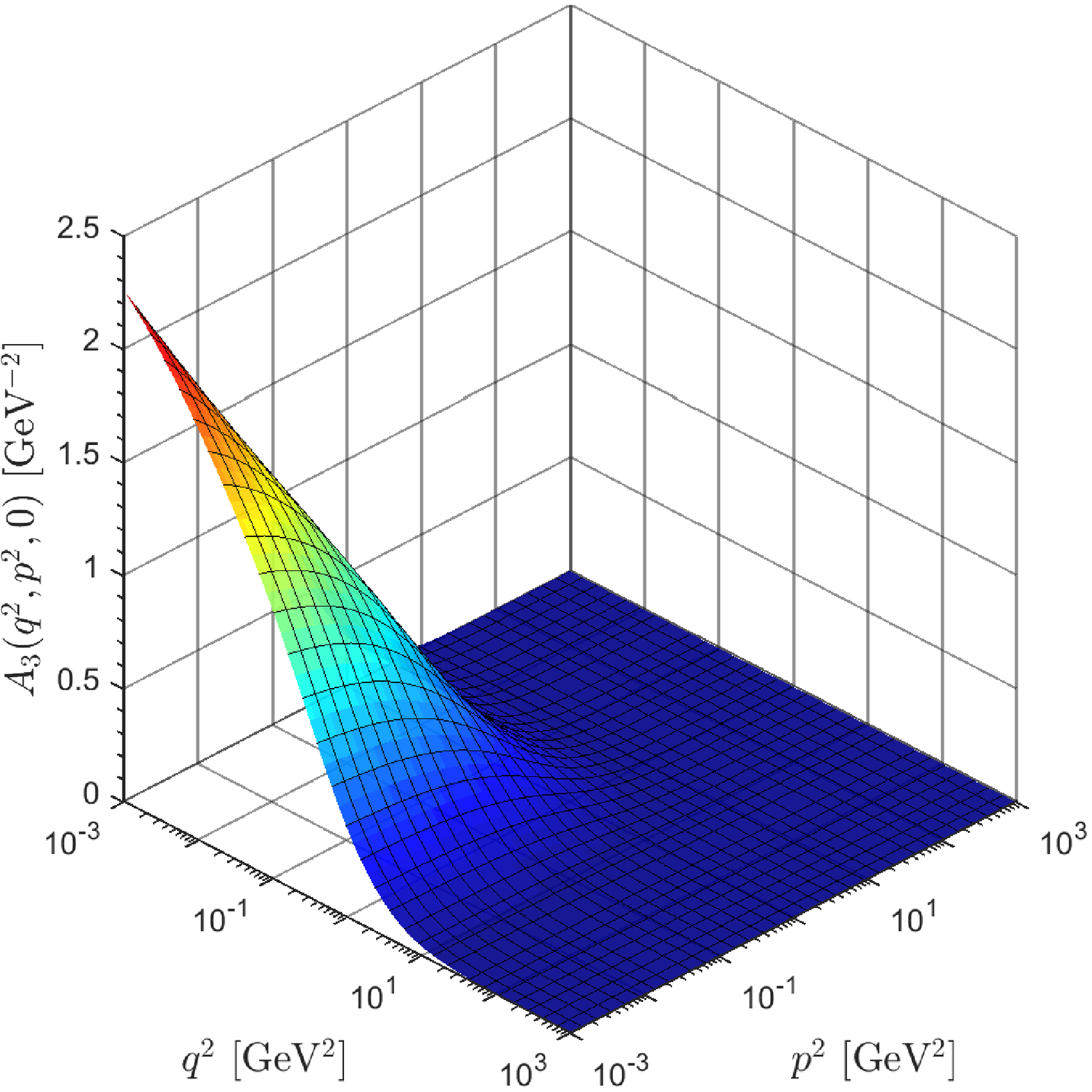}
\endminipage\hfill
\minipage{0.32\textwidth}%
  \includegraphics[width=\linewidth]{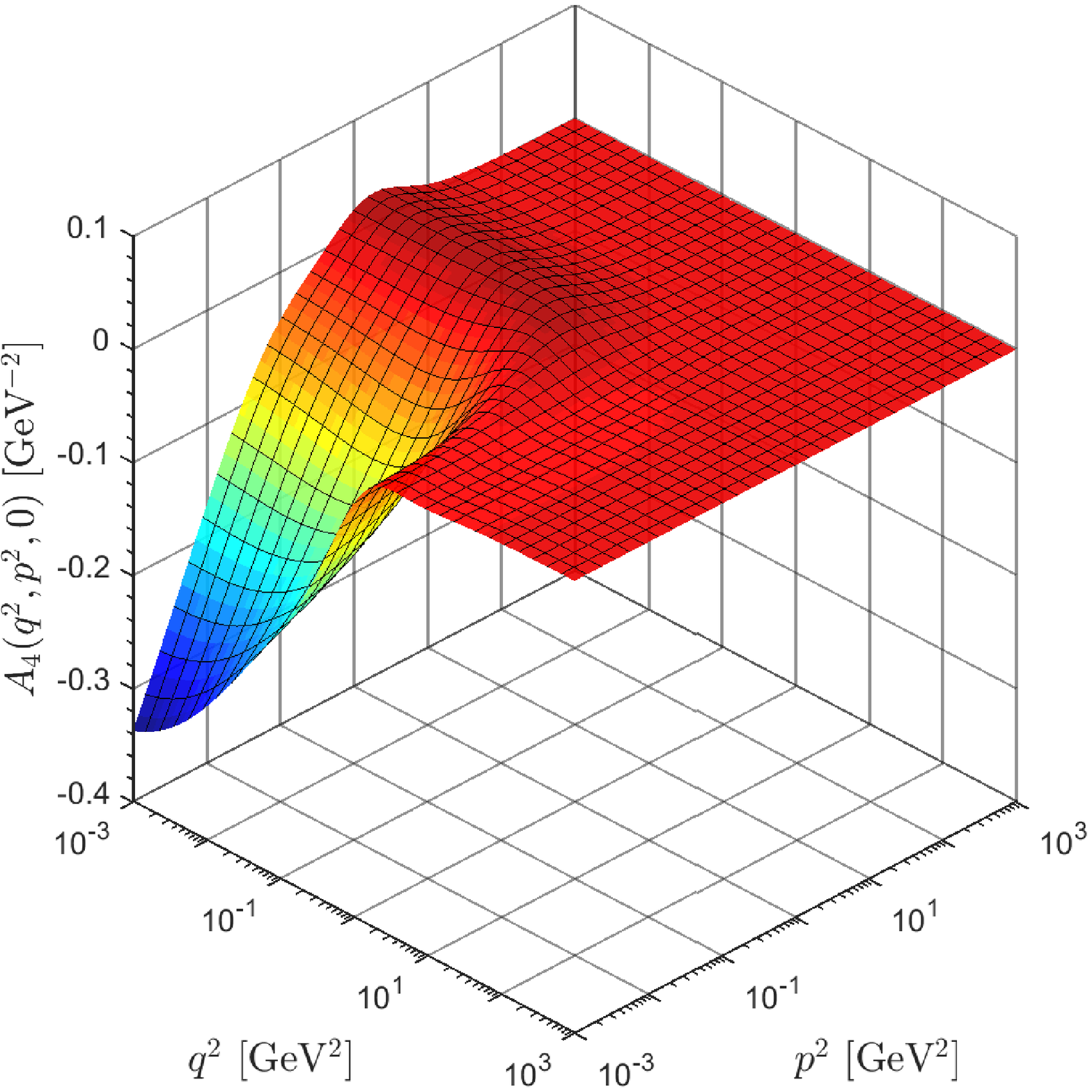}
\endminipage
\caption{The form factors of the ghost-gluon scattering kernel $A_1(q^2,p^2,\phi)$ (first panel), $A_3(q^2,p^2,\phi)$ (second panel), and  $A_4(q^2,p^2,\phi)$ (third panel) for $\phi = 0$ 
and $\alpha_s=0.22$.}
\label{fig:Ais}
\end{figure}

({\it iii}) The final ingredients needed for the evaluation of the BC solution
are the form factors $A_1$, $A_3$, and $A_4$ of the ghost-gluon kernel $H_{\nu\mu}(q,p,r)$, defined 
in \1eq{eq:H}. Their nonperturbative evaluation for general Euclidean momenta has been presented in~\cite{Aguilar:2018csq},
where the  one-loop dressed version of the 
SDE satisfied by $H_{\nu\mu}(q,p,r)$ was employed. In Fig.~\ref{fig:Ais} we show 
a representative case for $A_1$, $A_3$, and $A_4$, when the angle between the momenta $q$ and $p$ is fixed at $\phi=0$.
Note in particular that $A_1$ is finite within the entire range of its momenta, whereas $A_3$ and $A_4$ display a logarithmic
divergence in the deep infrared. 

\begin{figure}[t]
\begin{minipage}[b]{0.45\linewidth}
\centering
\includegraphics[scale=0.4]{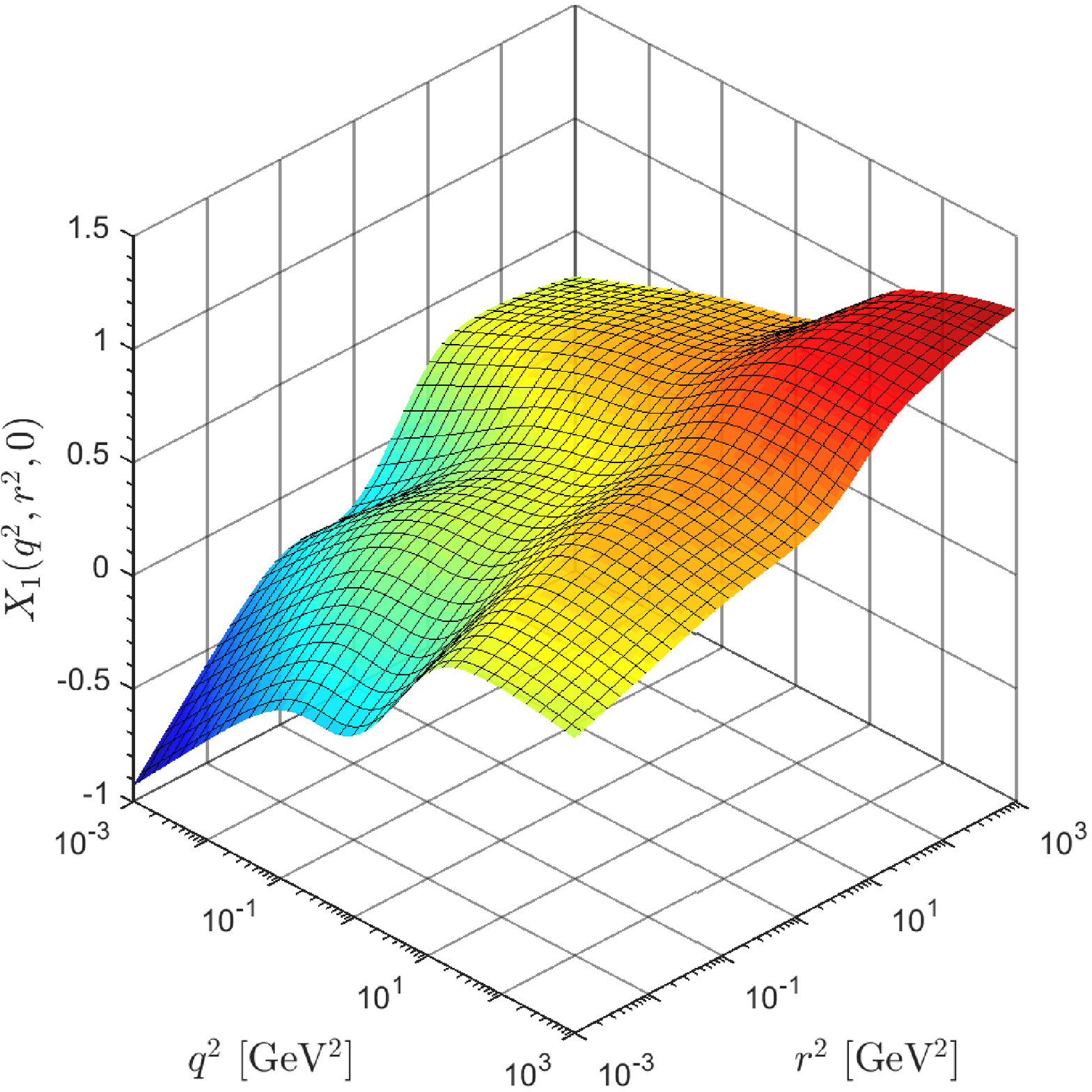}
\\
\vspace{0.5cm}
\includegraphics[scale=0.4]{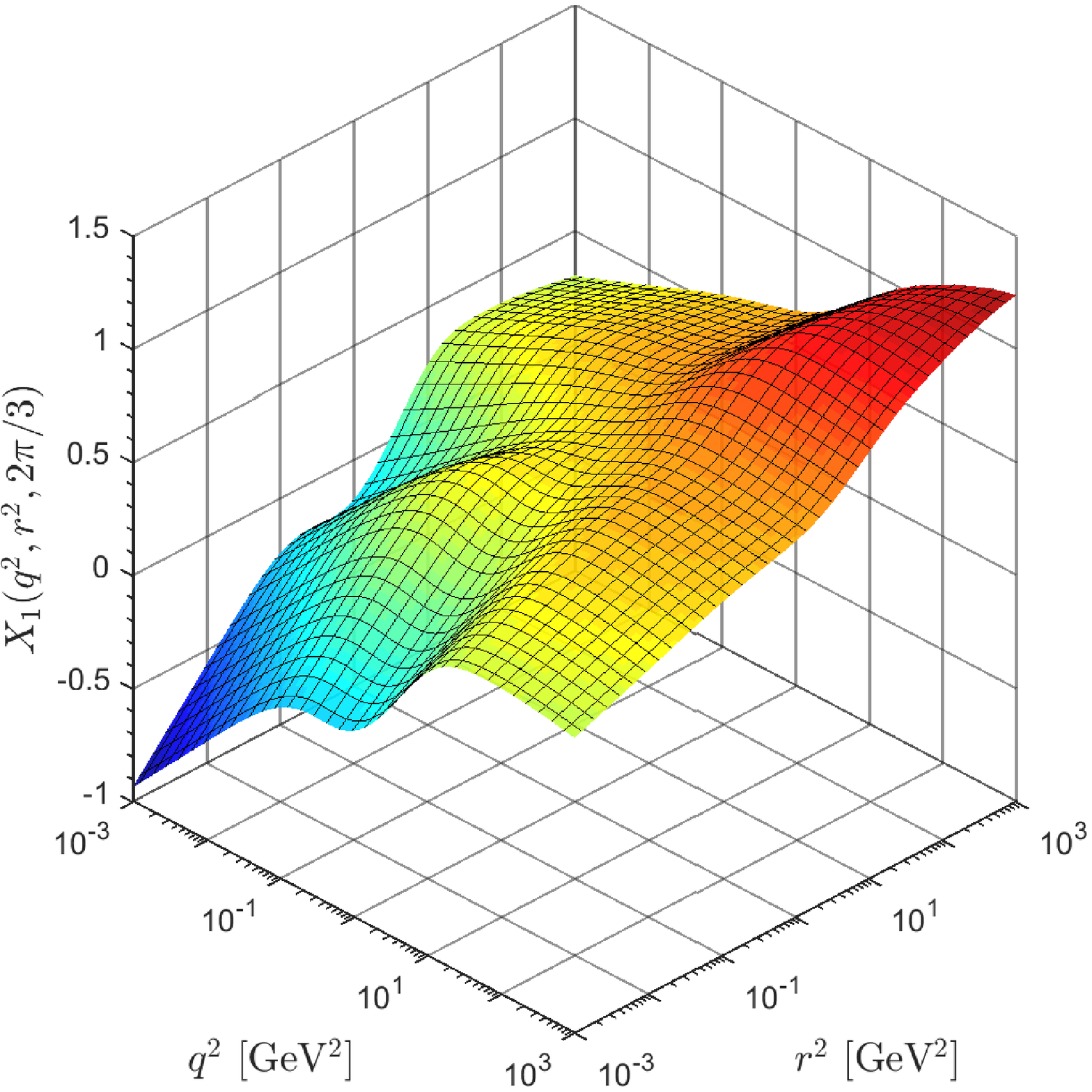}
\end{minipage}
\hspace{0.25cm}
\begin{minipage}[b]{0.45\linewidth}
\includegraphics[scale=0.4]{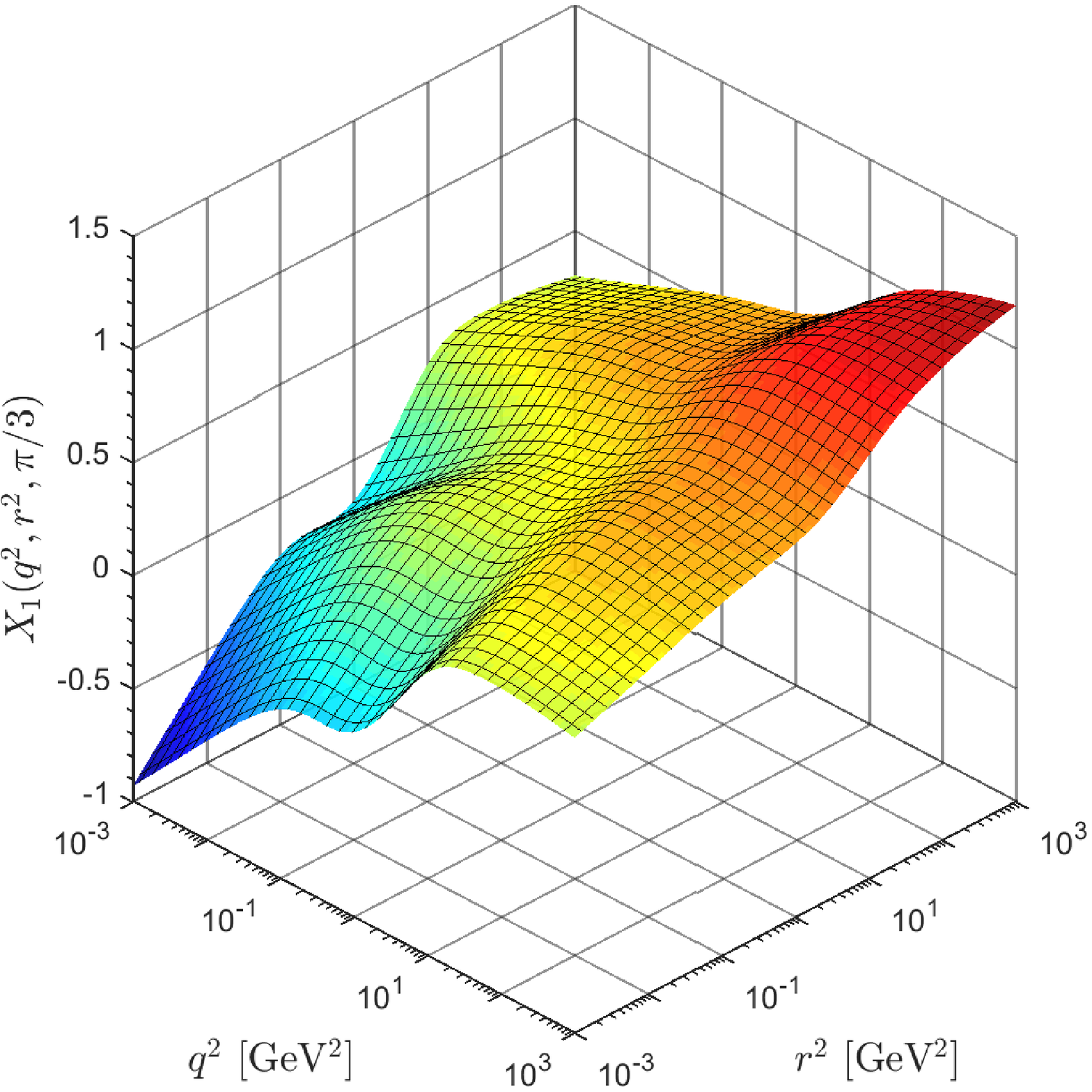}
\\
\vspace{0.5cm}
\includegraphics[scale=0.4]{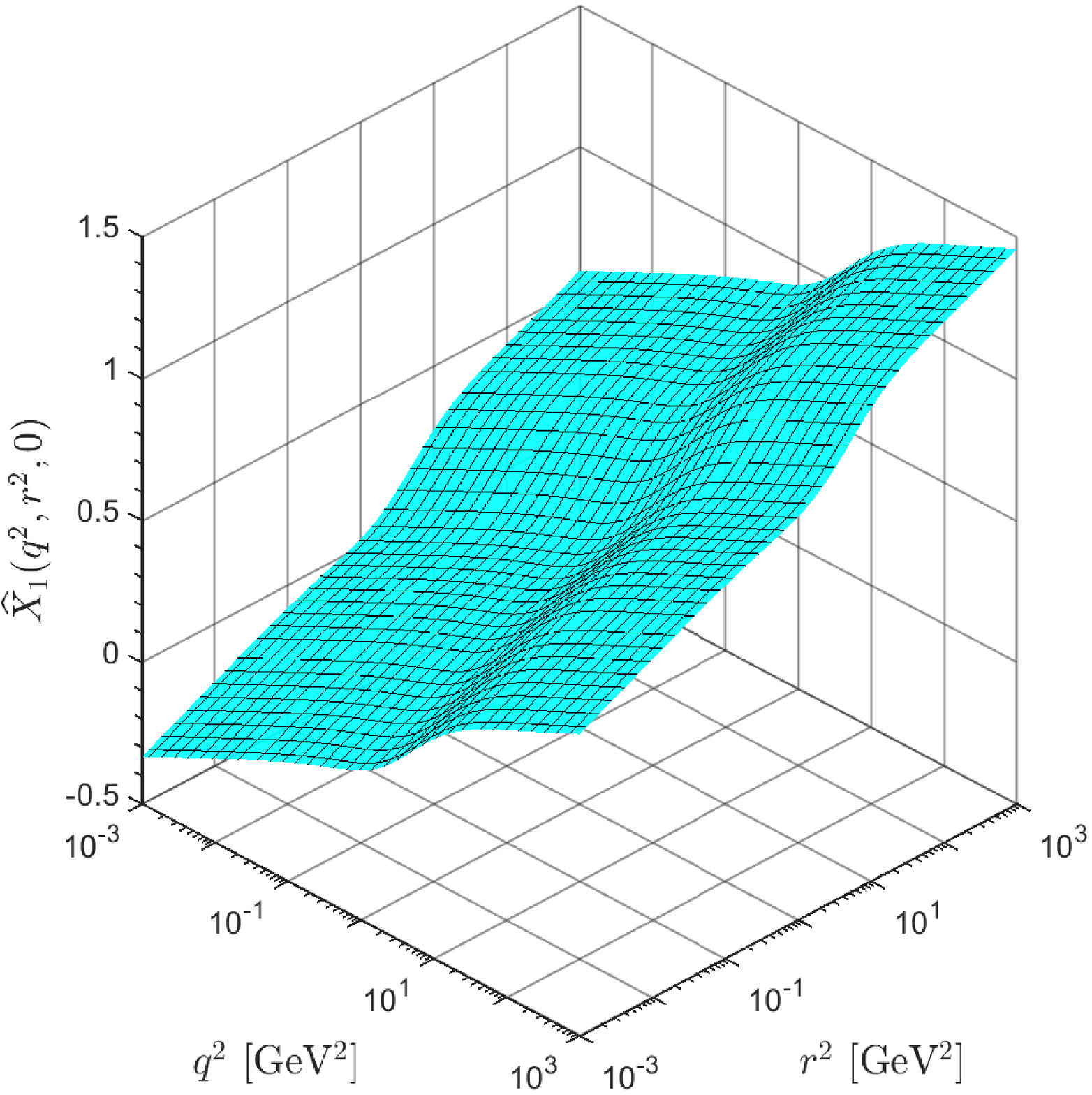}
\end{minipage}
\caption{$X_1(q^2,r^2,\theta)$ for $\theta = 0$ (top left), $\pi/3$ (top right), and  $2\pi/3$ (bottom left), together 
with the ``abelianized'' $\widehat{X}_1$ (bottom right).}
\label{fig:X1_gen_fig}
\end{figure}

\subsection{\label{subsec:Xires} The three-gluon form factors: general kinematics}

With the inputs introduced in the previous subsection, 
the form factors $X_1$, $X_2$, $X_3$ and $X_{10}$ may now be computed from Eqs.~\eqref{eq:X_sol}.

The evaluation of Eqs.~\eqref{eq:X_sol} is carried out for
general {\it Euclidean kinematics}; we will express 
the form factors as functions of
$q^2$,  $r^2$, and the angle $\theta$ formed between $q$ and $r$, namely $X_i(q,r,p)\to X_i(q^2,r^2,\theta)$. 
For the numerical computation of the relevant integrals we  use logarithmically
spaced grids for $q^2$ and $r^2$, with $96$
values for each, in the range \mbox{ $[ 5\times 10^{-5}\,\mbox{{GeV}}^2, 10^4\!,\mbox{{GeV}}^2]$}.
The corresponding  grid for the angle $\theta$ is uniformly spaced,
with $19$ values distributed within $[0, \pi]$. Moreover, 
the required interpolations of the  results for $A_1$, $A_3$, and $A_4$, obtained in~\cite{Aguilar:2018csq}, are performed 
using tensor products of B-splines~\cite{de2001practical}.

\begin{figure}[!h]
\begin{minipage}[b]{0.45\linewidth}
\centering
\includegraphics[scale=0.4]{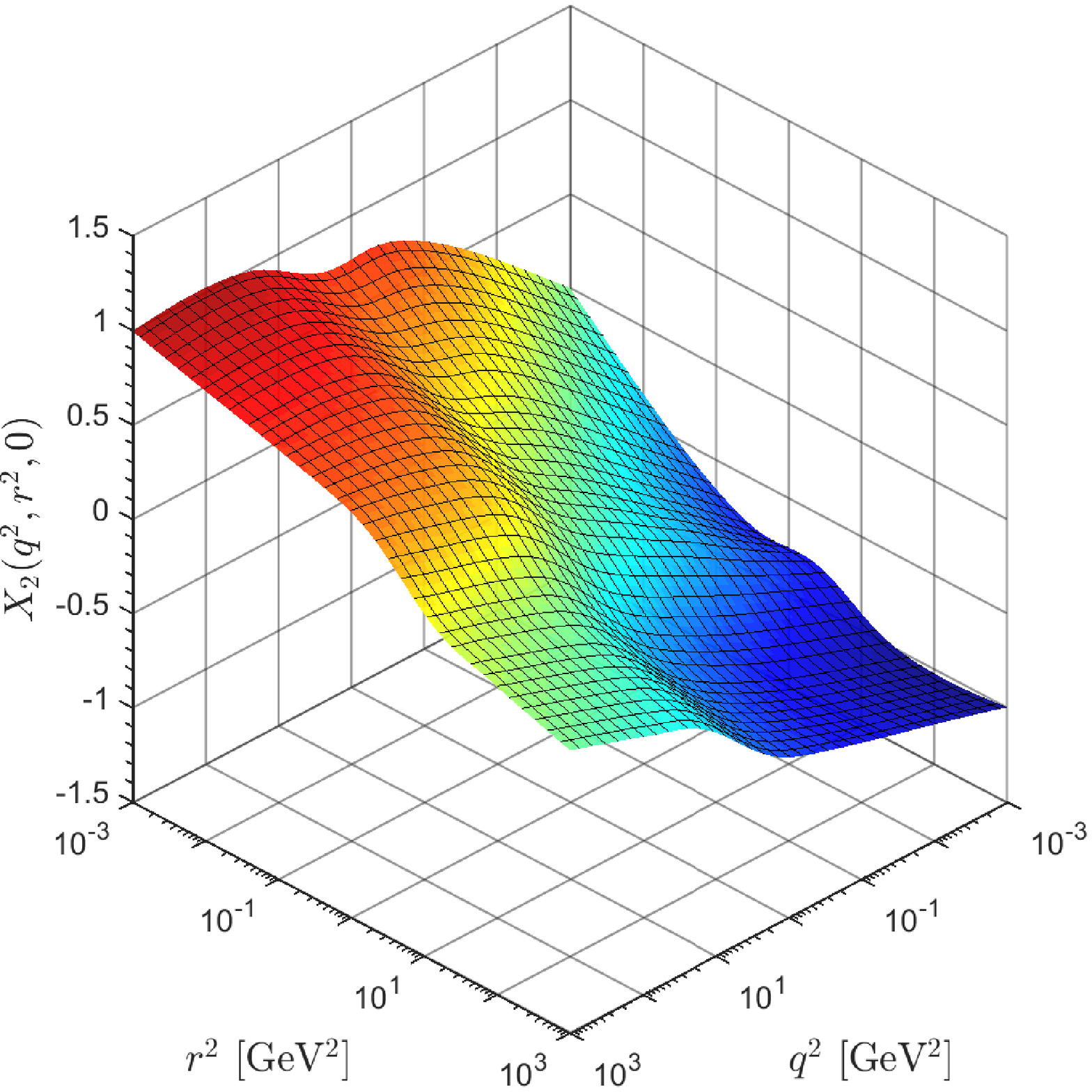}
\\
\vspace{0.5cm}
\includegraphics[scale=0.4]{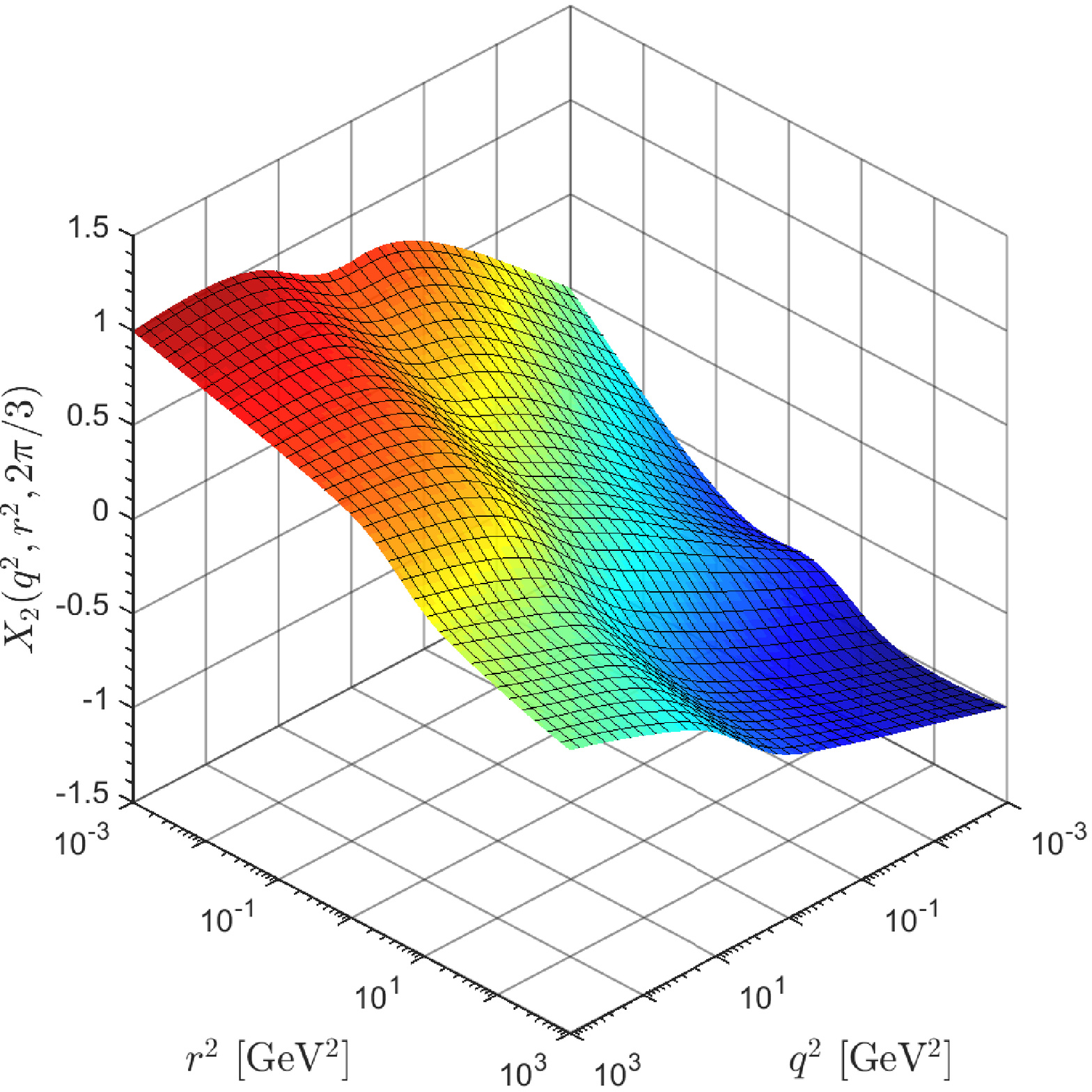}
\end{minipage}
\hspace{0.25cm}
\begin{minipage}[b]{0.45\linewidth}
\includegraphics[scale=0.4]{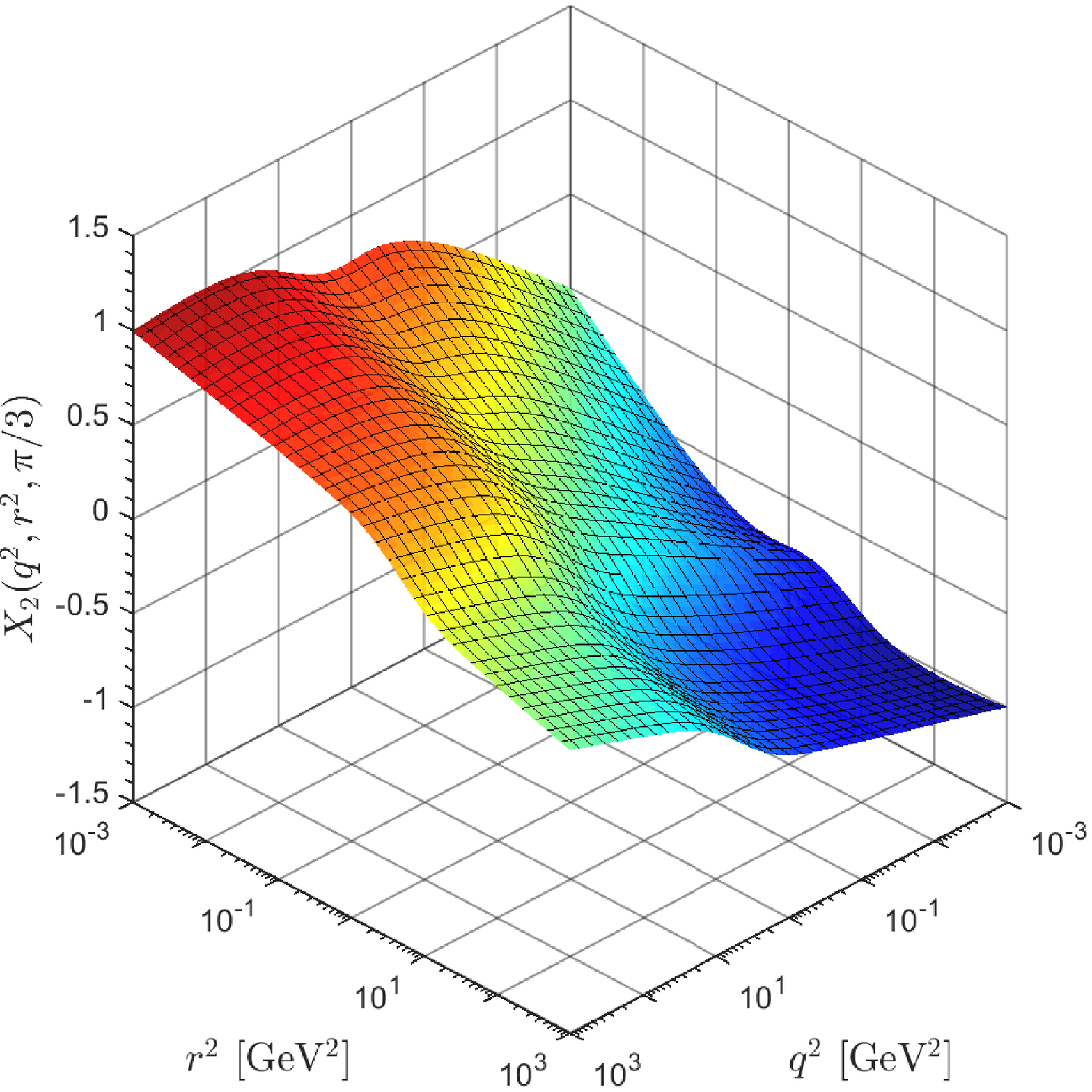}
\\
\vspace{0.5cm}
\includegraphics[scale=0.4]{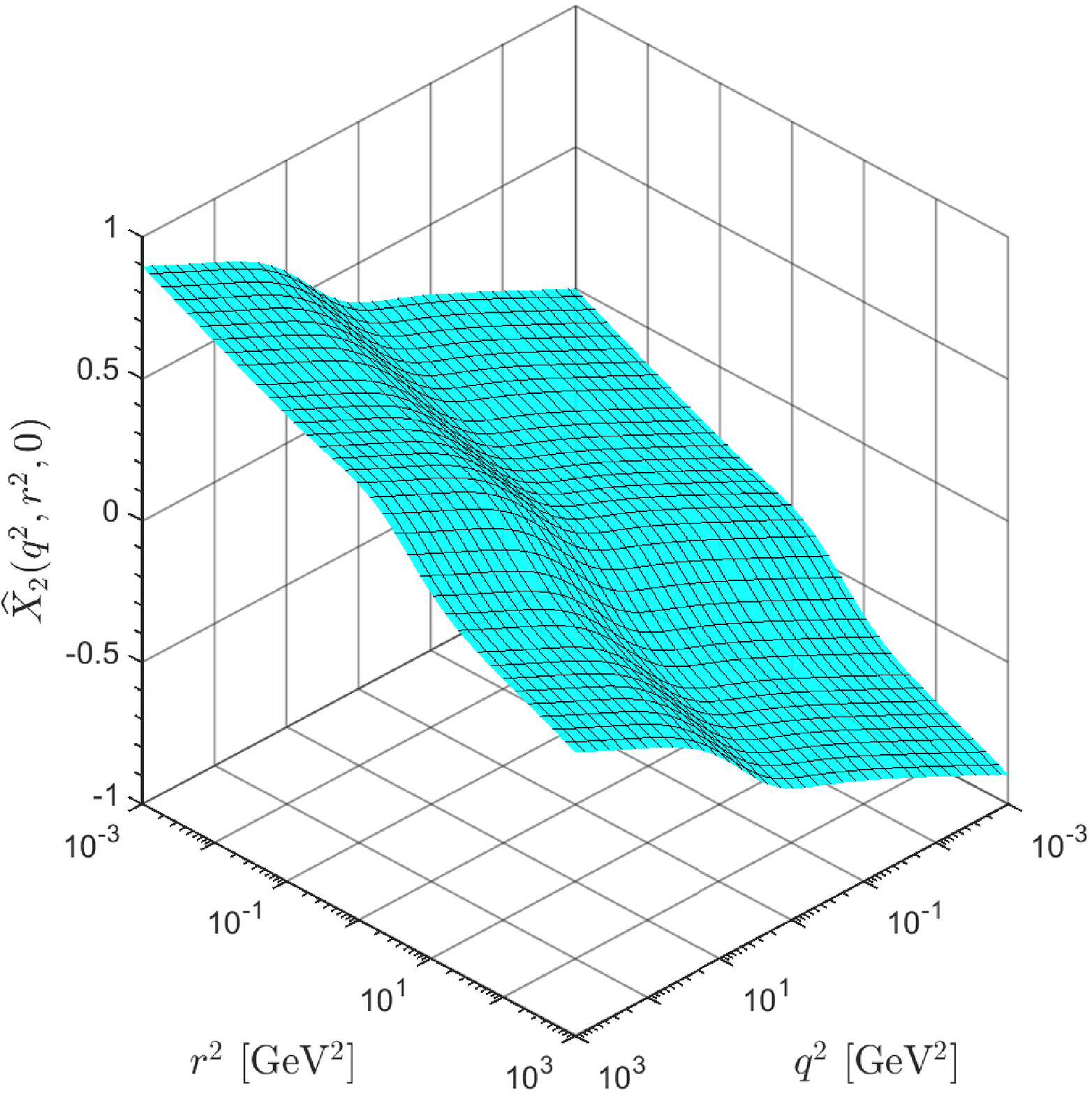}
\end{minipage}
\caption{$X_2(q^2,r^2,\theta)$ for $\theta = 0$ (top left), $\pi/3$ (top right), and  $2\pi/3$ (bottom left), together 
with the ``abelianized'' $\widehat{X}_2$ (bottom right). Note that, in order to better visualize the surfaces, 
the $q^2$ and $r^2$ axes have been rotated by $\pi/2$ with respect to the other 3D figures.}\label{fig:X2_gen_fig}
\end{figure}

\begin{figure}[!h]
\begin{minipage}[b]{0.45\linewidth}
\centering
\includegraphics[scale=0.4]{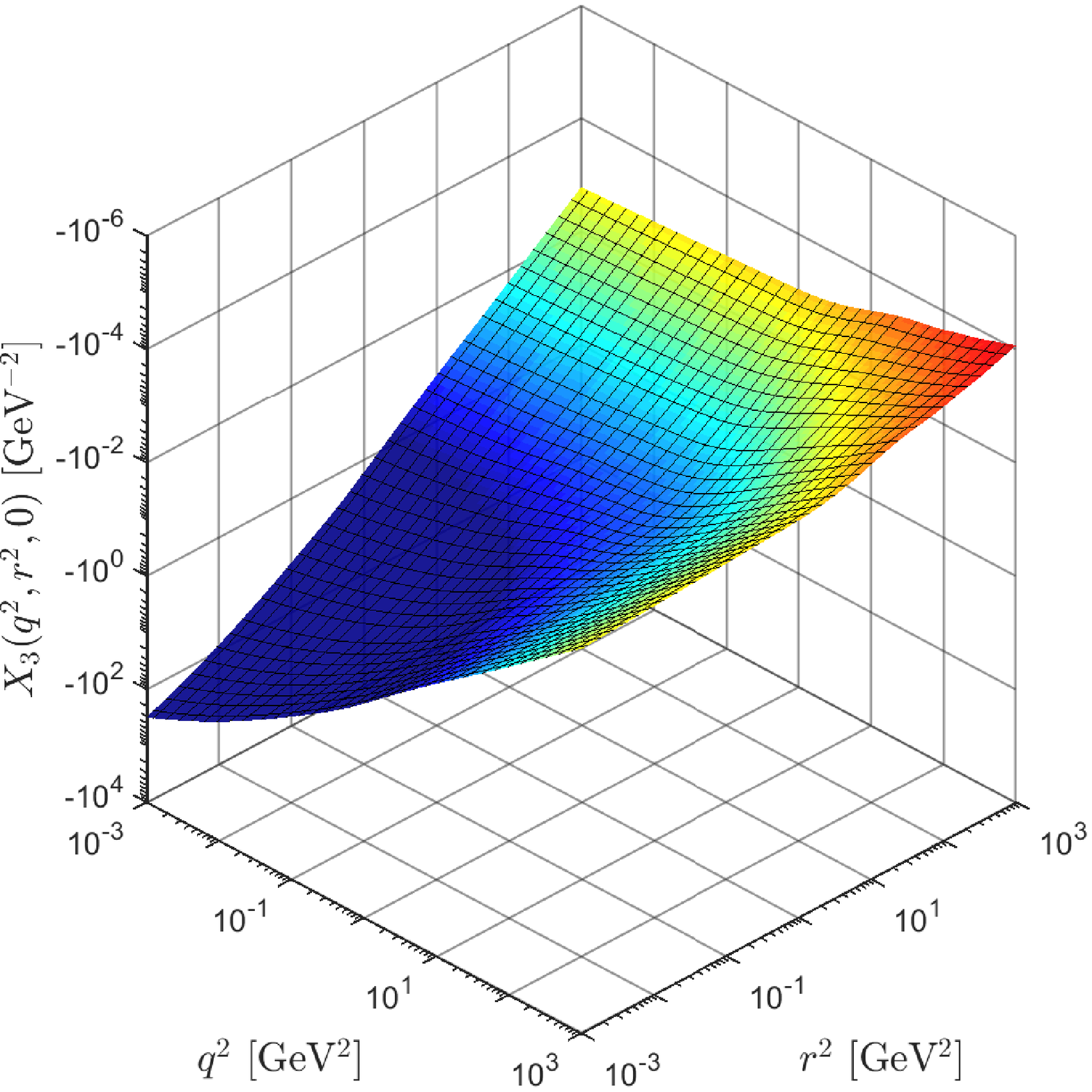}
\\
\vspace{0.5cm}
\includegraphics[scale=0.4]{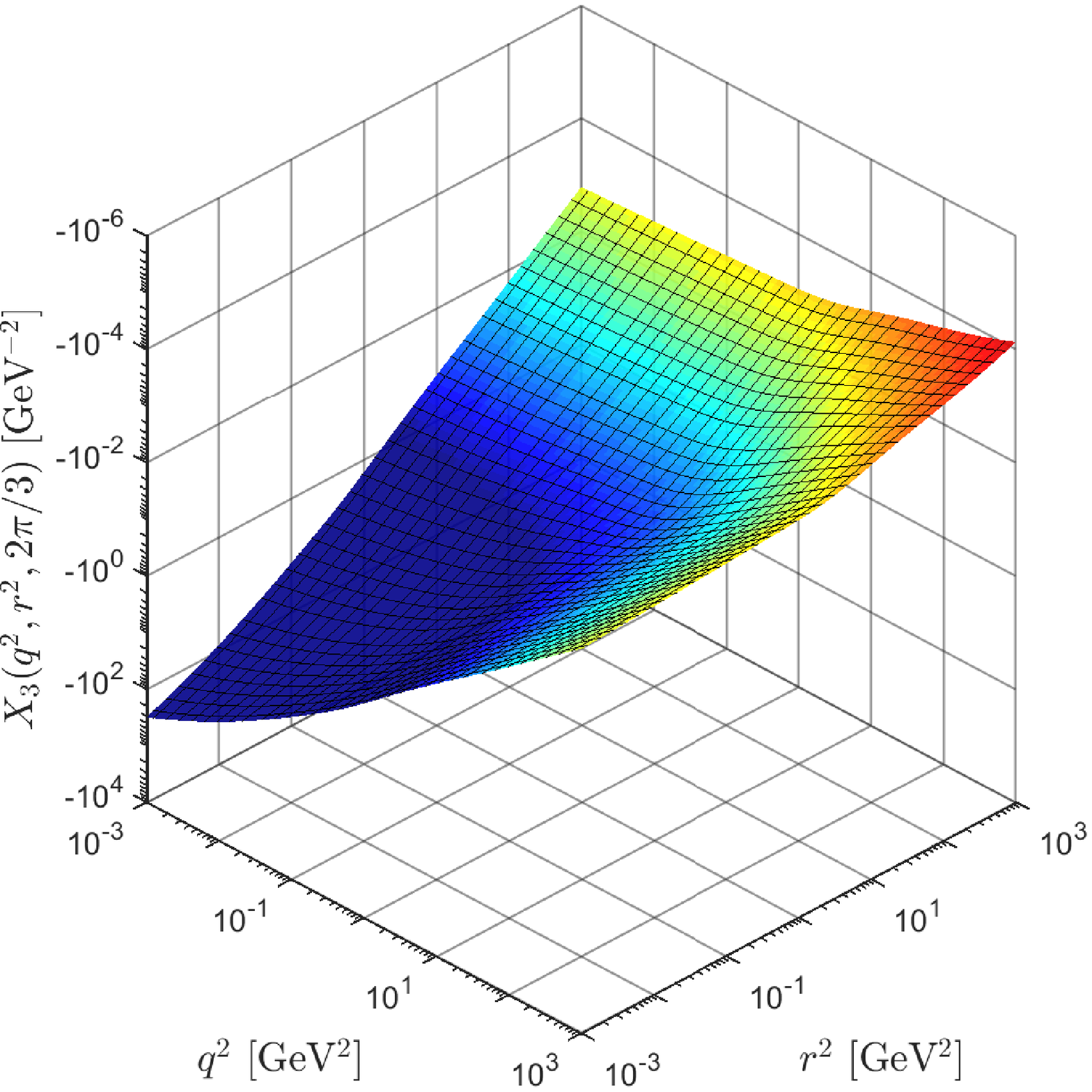}
\end{minipage}
\hspace{0.25cm}
\begin{minipage}[b]{0.45\linewidth}
\includegraphics[scale=0.4]{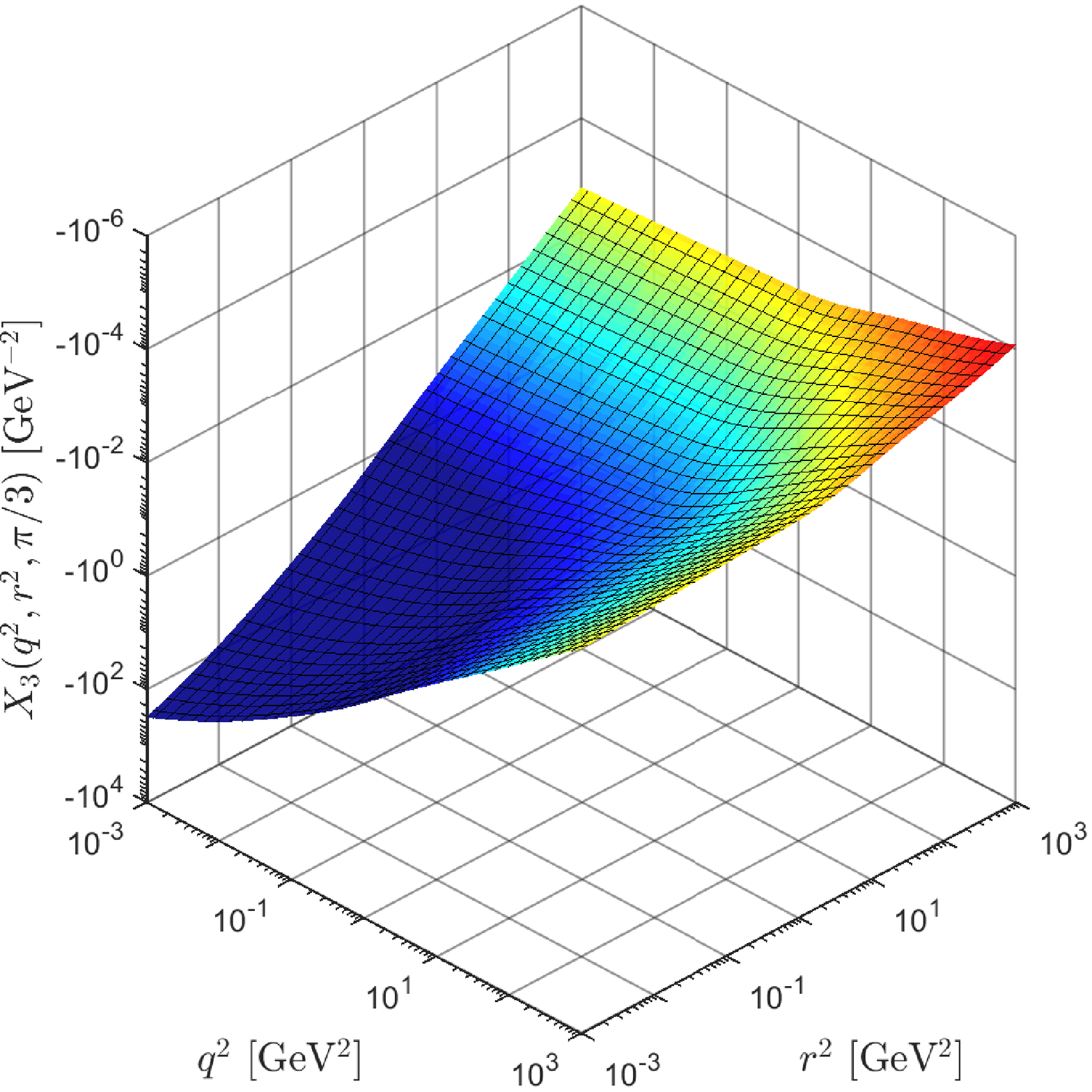}
\\
\vspace{0.5cm}
\includegraphics[scale=0.4]{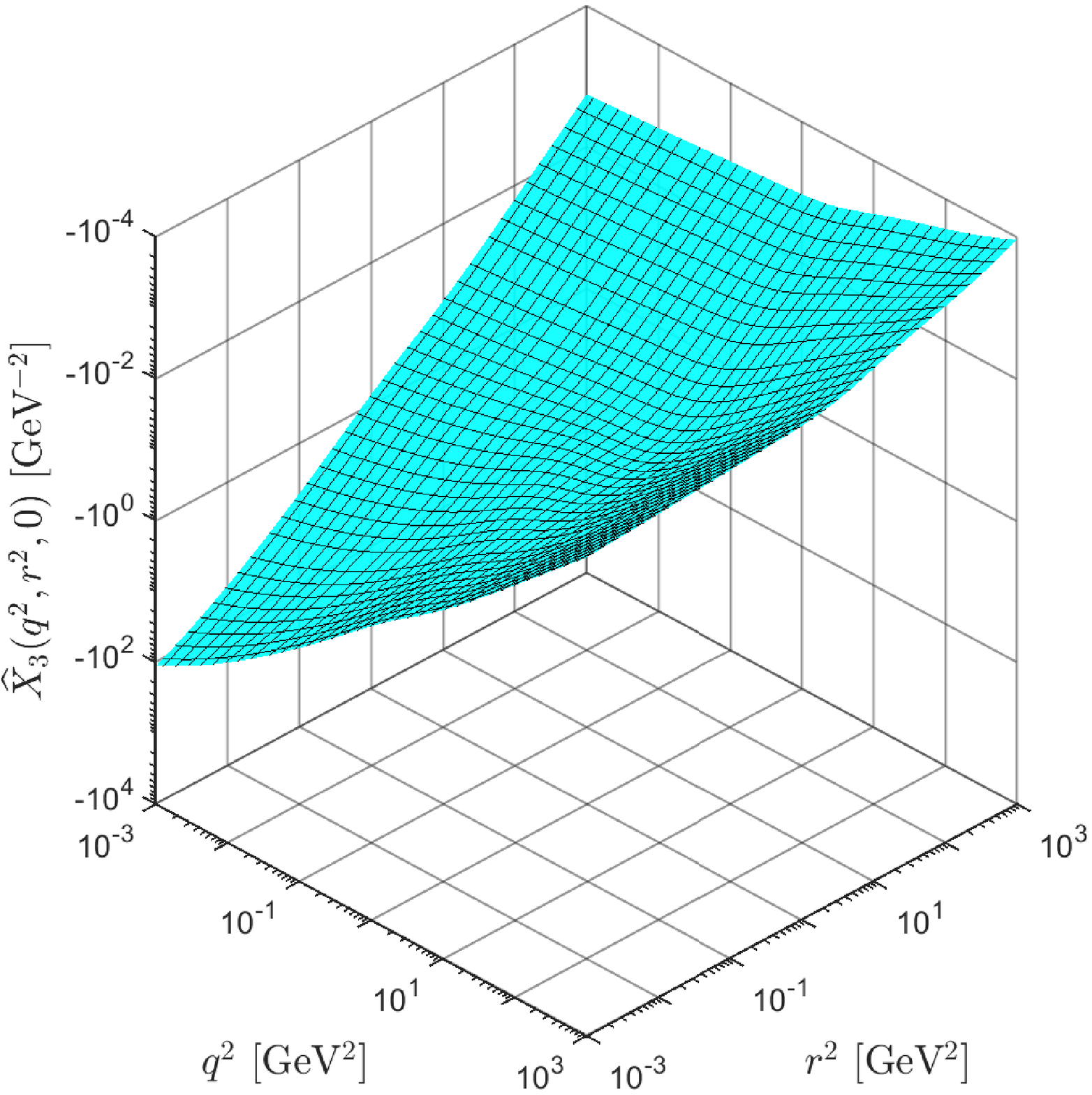}
\end{minipage}
\caption{ $X_3(q^2,r^2,0)$  for $\theta = 0$ (top left), $\pi/3$ (top right), and  $2\pi/3$ (bottom left), together 
with the ``abelianized'' $\widehat{X}_3$ (bottom right).}
\label{fig:X3num_gen_fig}
\end{figure}

\begin{figure}[!h]
\begin{minipage}[b]{0.45\linewidth}
\centering
\includegraphics[scale=0.4]{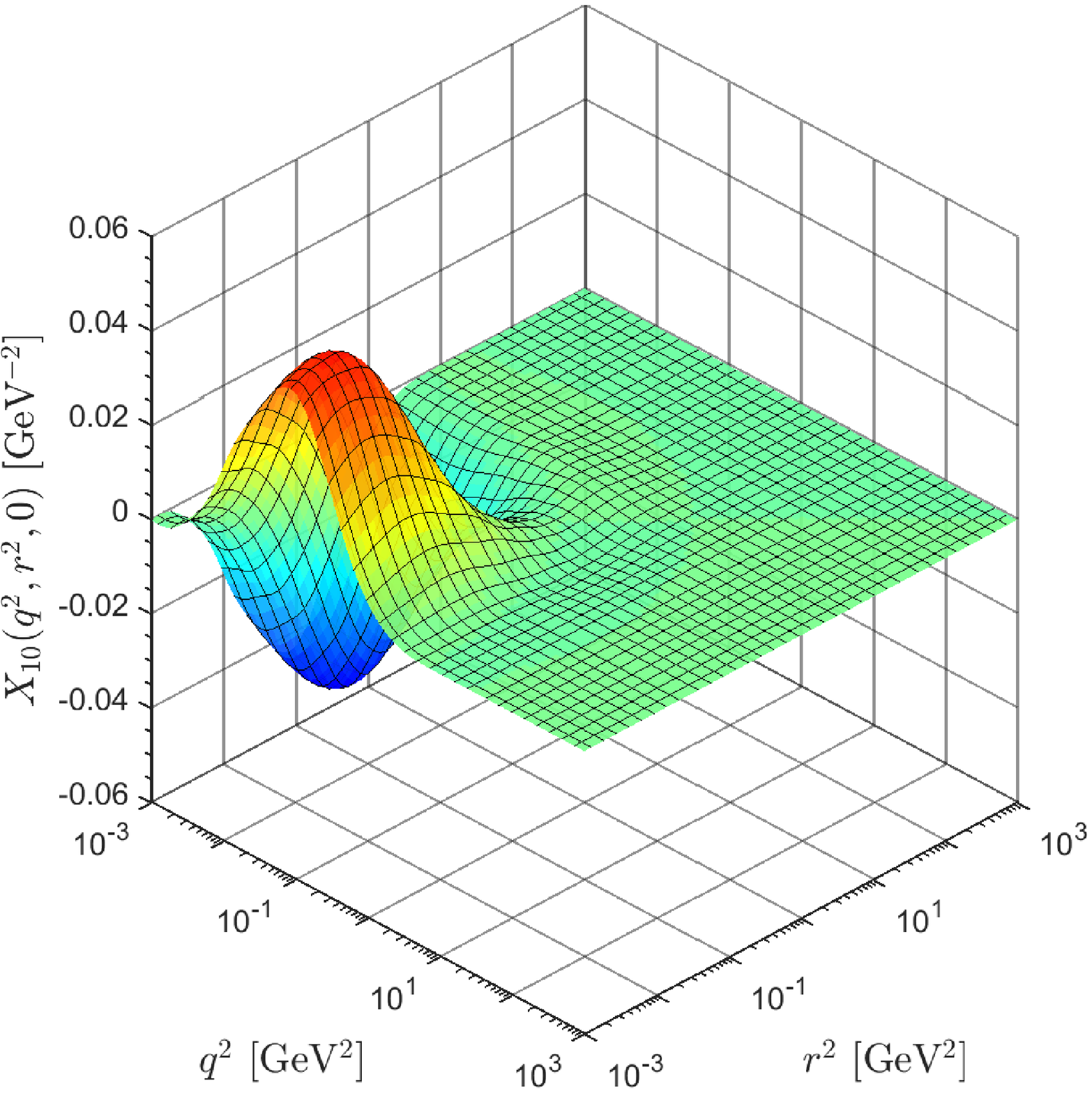}
\\
\vspace{0.5cm}
\includegraphics[scale=0.4]{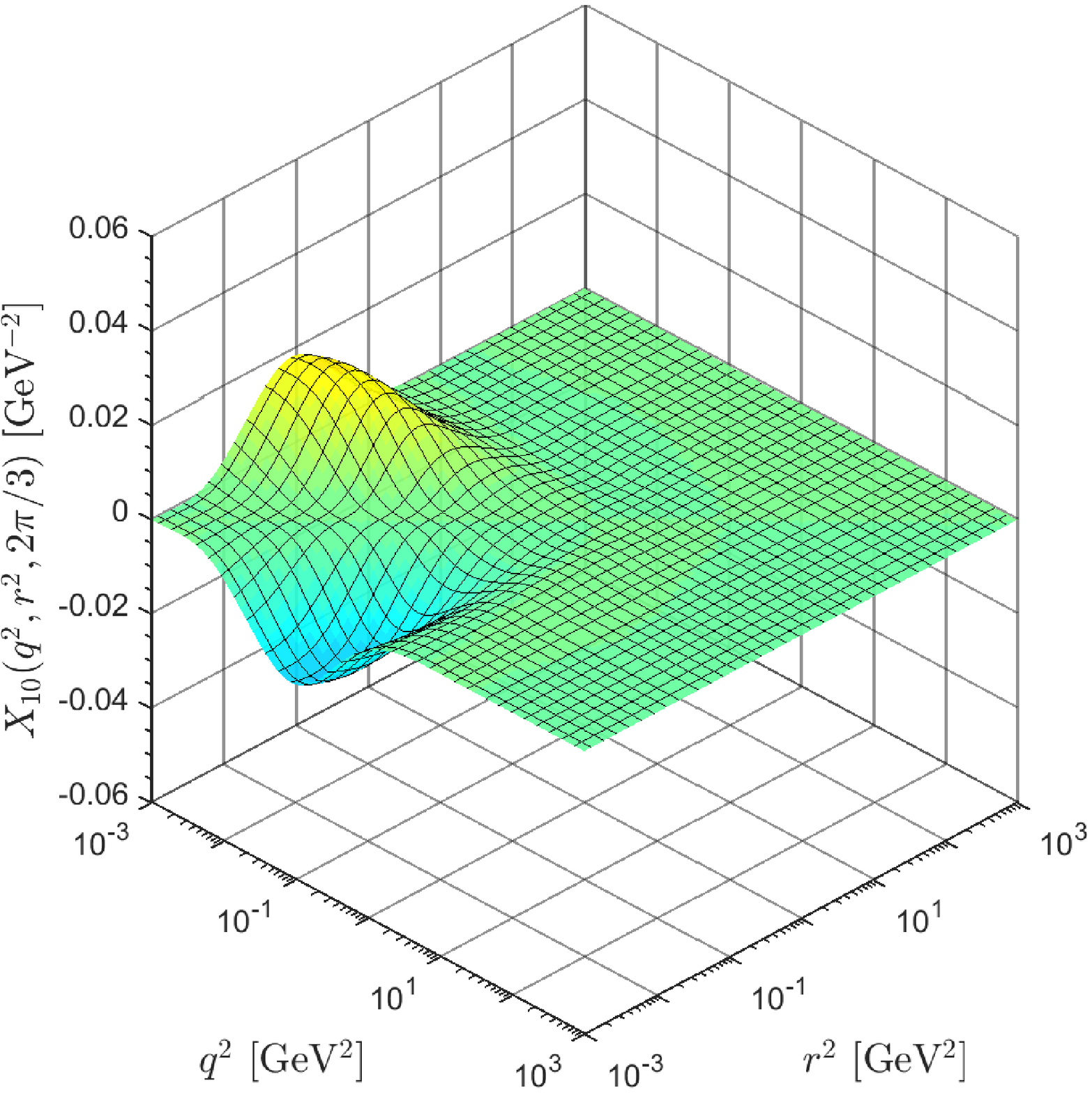}
\end{minipage}
\hspace{0.25cm}
\begin{minipage}[b]{0.45\linewidth}
\includegraphics[scale=0.4]{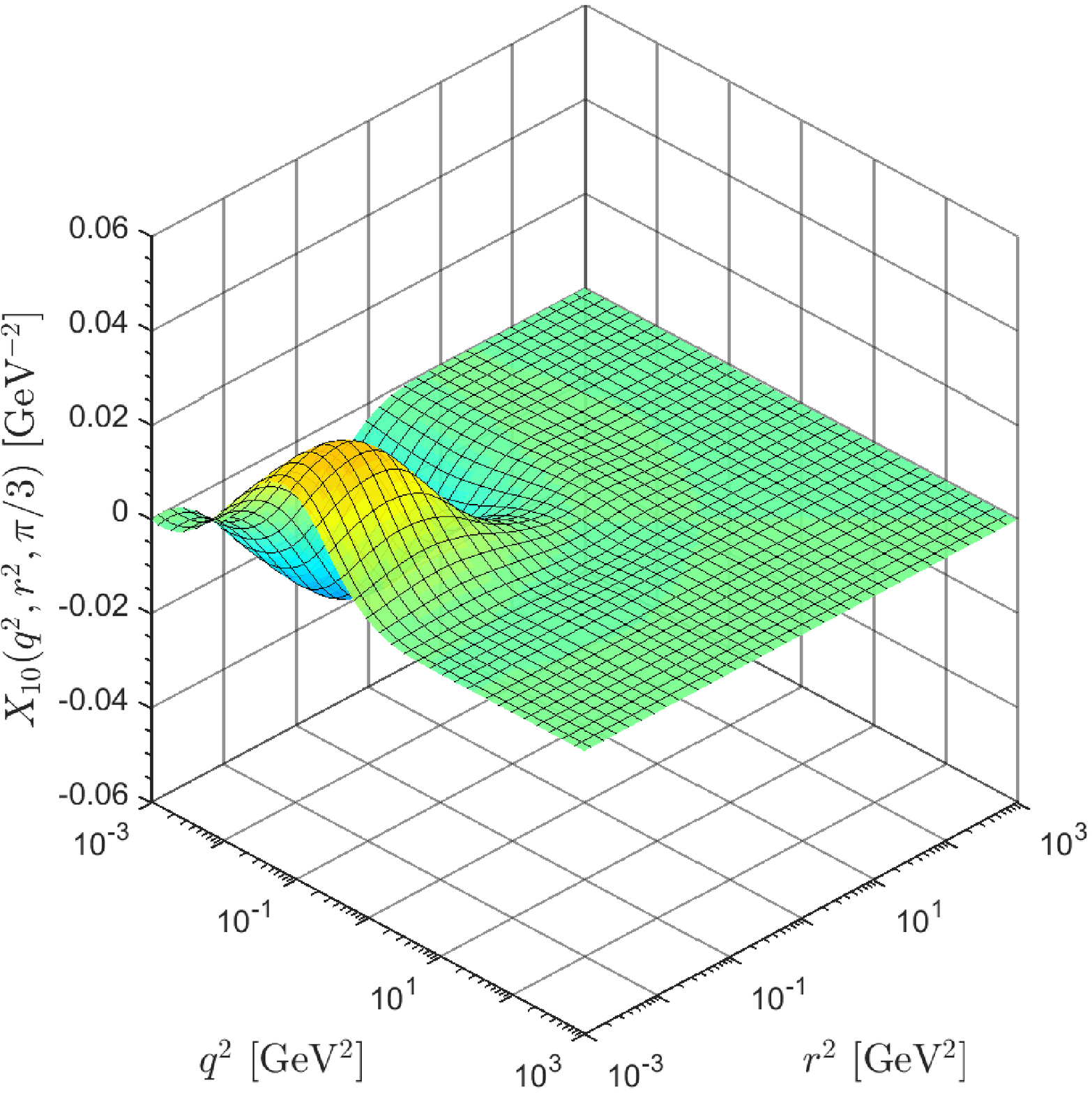}
\\
\vspace{0.5cm}
\includegraphics[scale=0.4]{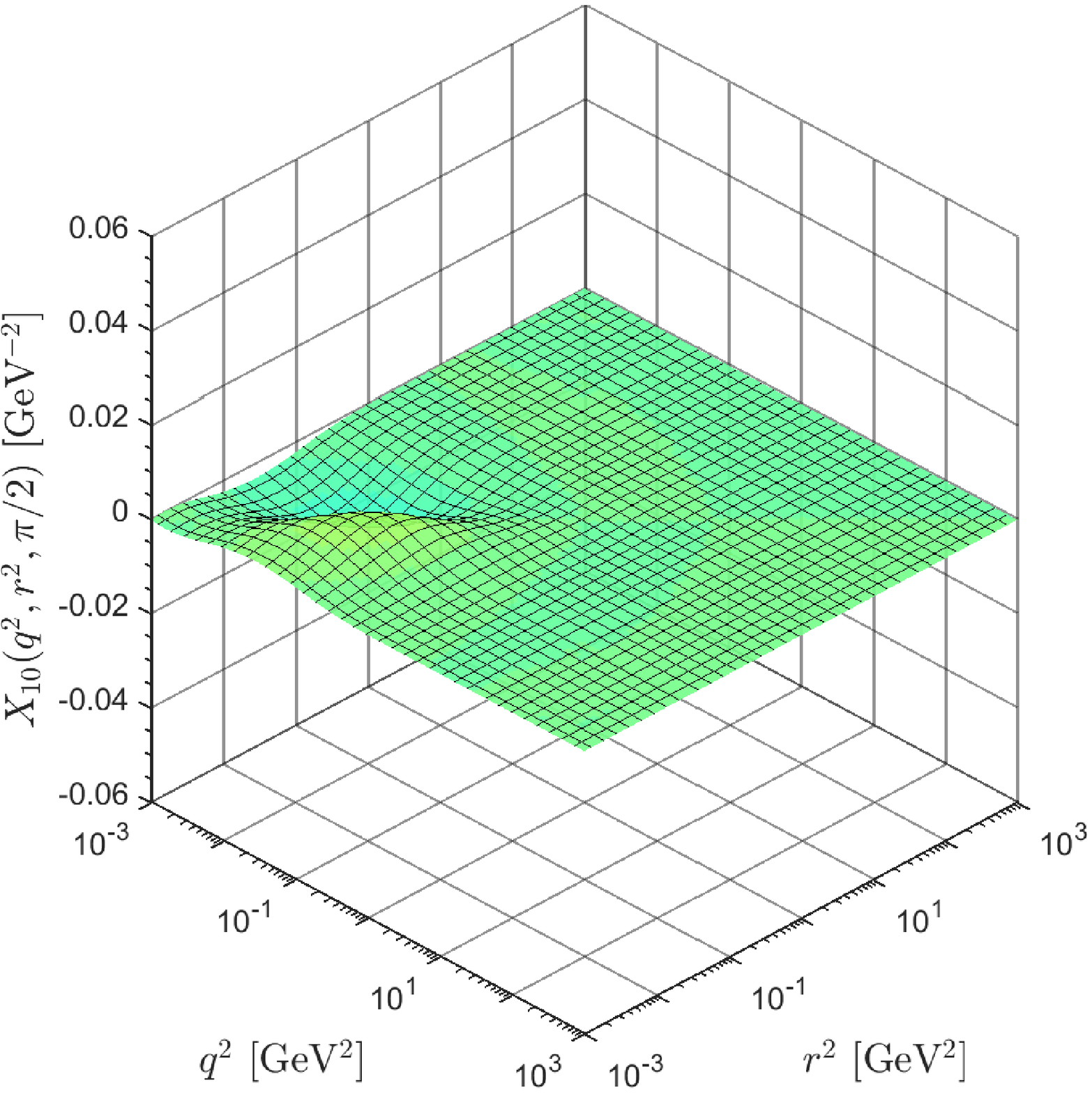}
\end{minipage}
\caption{$X_{10}(q^2,r^2,\theta)$ for $\theta = 0$ (top left), $\pi/3$ (top right)  $2\pi/3$ (bottom left), and $\pi/2$ (bottom right).}\label{fig:X10_gen_fig}
\end{figure}

The results for $X_1(q^2,r^2,\theta)$, $X_2(q^2,r^2,\theta)$, $X_3(q^2,r^2,\theta)$  and  $X_{10} (q^2,r^2,\theta)$ are shown in 
Figs.~\ref{fig:X1_gen_fig}-\ref{fig:X10_gen_fig}, respectively. In each of these figures, we present the corresponding form factor for three representative values of the angle: $\theta =0$ (top left panels), $\theta =\pi/3$ (top right panels), and $\theta =2\pi/3$ (bottom left panels).  In addition, we provide a visual impression of the impact that the
ghost sector has on the $X_i$ by plotting the corresponding
``abelianized'' quantities,  $\widehat{X}_i$, in the bottom right panels; these latter quantities are given by \1eq{eq:X10_mBC}, and are independent of the angle 
$\theta$.  Since $\widehat{X}_{10}=0$, we occupy its panel in 
Fig.~\ref{fig:X10_gen_fig}  with one additional configuration, namely $X_{10} (q^2,r^2,\theta=\pi/2)$.


The results exhibit the following 
features: ({\it i}) in the infrared, $X_1$, $X_2$, and $X_3$ depart considerably from their tree level values
(1, 0, and 0, respectively), while $X_{10}$, even though nonvanishing, is very suppressed;
({\it ii}) in the ultraviolet, all form factors approach their expected perturbative behavior; ({\it iii})
the patterns displayed by the $X_i$ are rather similar to those of the $\widehat{X}_i$,
but with small ``oscillations'' distributed around their main structures, owing to the contributions from the
ghost sector; ({\it iv}) in general, they display a mild dependence on the angle $\theta$.

It is important to emphasize that, while the form factors $X_1$, $X_2$, and $X_{10}$
diverge at most logarithmically in the infrared, under certain special kinematic circumstances 
$X_3$ displays a pole divergence. This, in turn, is the reason for employing double-log graphs for the surfaces shown in Fig.~\ref{fig:X3num_gen_fig}.

To appreciate this point,
note first that, due to the presence of the factor $(q^2 - r^2)$ in the denominator of Eq.~\eqref{eq:X_sol},
the computation of $X_3$ in the limit $q^2\to r^2:= Q^2$ requires the use of a limiting procedure, which amounts
to taking appropriate total or partial derivatives. Note that the equality $q^2=r^2$ may be realized for any value of the
angle $\theta$; momentum conservation restricts $p^2$ to satisfy
\be
p^2= 2Q^2 (1 + \cos \theta)\,,
\label{pcond}
\ee
with limiting cases $p^2=0$ and $p^2 = 4 Q^2$. 

In order to simplify the algebra without compromising the essence, let us
revert to the ``abelianized'' form of $X_3$ given in \1eq{eq:X10_mBC}, and let $q^2\to r^2:= Q^2$.
Then, one has that in Euclidean space
\be 
\widehat{X}_3(Q) = -\frac{d J(Q)}{dQ^2} \,,
\label{limX3}
\ee
which, after employing the functional form for $J(Q)$ given in \1eq{eq:J_logs}, reads  
\be 
\widehat{X}_3(Q) =   
-\frac{C_\mathrm{A}\alpha_s}{24\pi}\left[ \frac{1}{Q^2} \left(1 + \frac{\tau_1}{Q^2 + \tau_2}\right) - \frac{\tau_1}{(Q^2 + \tau_2)^2}\ln\left( \frac{Q^2}{\mu^2}\right) \right] + \cdots \,;
\label{derJ}
\ee
evidently,  the above expression contains a simple pole together with a subleading logarithmic divergence, while 
the ellipses denote terms that are finite as  $Q^2 \to 0$.

Regarding the behavior of $\widehat{X}_3(Q)$ observed above, the following remarks are in order.

({\it a}) Note that this particular divergence is not an artifact of the BC basis, which in~\cite{Ball:1980ax} was advocated   
to be free of kinematic singularities. In fact, it should be clear that the origin of the divergence is
{\it dynamical}, stemming from the presence of the ``unprotected'' logarithm, and hence, from the nonperturbative masslessness of the
ghost. If for instance the aforementioned logarithm had been omitted from the $J(Q)$, the answer in \1eq{derJ} would be perfectly finite;
and the same would be true if the argument of the logarithm had been saturated by a ``ghost mass'', whose generation, 
however, does {\it not} occur dynamically. 

({\it b}) The type of pole divergence found in \1eq{derJ} should be clearly distinguished from those appearing in $\Gp_{\alpha\mu\nu}(q,r,p)$.
Note, in particular, that the pole $1/q^2$ (or any other) is {\it explicitly} present, and leads to a divergence when $q^2 \to 0$,
while $r^2 = p^2 \neq 0$. In fact, unlike $\Gp_{\alpha\mu\nu}(q,r,p)$, the direct substitution of $q^2=0$ into the $\widehat{X}_3$ of \1eq{eq:X10_mBC}
yields simply ${\widehat X}_3(r^2) = -\frac{ [ J(r) - J(0) ] }{r^2}$ (in Euclidean space), which is only logarithmically divergent [due to $J(0)$] as long as 
$r^2\neq 0$. 

({\it c}) The above arguments hold also for the full (non-Abelian) $X_3(q,r,p)$, given that
$F(q)$ is completely finite, while the $A_i(q,p,r)$ are at most logarithmically divergent in the infrared, as was found in~\cite{Aguilar:2018csq}.

({\it d}) A concrete manifestation of the divergences captured by \1eq{derJ}
will be encountered shortly in Sec.~\ref{limits}, in the context of the ``totally symmetric'' and ``asymmetric'' configurations,
which fulfill the kinematic circumstances described above, being both special cases of \1eq{pcond}, for $\theta = 2\pi/3$ and $\theta = \pi$,
respectively.

Let us next turn  to $X_{10}$; this particular form factor  vanishes identically at both
tree and one-loop levels~\cite{Davydychev:1996pb}. As we can see in Fig.~\ref{fig:X10_gen_fig}, 
$X_{10}(q^2,p^2,\theta)$ does {\it not} vanish nonperturbatively; note, however, 
that it is extremely suppressed in comparison with the other $X_i$,
tending rapidly to zero whenever any of its momenta becomes large.

\begin{figure}[ht]
\begin{minipage}[b]{0.45\linewidth}
\centering
\includegraphics[scale=0.4]{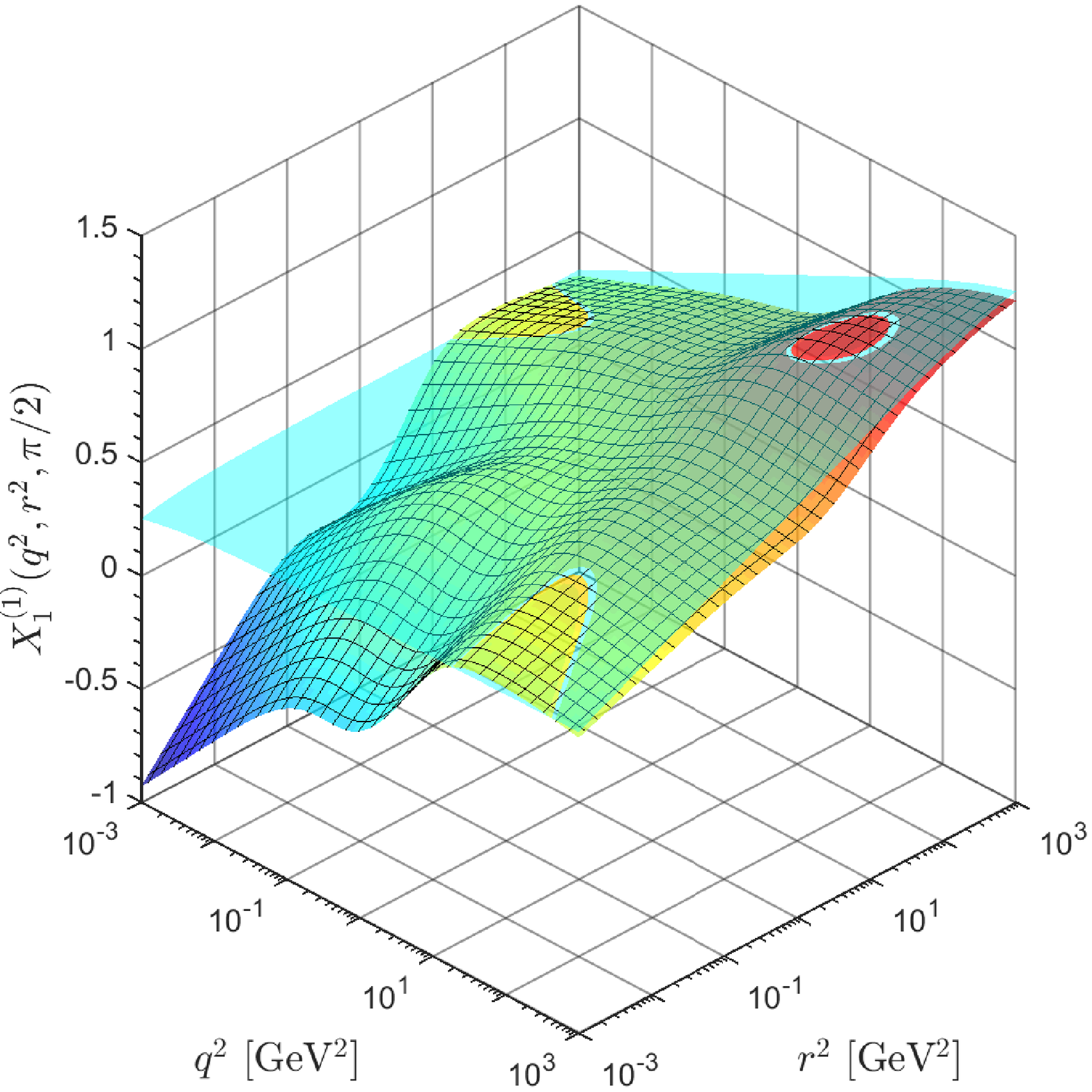}
\end{minipage}
\hspace{0.25cm}
\begin{minipage}[b]{0.45\linewidth}
\includegraphics[scale=0.32]{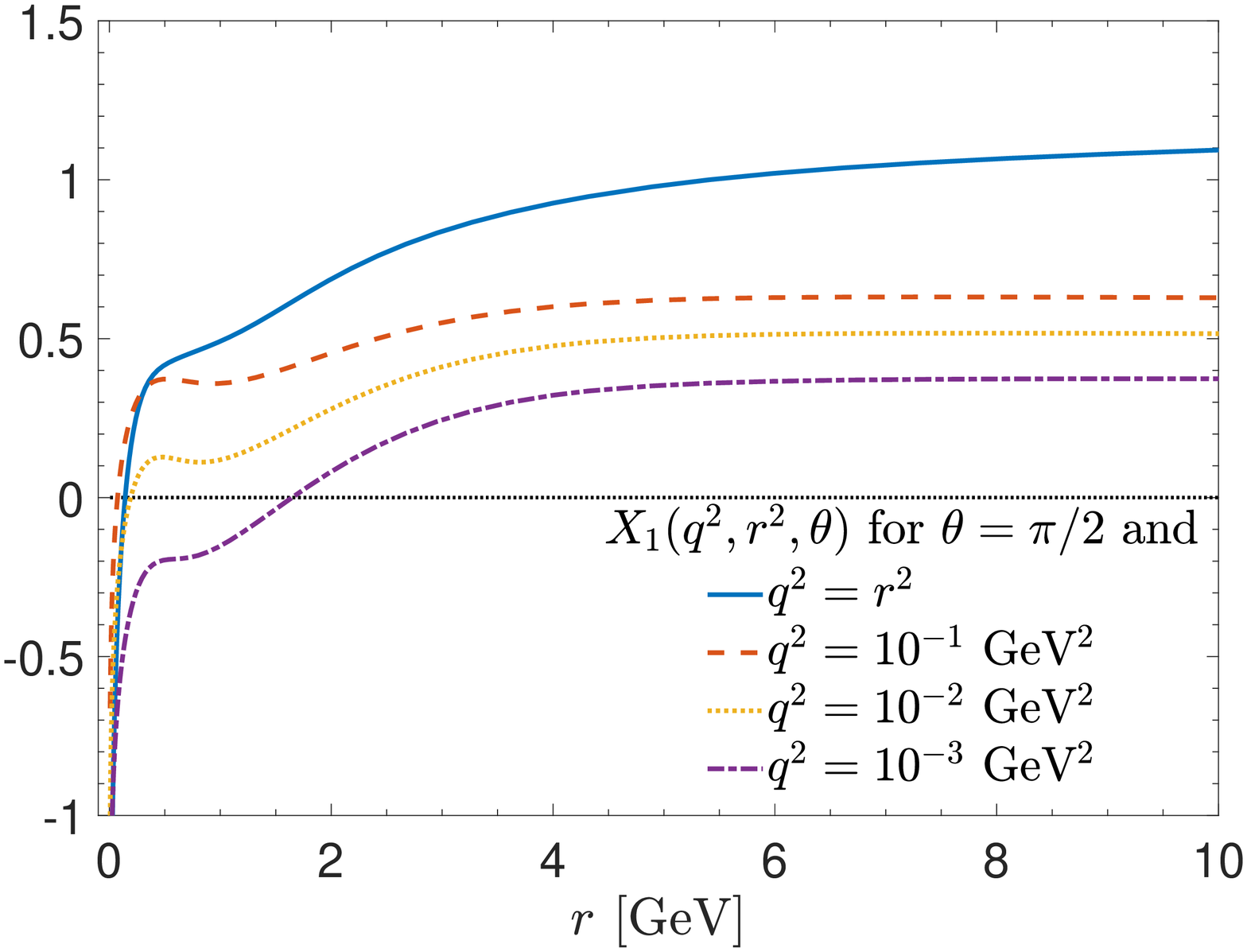}
\end{minipage}
\caption{
Comparison of the nonperturbative (colored surface) and the one-loop (cyan surface) results  for  $X_1(q^2,r^2,\pi/2)$  
(left panel). Special kinematic limits  of $X_1$ 
for a fixed values of $q^2$ when $\theta=\pi/2$.} 
\label{fig:3d_one}
\end{figure}

We end this subsection by comparing the infrared suppression of the 
nonperturbative  $X_1$ with that obtained from a direct one-loop calculation.
Specifically, 
in  the left  panel of  Fig.~\ref{fig:3d_one} we compare our 
result for  $X_1(q^2,r^2,\pi/2)$ (colored surface) with the corresponding one-loop expression,
given by Eq.~\eqref{eq:X1_ortho_pert} (cyan surface). Evidently,
the colored surface is considerably more ``tilted'' towards the infrared 
region,  due to the  presence of the zero crossing.  It is also 
interesting to observe that  $X_1$ reaches its maximum value along the curve projected on the  ``diagonal'' plane\footnote{The one-loop expression for this slice is given by Eq.~\eqref{ortho-sym}.},
where $q^2=r^2$, and then drops in all directions. To 
appreciate this effect more clearly, in the right panel of Fig.~\ref{fig:3d_one} we selected three additional 
slices of the 3D plot; indeed, 
the symmetric limit corresponds to the highest kinematic configuration of $X_1(q^2,r^2,\pi/2)$.

\subsection{\label{otherxi} Evaluation of the remaining form factors: an example}

In principle, 
if the functional dependence of $X_1$, $X_2$, and $X_3$ on $q$, $r$, and $p$ is known, 
the remaining $X_i$ can be obtained by invoking the
Bose symmetry relations of Eqs.~\eqref{eq:Xi_bose};
for instance, $X_4(q,r,p)$ can be obtained by permuting the arguments of $X_1(q,r,p)$ to $X_1(r,p,q)$.
In practice, however, what one has is the values 
of $X_1(q,r,p)$ tabulated for a grid of points for $q^2$, $r^2$ and $\theta$ (the angle between $q$ and $r$),
which we represent as $X_1(q^2,r^2,\theta)$. In order to evaluate the data point $X_4(q^2,r^2,\theta) = X_1(r^2,p^2,\varphi)$, where $\varphi$ is the angle between $r$ and $p$, one invokes momentum conservation to relate $p^2 = q^2 + r^2 + 2 q\cdot r$. Similarly, one finds for the angle
\be 
\varphi = \cos^{-1} \left( \frac{r\cdot p}{|r||p|} \right) = \cos^{-1} \left[ - \frac{( |r| + |q|\cos\theta )}{\sqrt{q^2 + r^2 + 2 |q||r|\cos\theta}} \right]\,.
\label{var}
\ee
Then, one carries out a three-dimensional interpolation, using, for example,
tensor products of B-splines~\cite{de2001practical}, and obtains the value of $X_1$ at $(r^2,p^2,\varphi)$. 

\begin{figure}[ht]
\begin{minipage}[b]{0.45\linewidth}
\centering
\includegraphics[scale=0.4]{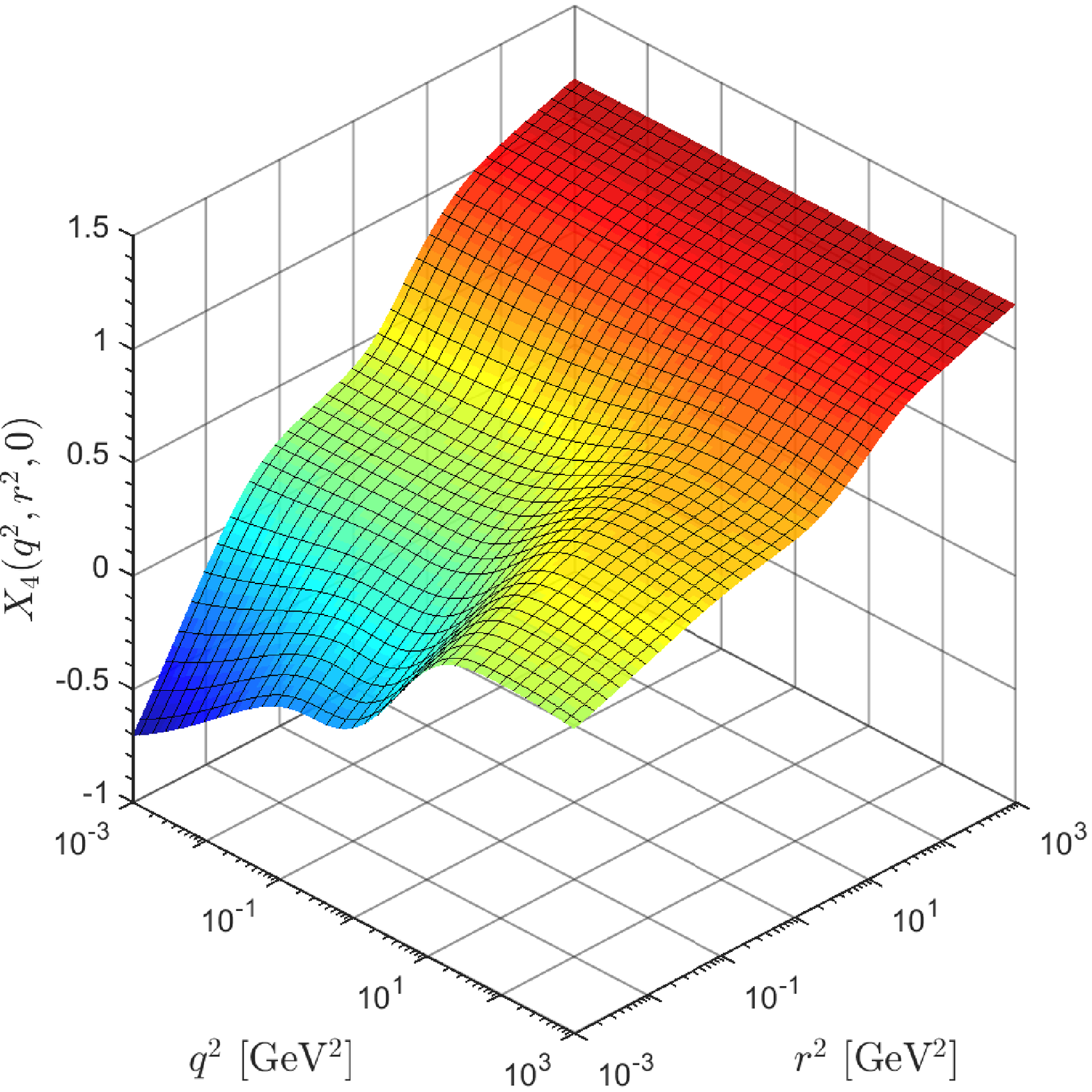}
\end{minipage}
\hspace{0.25cm}
\begin{minipage}[b]{0.45\linewidth}
\includegraphics[scale=0.4]{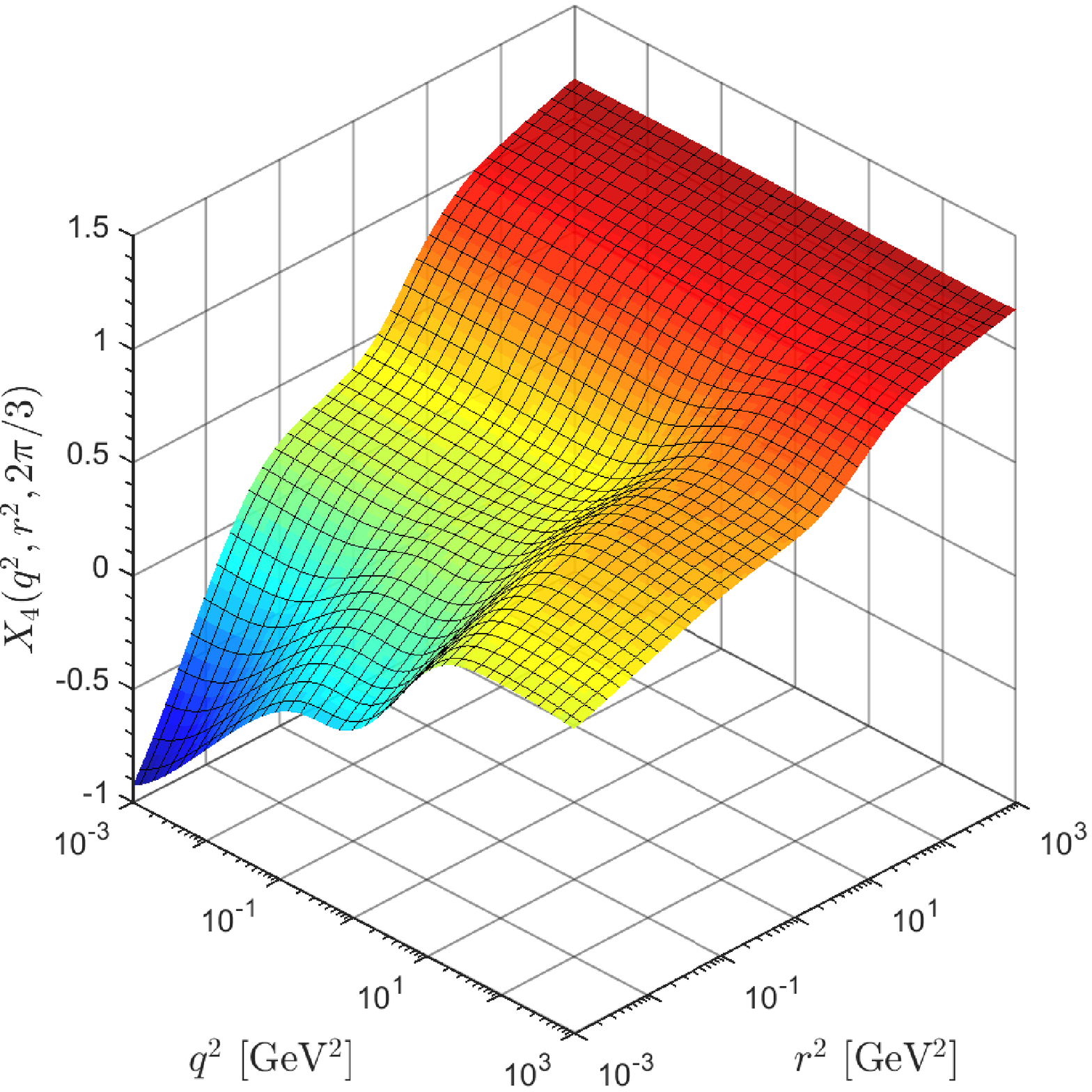}
\end{minipage}
\caption{ The form factor $X_4(q^2,r^2,\theta)$ for $\theta = 0$ (left) and  $2\pi/3$ (right) obtained from  $X_1(r^2,p^2,\varphi)$ using the Eq.~\eqref{var}.}
\label{fig:X4_gen_fig}
\end{figure}

In Fig.~\ref{fig:X4_gen_fig} we show the result of the exercise described above for $X_4(q^2,r^2,\theta)$,
for two representative values of $\theta$; notice that $X_4$ is {\it not} symmetric under the exchange of $q$ and $r$,
a fact that is clearly reflected in the shape of the surfaces obtained.

\subsection{\label{limits} Special kinematics limits}

In this subsection we extract two special kinematic configurations from the general solutions for $X_i$ reported above,
and compare them with the corresponding one-loop results, given in Appendix~\ref{app:pert}. 

{\it(i)} First we consider the \emph{totally symmetric limit}, obtained when 
\bea
q^2 = p^2= r^2 = Q^2\,, \quad q\cdot p = q\cdot r = p\cdot r = -\frac{1}{2}Q^2\,, \quad \theta= 2\pi/3\,;
\label{defsym}
\eea
the form factors in this configuration will be denoted by $X_i(Q)$.

Recalling that $X_2$ and $X_{10}$ are anti-symmetric under the exchange of at least two of their arguments
[see Eq.~\eqref{eq:Xi_more_bose}], it is clear that, in this particular configuration, they both vanish identically.
Therefore, we consider only $X_1(Q)$ and $X_3(Q)$, which 
may be obtained as a projection on the plane $q^2 = r^2$ of their 3D surfaces, shown 
on the bottom left sides of Figs.~\ref{fig:X1_gen_fig} and~\ref{fig:X3num_gen_fig}, respectively.

\begin{figure}[t!]
\begin{minipage}[b]{0.45\linewidth}
\centering
\includegraphics[scale=0.32]{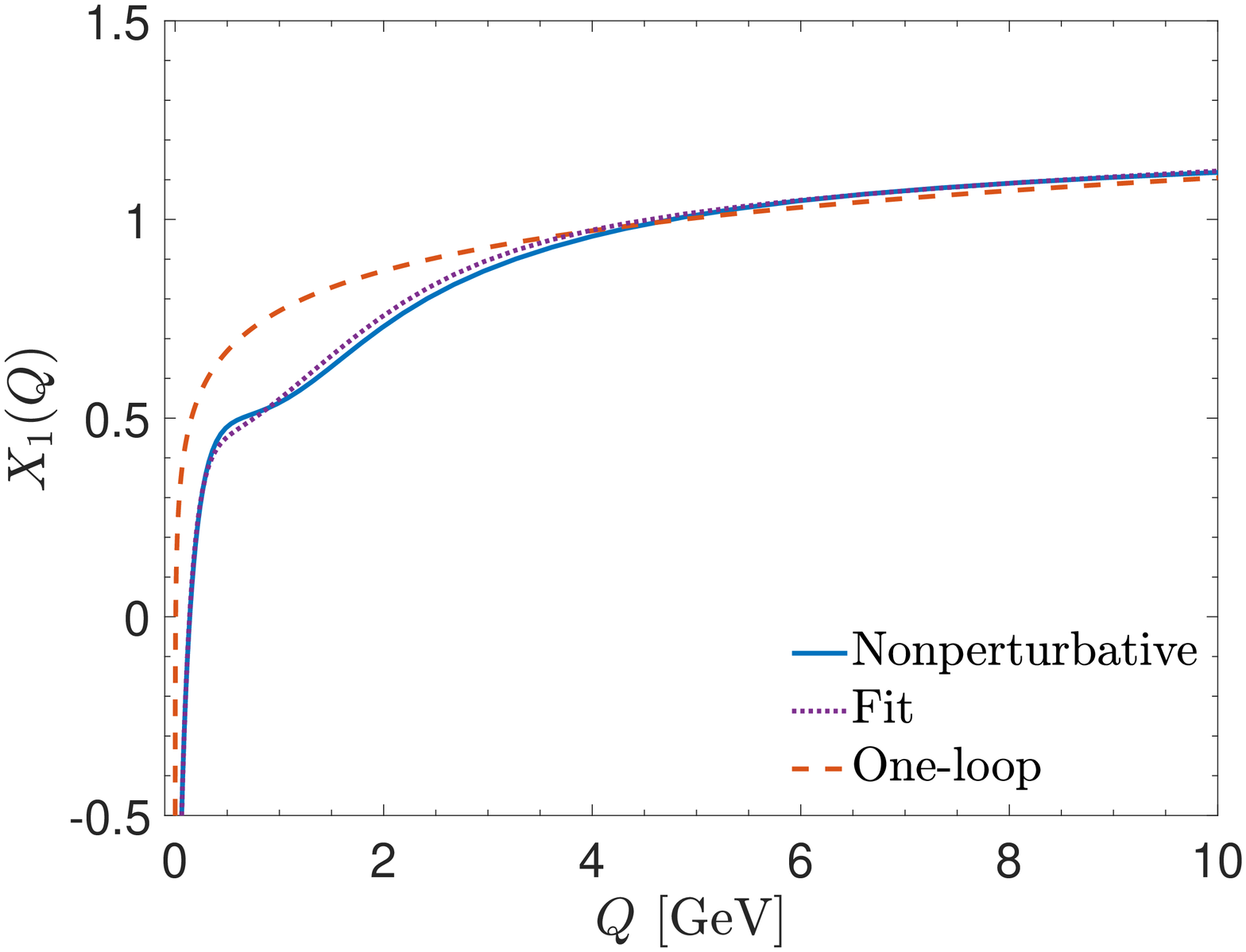}
\end{minipage}
\hspace{0.25cm}
\begin{minipage}[b]{0.45\linewidth}
\includegraphics[scale=0.32]{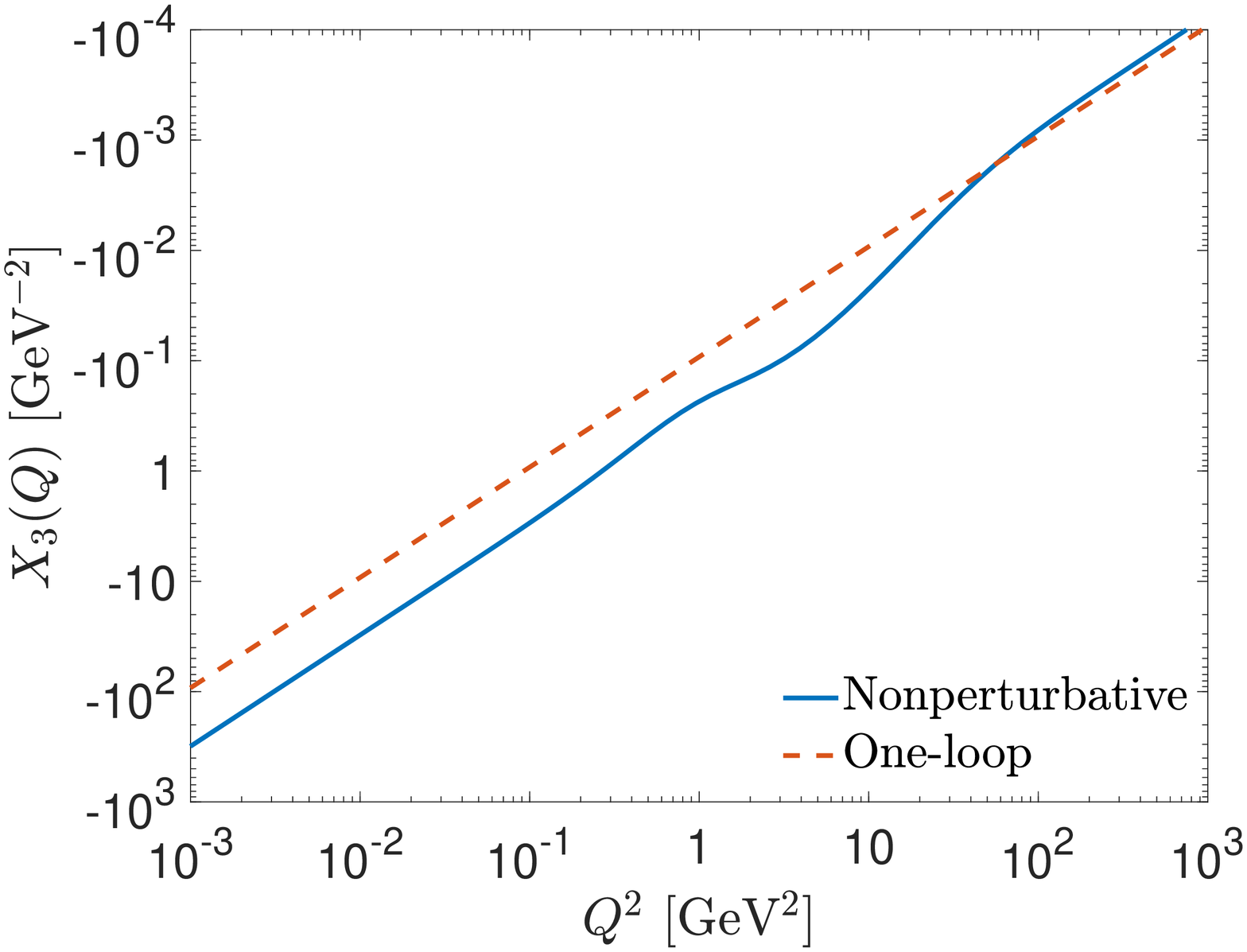}
\end{minipage}
\caption{
Comparison of the nonperturbative  form factors  $X_1(Q)$ (left panel) and  $X_3(Q)$ (right panel) with their one-loop counterparts, given by Eqs.~\eqref{eq:3g_pert} (red dashed), in the totally symmetric limit. In the left panel we also plot the fit for $X_1(Q)$ given by Eq.~\eqref{fit_X1} with $\gamma=0$ (purple dotted).} 
\label{fig:X1_pert}
\end{figure}

In the left panel of Fig.~\ref{fig:X1_pert}, we can see that $X_1(Q)$ displays a notable discrepancy from its 
one-loop behavior [see \1eq{eq:3g_pert}] in the window of momenta  \mbox{$0.5\,\mbox{GeV} \le Q \le 3.0 \,\mbox{GeV}$}.
In that range, $X_1(Q)$ suffers an abrupt change of curvature, which, at first sight, might be considered as a numerical artifact.
However, as we will see in Sec.~\ref{sec:lattice_comp}, this ``bending'' is crucial for reproducing a characteristic ``knee''
that appears in the lattice data in the same region of momenta.

In the same figure we also show a physically motivated fit for $X_1(Q)$ (purple dotted line), which is in good agreement with our nonperturbative result and 
recovers the one-loop result in the ultraviolet. Specifically, 
\begin{align}
X_1(Q) =& 1 + \frac{ C_\mathrm{A} \alpha_s }{96\pi}\left[ 1 + \frac{\kappa_1}{ 1 + \left(Q^2/\kappa_2\right)^{1 + \kappa_3} } \right]\bigg\lbrace 33\ln\left[ \frac{ Q^2 + \rho_{\ell} m^2(Q^2) }{\mu^2} \right] + \ln\left(\frac{Q^2}{\mu^2}\right)\bigg\rbrace \nonumber\\
& + \frac{ C_\mathrm{A} \alpha_s }{16\pi}(1 - \mathrm{I} ) \,,
\label{fit_X1}
\end{align}
where $m^2(Q^2)$ is given by Eq.~\eqref{eq:gmass} and the corresponding value of $\gamma$ should be used, while $\mathrm{I}$ is given in Eq.~\eqref{eq:Icel}.  The fitting parameters for the case where \mbox{$\gamma = 0$} are \mbox{$\kappa_1 = 135.3$}, \mbox{$\kappa_2 = 0.086\,\text{ GeV}^2$}, 
\mbox{$\kappa_3 = 0$}, and \mbox{$\rho_{\ell} = 140.4$}.
%

As for $X_3(Q)$, the present kinematic limit 
is precisely of the type considered in Sec.~\ref{subsec:Xires},
leading to \1eq{derJ}; note that the substitution of $\theta= 2\pi/3$ into \1eq{pcond} yields indeed $p^2=Q^2$.
Given that both  $X_3(Q)$ and its perturbative counterpart [see \1eq{eq:3g_pert}] 
diverge as $1/Q^2$ in the infrared, they are displayed in the log-log plot shown on the right panel of Fig.~\ref{fig:X1_pert}.
The coincidence with the perturbative result is quite satisfactory in the ultraviolet, but a considerable
departure is observed as one moves towards the infrared, where the two curves run nearly ``parallel'' to each other,
with the nonperturbative $X_3(Q)$ (blue curve) being about a factor of 4-5 larger. Evidently, even though both curves
are dominated by the pole $1/Q^2$, the values of their corresponding residues are rather different.

{\it(ii)} The \emph{asymmetric limit}, defined when
\bea
p = 0\,, \quad  r = - q\,, \quad \theta=\pi\,;
\label{defasym}
\eea
in what follows we will express our results for this configuration in terms of the momentum $q$, {\ie} $X_i(q^2,q^2,\pi)$.
In this configuration, the tensorial structure of $\GL^{\alpha\mu\nu}(q,r,p)$
reduces to that given in Eq.~\eqref{eq:asym_tensors}. 

\begin{figure}[!h]
\begin{minipage}[b]{0.45\linewidth}
\centering
\includegraphics[scale=0.32]{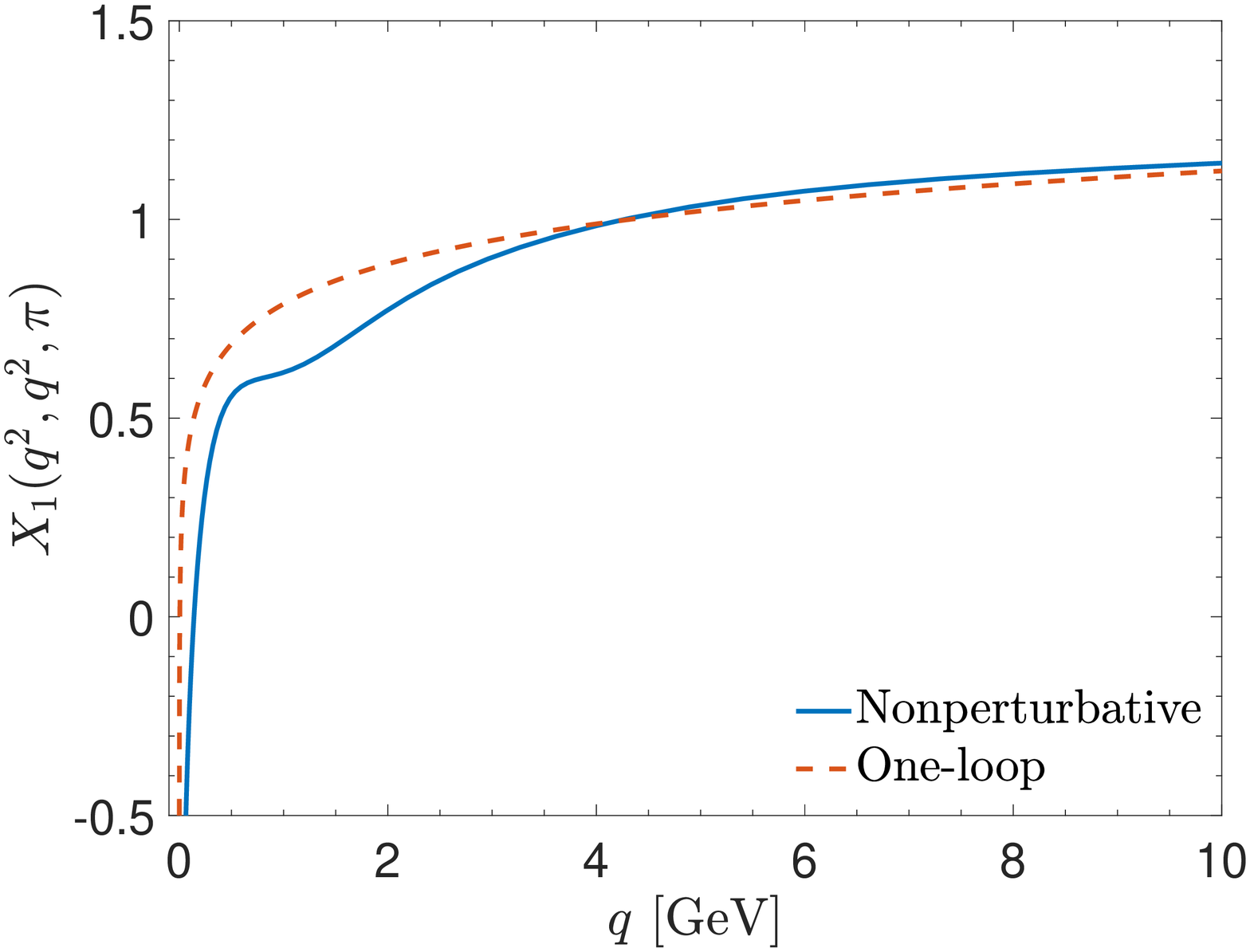}
\end{minipage}
\hspace{0.25cm}
\begin{minipage}[b]{0.45\linewidth}
\includegraphics[scale=0.32]{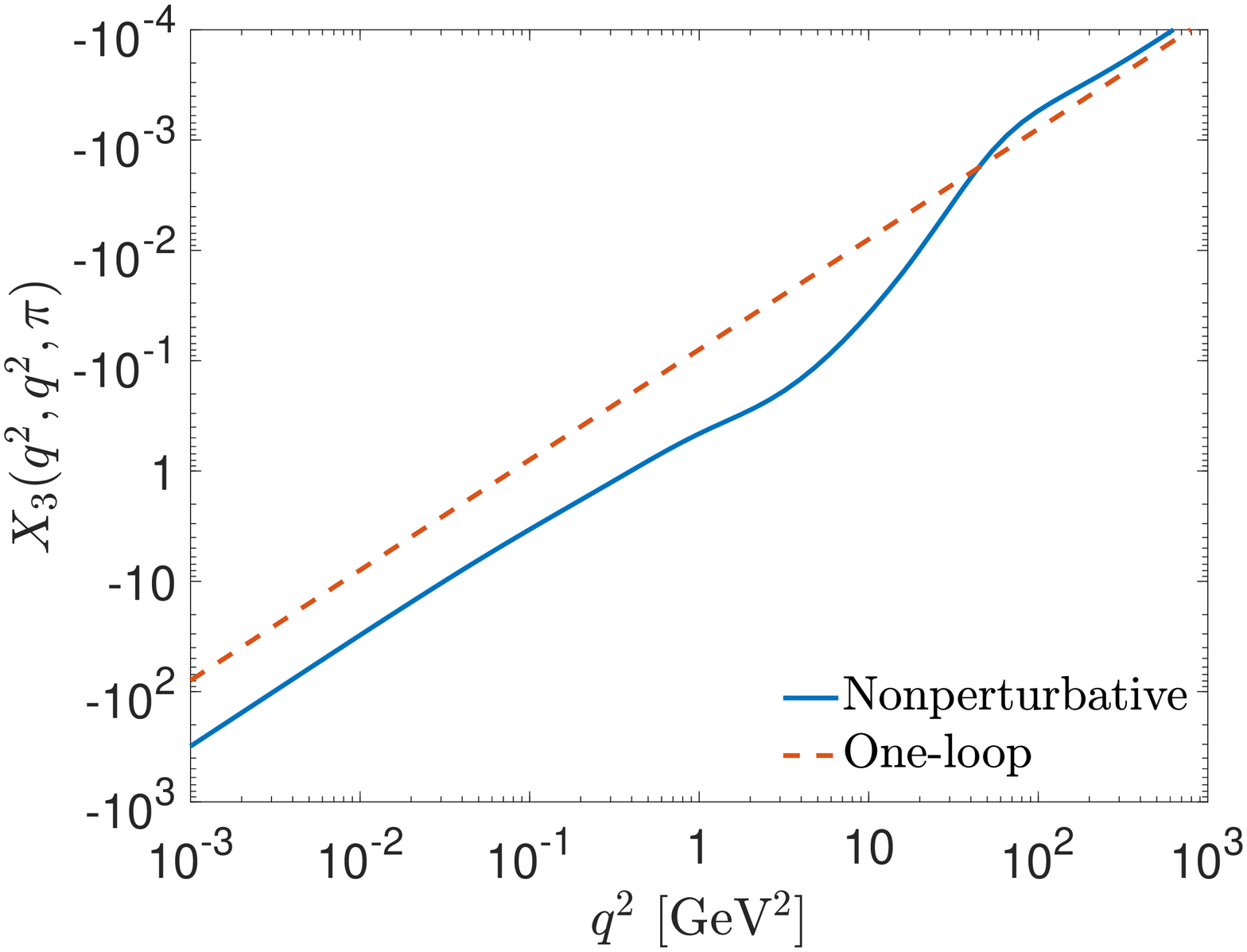}
\end{minipage}
\caption{Comparison of the nonperturbative  form factors  $X_1^{(1)}(q^2,q^2,\pi)$ (left panel) and  $X_3^{(1)}(q^2,q^2,\pi)$  (right panel) with their one-loop counterparts given by Eqs.~\eqref{eq:X1_and_X3_asym} (red dashed) in the  asymmetric limit.}
\label{fig:X1_perta}
\end{figure}

In Fig.~\ref{fig:X1_perta}
we show $X_1^{(1)}(q^2,q^2,\pi)$ (left panel) and $X_3^{(1)}(q^2,q^2,\pi)$ (right panel), which are 
clearly very similar to those  obtained in the symmetric limit.
More specifically, $X_1^{(1)}(q^2,q^2,\pi)$ deviates mildly from its one-loop behavior,
displaying the characteristic bending in the same range of momenta, 
while $X_3^{(1)}(q^2,q^2,\pi)$ diverges again as a pole, corresponding to the case where, for $\theta=\pi$,  \1eq{pcond} yields $p^2=0$.
As expected, in the ultraviolet regime both form factors tend towards
the behavior predicted by the one-loop result given in Eq.~\eqref{eq:X1_and_X3_asym}.

\section{\label{sec:comp} Comparison with previous results}

In this section we present a direct comparison between  our results and those obtained from  {\it (i)} the SDE analysis of~\cite{Blum:2014gna,Huber:2018ned}, and {\it (ii)} the lattice simulation of~\cite{Athenodorou:2016oyh}.

For the purposes of this section, it is convenient to introduce the following \emph{transversely projected}
counterparts of $\fatg_{\alpha\mu\nu}(q,r,p)$, $\Gnp_{\alpha\mu\nu}(q,r,p)$, and
${\Gamma}^{(0)}_{\alpha\mu\nu}(q,r,p)$, defined as
\bea
{\overline\fatg}_{\alpha\mu\nu}(q,r,p) &:=& P_{\alpha'\alpha}(q)P_{\mu'\mu}(r)P_{\nu'\nu}(p){\fatg}^{\alpha'\mu'\nu'}(q,r,p) \,,
\nonumber\\     
{\overline\Gnp}_{\alpha\mu\nu}(q,r,p) &:=& P_{\alpha'\alpha}(q)P_{\mu'\mu}(r)P_{\nu'\nu}(p)\Gnp^{\alpha'\mu'\nu'}(q,r,p)\,, 
\nonumber\\
{\overline \Gamma}^{(0)}_{\alpha\mu\nu}(q,r,p) &:=& P_{\alpha'\alpha}(q)P_{\mu'\mu}(r)P_{\nu'\nu}(p) \Gamma^{(0)\alpha'\mu'\nu'}(q,r,p)\,.
\label{projvert}
\eea
Note that, by virtue of \1eq{eq:transvp}, we have the important relation
\be
{\overline\fatg}_{\alpha\mu\nu}(q,r,p) = {\overline\Gnp}_{\alpha\mu\nu}(q,r,p) \,.
\label{drop}
\ee

Next we introduce the general projector $L(q,r,p)$, given by
\bea
L(q,r,p) &=& \frac{W^{\alpha\mu\nu}(q,r,p){\overline\fatg}_{\alpha\mu\nu}(q,r,p)}
{W^{\alpha\mu\nu}(q,r,p)W_{\alpha\mu\nu}(q,r,p)}
\nonumber\\
 &=& \frac{W^{\alpha\mu\nu}(q,r,p){\overline\Gnp}_{\alpha\mu\nu}(q,r,p)}
{W^{\alpha\mu\nu}(q,r,p)W_{\alpha\mu\nu}(q,r,p)}\,,
\label{eq:GammaSym_proj1}
\eea
where in the second step we have used \1eq{drop}. The precise form of the tensor $W^{\alpha\mu\nu}(q,r,p)$
will depend on the particular circumstances considered.

\subsection{\label{previous} Comparison with SDE-derived results}

Next, we compare our results
with those in~\cite{Blum:2014gna,Huber:2018ned}. In that work, 
an approximate version of the SDE governing
the \emph{transversely projected} three gluon vertex was derived. To make contact, 
we consider the $L(q,r,p)$ of 
Eq.~\eqref{eq:GammaSym_proj1}, and carry our the substitution $W_{\alpha\mu\nu}(q,r,p)\to W^{\rm \s{SDE}}_{\alpha\mu\nu}(q,r,p)$, 
where
\be 
W^{\rm \s{SDE}}_{\alpha\mu\nu}(q,r,p) = {\overline \Gamma}^{(0)}_{\alpha\mu\nu}(q,r,p)\,,
\label{wsde}
\ee
denoting the resulting expression by $L^{\rm \s{SDE}}(q,r,p)$.

Expanding $\Gnp^{\alpha\mu\nu}(q,r,p)$ in the basis of \2eqs{eq:3g_sti_structure}{eq:3g_tr_structure}
and substituting into \1eq{eq:GammaSym_proj1}, 
one may express $L^{\rm \s{SDE}}(q,r,p)$ in terms of 
the various $X_i$ and $Y_i$. Here, we do not report the general expression for $L^{\rm \s{SDE}}(q,r,p)$, but
consider instead the following representative kinematic limits:
  
\begin{itemize}
\item[($i$)] \textit{ Totally symmetric configuration:} Fixing
the momenta and the angle $\theta$ according to Eq.~\eqref{defsym}, one
obtains 
\be\label{eq:huber_sym}
L^{\rm \s{SDE}}(Q) = X_1(Q) - \frac{10}{11}Q^2X_3(Q) + \frac{5}{11}Q^4Y_1(Q) - \frac{4}{11}Q^2 Y_4(Q) \,. 
\ee

\item[($ii$)] \textit{ Orthogonal-symmetric:} In this configuration, the momenta $q$ and $r$ are orthogonal and have equal magnitudes, \emph{i.e.} $\theta = \pi/2$ and $q^2 = r^2$, which also implies that $p^2 = 2r^2$. In this case, the corresponding projection yields
\begin{align}
L^{\rm \s{SDE}}(r^2,r^2,\pi/2) =& \frac{1}{7}[ X_1(r^2,r^2,\pi/2) + 6 X_1(2r^2,r^2,3\pi/4) \nonumber \\
& - r^2 X_3(r^2,r^2,\pi/2) - 8 r^2 X_3(2r^2,r^2,3\pi/4)   \label{eq:huber_ortho_eq}\\
& + r^4 Y_1(r^2,r^2,\pi/2) + 4 r^4 Y_1(2r^2,r^2,3\pi/4) - 3 r^2 Y_4(r^2,r^2,\pi/2) ] \,.
\nonumber 
\end{align}
\item[($iii$)] \textit{Asymmetric limit:}  Fixing the momenta according to Eq.~\eqref{defasym}, we obtain  
\be\label{eq:huber_asym}
L^{\rm \s{SDE}}(q) =  X_1(q^2,q^2, \pi) - q^2X_3(q^2,q^2, \pi)\,.
\ee
Note that the above kinematic configuration also corresponds to the so-called  ``orthogonal soft'' limit,
obtained in~\cite{Blum:2014gna,Huber:2018ned}~\footnote{The orthogonal soft configuration defined in~\cite{Blum:2014gna} corresponds to the limit $q\to 0$ and $\theta=\pi/2$.}. To establish their equivalence, first notice
that  Eq.~\eqref{eq:GammaSym_proj1}, with the $W^{\rm \s{SDE}}_{\alpha\mu\nu}(q,r,p)$ 
given by Eq.~\eqref{wsde}, is symmetric under $p \leftrightarrow q$. Therefore, 
the limits of vanishing $p$ or $q$ lead to the same result.  In addition, 
when  $q\to 0$, evidently $|q||r|cos\theta =0$, and any dependence on the angle $\theta$ is washed out. Thus, \mbox{$L^{\rm \s{SDE}}(q^2,q^2, \pi)= L^{\rm \s{SDE}}(0,r^2,\pi/2)$}.

\end{itemize}

\begin{figure}[!t]
\begin{minipage}[b]{0.45\linewidth}
\centering
\includegraphics[scale=0.32]{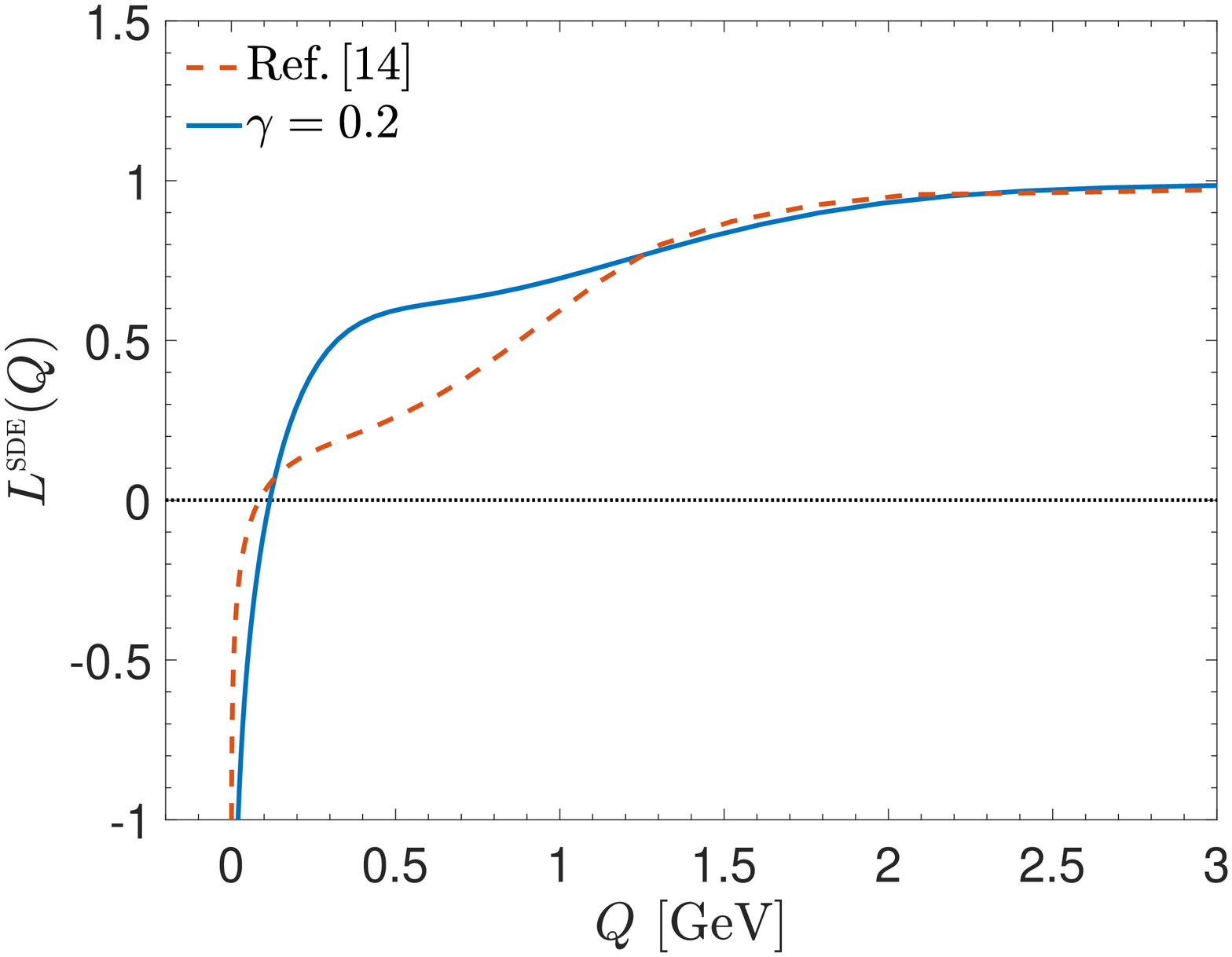}
\end{minipage}
\hspace{0.25cm}
\begin{minipage}[b]{0.45\linewidth}
\includegraphics[scale=0.32]{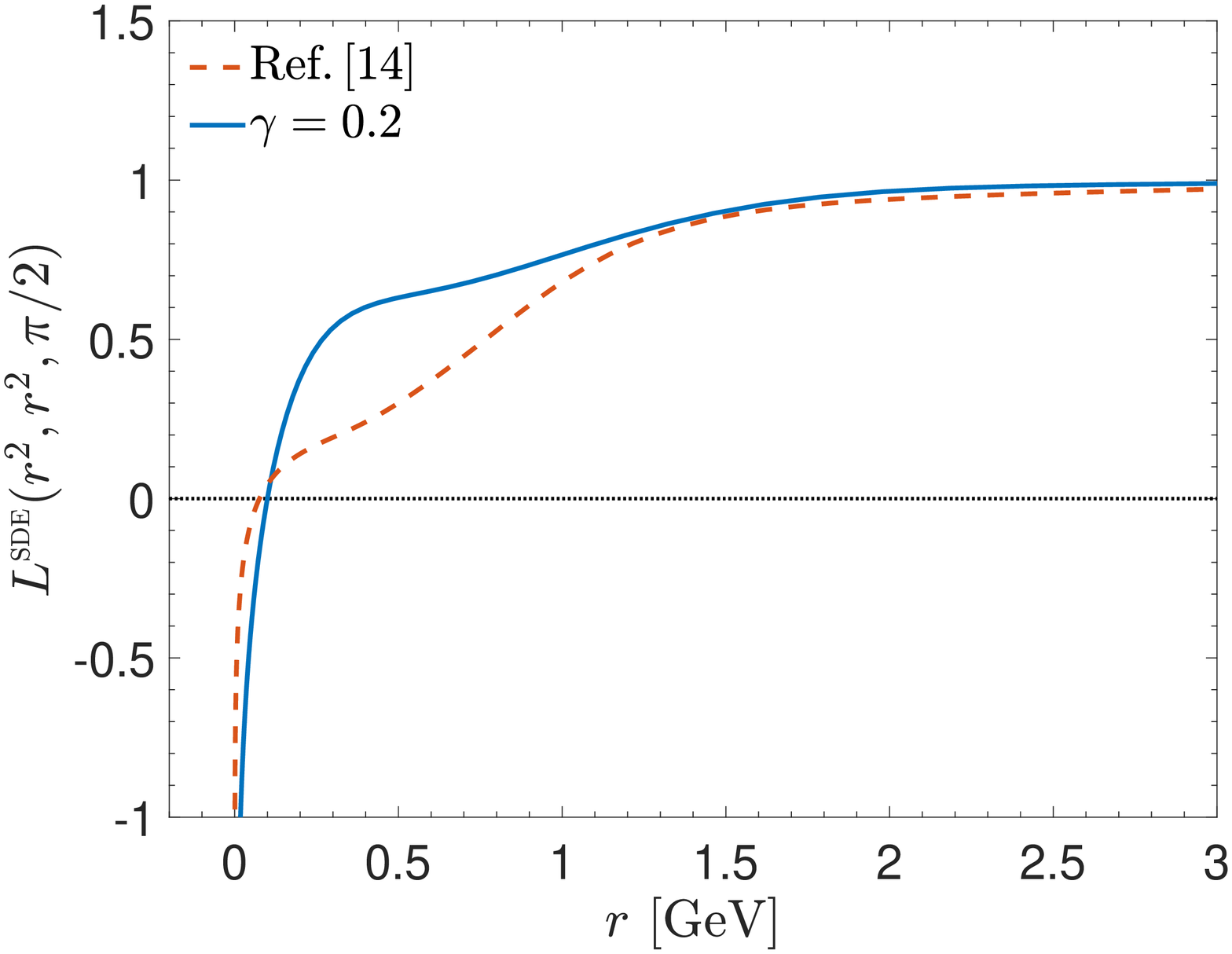}
\end{minipage}
\includegraphics[scale=0.32]{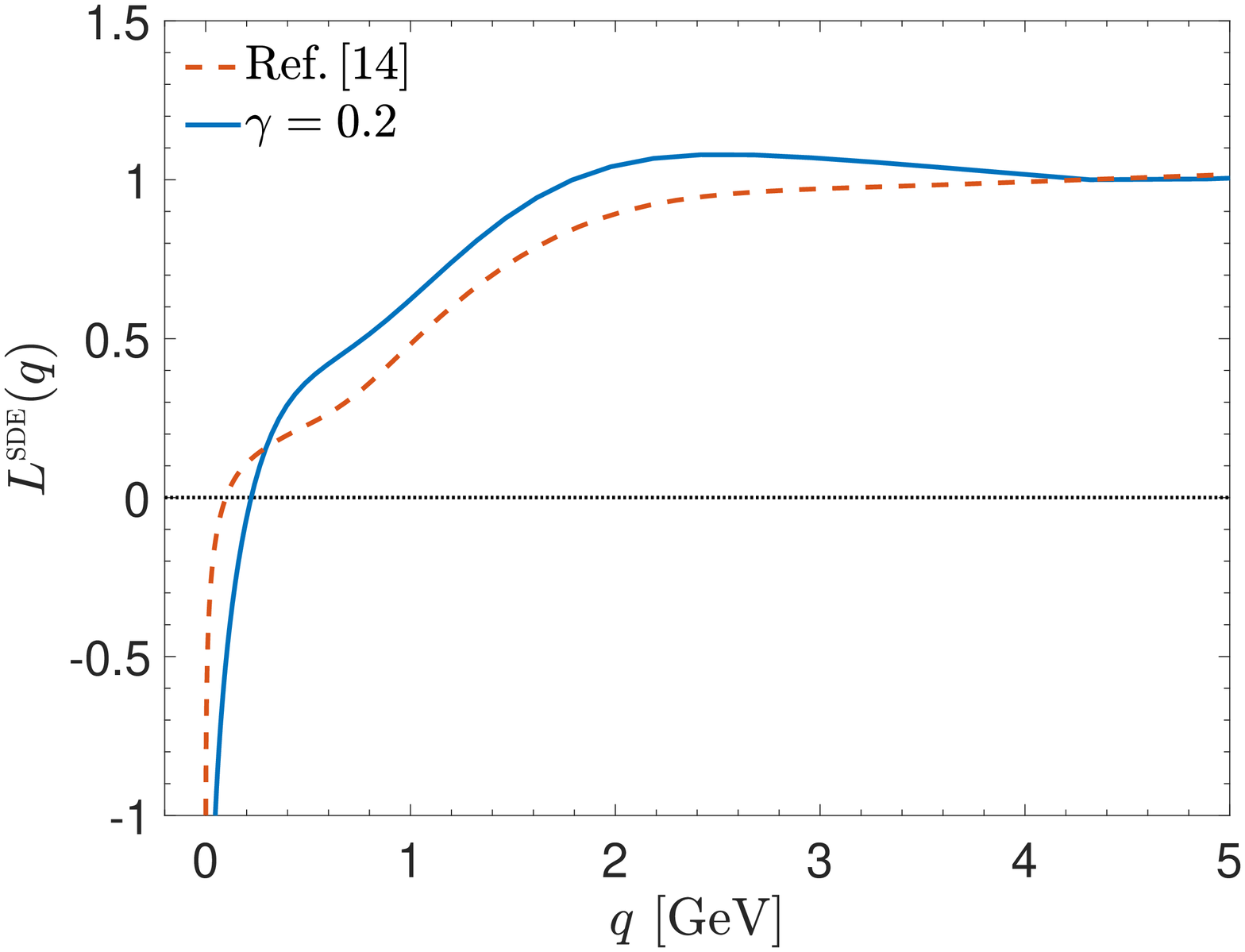}
\caption{
Comparison of our results for $L^{\rm \s{SDE}}$  with those of~\cite{Blum:2014gna,Huber:2018ned} for three kinematics: {\it(i)} the totally symmetric $L^{\rm \s{SDE}}(Q)$ (top left panel), {\it(ii)} the orthogonal-symmetric $L^{\rm \s{SDE}}(r^2,r^2,\pi/2)$
(top right panel), and {\it(iii)} asymmetric $L^{\rm \s{SDE}}(q)$ (bottom panel) configurations.}
 \label{fig:comprevious}
\end{figure}

An interesting property of the asymmetric configuration is
the fact that $L^{\rm \s{SDE}}(q)$  depends only on the $X_i$, while  the other two limits depend on both the $X_i$ and 
the $Y_i$.   Since our approach
does not allow the determination of $\GT(q,r,p)$, in what follows we set $Y_i = 0$ in Eqs.~\eqref{eq:huber_sym}~and~\eqref{eq:huber_ortho_eq}.

Evidently, the omission of the transverse form factors $Y_i$ in the evaluation of $L^{\rm \s{SDE}}(Q)$  and $L^{\rm \s{SDE}}(r^2,r^2,\pi/2)$  introduces an error, whose size in the infrared is difficult
to estimate without a concrete calculation. At this point we may only 
report the perturbative behavior of the terms  omitted 
from Eq.~\eqref{eq:huber_sym}, using the one-loop expressions for $Y_1(Q)$ and $Y_4(Q)$ given in 
\1eq{eq:3g_pert}. In particular, one finds that the one-loop combination amounts only to a small constant, namely 
\be
 \frac{5}{11}Q^4 Y_1^{(1)}(Q) - \frac{4}{11}Q^2Y_4^{(1)}(Q) = 0.039\,.
\label{transsym}
\ee

Unfortunately, the perturbative calculation for the transverse terms omitted in 
Eq.~\eqref{eq:huber_ortho_eq} is more cumbersome, since it mixes
the $Y_i$ in two kinematic limits.

In order to compare our results with the SDE calculations of~\cite{Blum:2014gna}, we first rescale the results appropriately, in order to ensure that they are renormalized at the same point. Specifically, we define a multiplicative renormalization constant $z_3$ for both sets of results, such that tree level value is restored at the
symmetric point, \emph{i.e.}, $z_3 L^{\rm \s{SDE}}(\mu) = 1$; subsequently, 
we rescaled $L^{\rm \s{SDE}}(r^2,r^2,\pi/2)$ and $L^{\rm \s{SDE}}(q)$ by the \emph{same} factor.

In Fig.~\ref{fig:comprevious}, we compare our results 
for $L^{\rm \s{SDE}}(Q)$  (left panel), $L^{\rm \s{SDE}}(r^2,r^2,\pi/2)$ (right panel), 
and  $L^{\rm \s{SDE}}(q)$ (bottom panel)  with those obtained  in~\cite{Blum:2014gna,Huber:2018ned}.  The general profiles of the curves are qualitatively similar, in the three kinematic limits, although considerable differences are observed at intermediate momenta.
Interestingly enough, the positions of the corresponding zero crossings practically
coincide in all configurations.

\subsection{\label{sec:lattice_comp} Comparison with the lattice}

Following the analysis presented in~\cite{Athenodorou:2016oyh}, we consider two particular cases of the $L(q,r,p)$ defined in \1eq{eq:GammaSym_proj1}.

First, for the symmetric configuration, we construct $\Glatt(Q)$ by setting 
$W_{\alpha\mu\nu}(q,r,p)\to W^{\rm sym}_{\alpha\mu\nu}(q,r,p)$, where 
\be
W^{\rm sym}_{\alpha\mu\nu}(q,r,p) = {\overline \Gamma}^{(0)}_{\alpha\mu\nu}(q,r,p)
+ \frac{1}{2r^2}(r-p)_\alpha (p-q)_\mu (q-r)_\nu  \,,
\label{wsym}
\ee
implementing subsequently the limit of \1eq{defsym}.

Second, for the asymmetric configuration, we evaluate $L^{\rm asym}(q)$ by replacing $W_{\alpha\mu\nu}(q,r,p)\to W^{\rm asym}_{\alpha\mu\nu}(q,r,p)$, where 
\be
W^{\rm asym}_{\alpha\mu\nu}(q,r,p) = 2q_{\alpha}P_{\mu\nu}(q)\,, 
\label{awsym}
\ee
taking finally the limit of \1eq{defasym}.

Expanding again ${\Gnp}_{\alpha\mu\nu}(q,r,p)$ in the basis of \2eqs{eq:3g_sti_structure}{eq:3g_tr_structure}, one finds that, in the symmetric configuration, Eq.~\eqref{eq:GammaSym_proj1} reduces to
\be  
\label{eq:GammaSym_Xi}
\Glatt(Q) = X_1(Q) - \frac{Q^2}{2} X_3(Q) + \frac{Q^4}{4} Y_1(Q) - \frac{Q^2}{2} Y_4(Q) \,, 
\ee
while for the asymmetric case, 
\be   
\label{eq:asyGamma}
L^{\rm asym}(q) =  X_1(q^2,q^2, \pi) - q^2X_3(q^2,q^2, \pi)\,. 
\ee
Thus, the combinations of form factors given in \2eqs{eq:GammaSym_Xi}{eq:asyGamma}  are precisely those considered in the lattice
simulations of~\cite{Athenodorou:2016oyh,Boucaud:2017obn}.
Notice that $L^{\rm asym}(q)$ given by Eq.~\eqref{eq:asyGamma} coincides
with the  projection $L^{\rm \s{SDE}}(q)$ of Eq.~\eqref{eq:huber_asym}.

Since our approach furnishes no information on $\GT(q,r,p)$, we will consider  the {\it approximate} version of Eq.~\eqref{eq:GammaSym_Xi} where  $Y_i = 0$, as was done in the previous subsection, More specifically, 
\be\label{eq:GammaSym_Xi_s}
\Glatt(Q) = X_1(Q) - \frac{Q^2}{2} X_3(Q) \,,
\ee
which will be used in the comparison with the lattice data. Let us simply mention that
the one-loop evaluation of the omitted terms gives rise to a small numerical constant,  
\be
 \frac{Q^4}{4} Y_1^{(1)}(Q) - \frac{Q^2}{2}Y_4^{(1)}(Q) = 0.08\,.
\label{transcont}
\ee
Note that, in order to perform a meaningful comparison, one must take into account the fact that
the lattice results of~\cite{Athenodorou:2016oyh} have been renormalized in a scheme which enforces independently that \mbox{$\Glatt(\mu) = 1$}
and \mbox{$L^{\rm asym}(\mu)=1$} at the renormalization point \mbox{$\mu = 4.3 \text{ GeV}$}.
Instead, we have computed the $X_i$ in the Taylor scheme, for both kinematic limits [see discussion at the end of Sec.~\ref{sec:BCS}].
To account for the difference introduced by the use of two distinct 
renormalization prescriptions, we rescale Eqs.~\eqref{eq:asyGamma} and~\eqref{eq:GammaSym_Xi_s} by a
{\it finite } renormalization constant, to be denoted by $z_3$, \emph{i.e.}
\bea 
\Glatt(Q) &\to & z_3^{\rm sym} \Glatt(Q) \,, \nonumber \\
L^{\rm asym}(q) &\to & z_3^{\rm asym}  L^{\s{\rm asym}}(q) \,.
\label{eq:re-renorm}
\eea
The numerical values of $z_3^{\rm sym}$ and $z_3^{\rm asym}$ are determined by requiring that $\Glatt(\mu)$
and $L^{\rm asym}(\mu)$ reduce to tree level, respectively. As expected on general grounds~\cite{Celmaster:1979km}, 
the discrepancy from unity is quite small; in particular, the choices 
of \mbox{$z_3^{\rm sym} = 0.95$} and  \mbox{$z_3^{\rm asym} = 0.93$} restore $\Glatt(\mu) = 1$ and $L^{\rm asym}(\mu)=1$, respectively.

\begin{figure}[!h]
\begin{minipage}[b]{0.45\linewidth}
\centering
\includegraphics[scale=0.32]{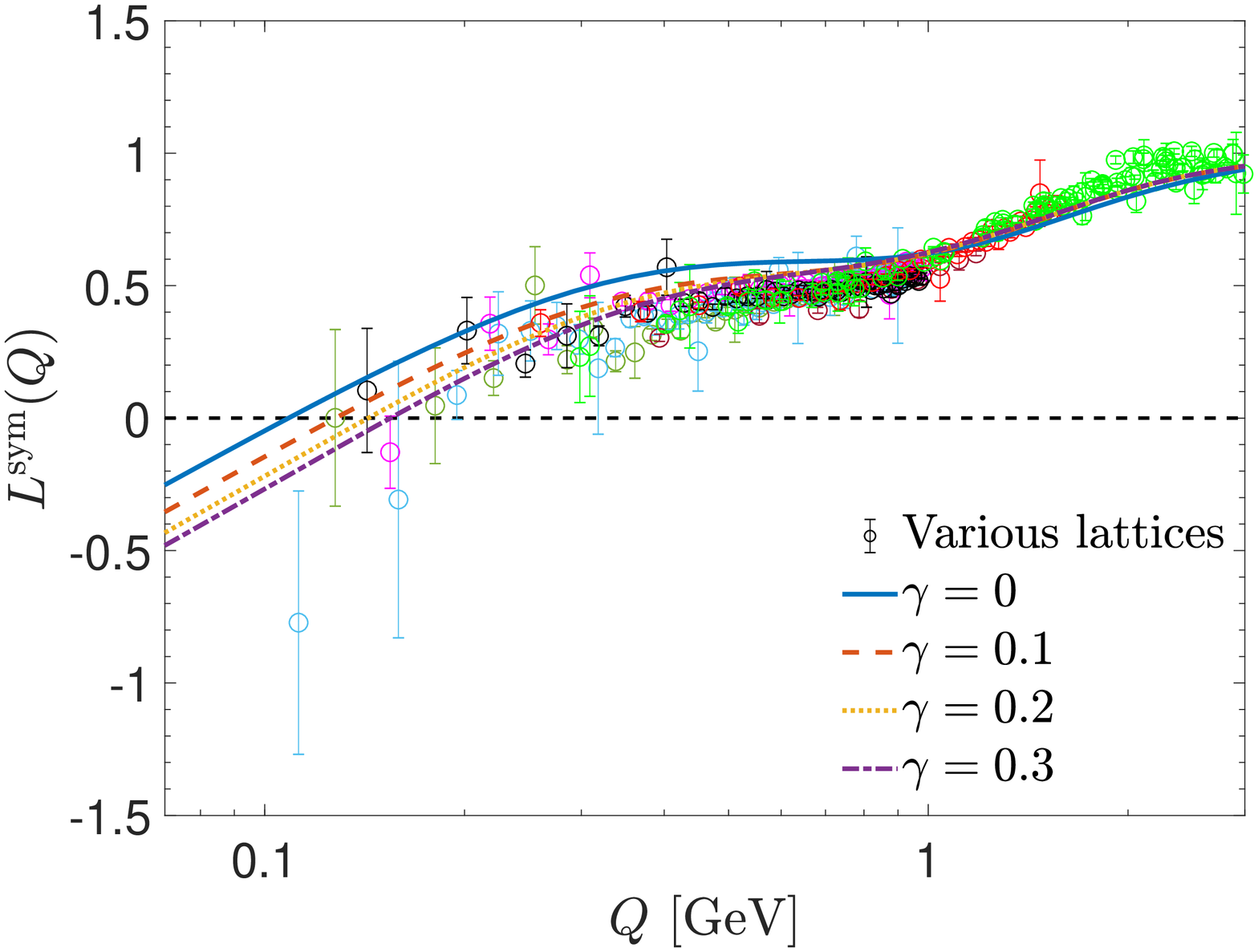}
\end{minipage}
\hspace{0.25cm}
\begin{minipage}[b]{0.45\linewidth}
\includegraphics[scale=0.32]{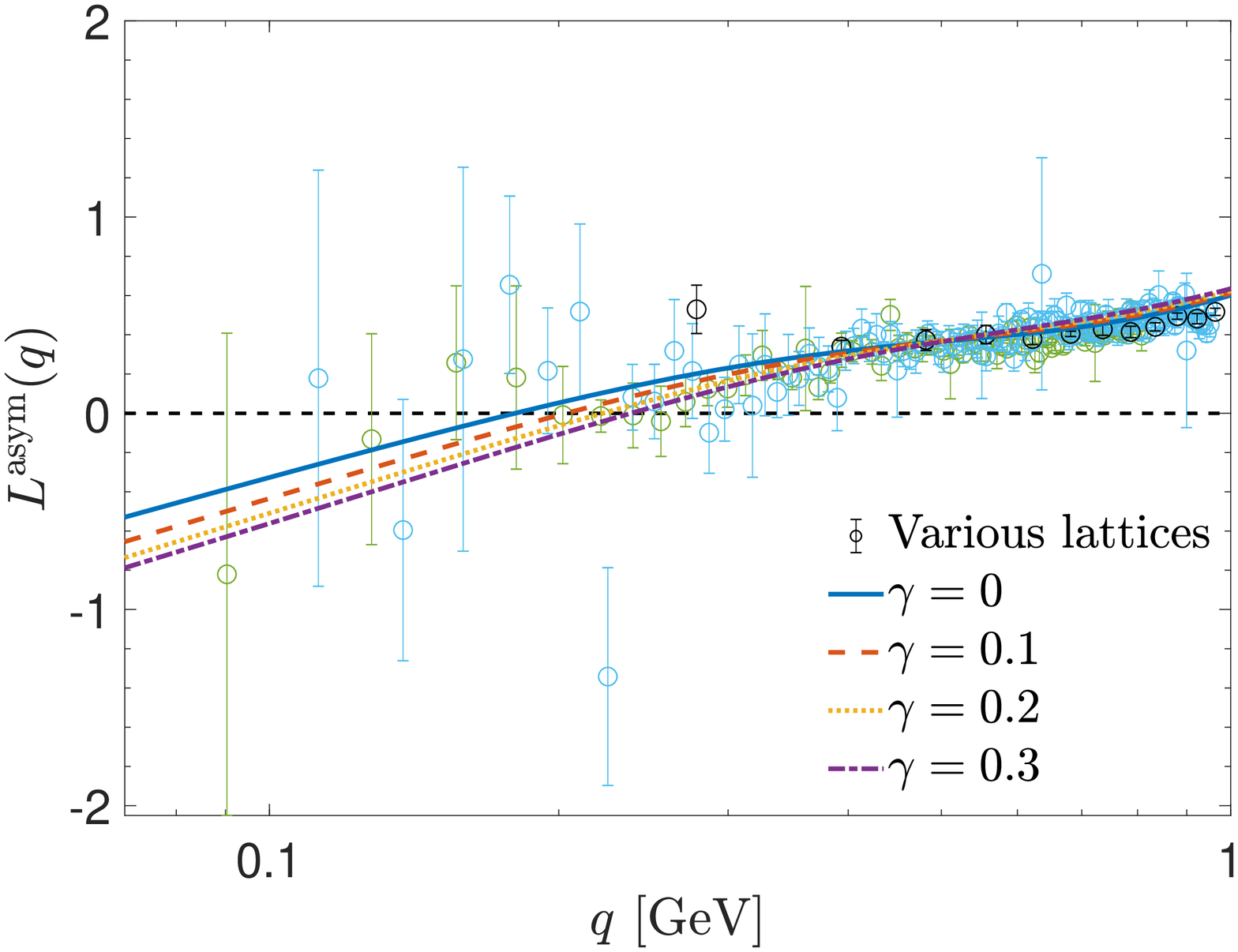}
\end{minipage}
\caption{Comparison of our results for the $\Glatt(Q)$ (left panel) given by Eq.~\eqref{eq:GammaSym_Xi_s} and for $L^{\rm asym}(q)$ (right panel) of Eq.~\eqref{eq:asyGamma} with the lattice data~\cite{Athenodorou:2016oyh} (circles). The curves were obtained varying the exponent $\gamma$ entering into the definition of the gluon mass given by Eq.~\eqref{eq:gmass}.}
\label{fig:Gamma_sym}
\end{figure}

In Fig.~\ref{fig:Gamma_sym} we compare the lattice data of~\cite{Athenodorou:2016oyh} with 
our results for $\Glatt(q)$ (left panel) and $L^{\rm asym}(q)$ (right panel), obtained for the different $J(q)$
shown in Fig.~\ref{fig:jm}; evidently, in both cases the agreement is rather good.
In fact, 
observe that, in the symmetric case, the lattice data display a change in the curvature (in the form of a ``knee'') around  $Q = 1 \text{ GeV}$. It is interesting to notice that, independently of the $J(q)$ employed, 
our results for $\Glatt(q)$  always exhibit this particular feature, which is clearly related to the abrupt bending observed at the level of the $X_1(Q)$ in the  Fig.~\ref{fig:X1_pert}.   Moreover, both the lattice data and our results present
the characteristic zero crossing, at momenta to be denoted by $q_0^{\rm sym}$ and $q_0^{\rm asym}$,
respectively, 
whose positions are located within the interval $[109,237] \text{ MeV}$ [see Table~\ref{table_crossing}].  Notice
that, within our approach, the precise location of the crossing $q_0$ 
depends on the value of the parameter $\gamma$, which controls the running of $m^2(q^2)$ in \1eq{eq:gmass}.

It is interesting to compute the amount by which $q_0^{\rm sym}$ and $q_0^{\rm asym}$ get shifted with
respect to $q_0^{\rm J}$, shown in the Fig.~\ref{fig:jm};  in Table~\ref{table_crossing}
we collect these numbers, in order to facilitate a direct comparison. As one may see, for all values of $\gamma$ the crossing of $\Glatt(Q)$ happens at a momentum about $23\%$ smaller than the value obtained from $J(q)$.
In the case of the $L^{\rm asym}(q)$, the change in the sign occurs at a momentum that
is located  $17\%-28\%$ more towards the ultraviolet with respect to $q_0^{\rm J}$.

\vspace{0.5cm}
\begin{table}[!h]
\begin{tabular}{|c|c|c|c|}
\hline
\hline
$\quad \gamma \quad $ &\; $q_0^{\rm J}$\;& $q_0^{\rm sym}$\;& $q_0^{\rm asym}$ \\ 
               &\; [in MeV]\;& [in MeV]\; & [in MeV] \\ \colrule
0&$140 $& $109 $ & $180$\\ 
\hline
0.1& $ 166 $ & $128 $& $204$ \\ 
\hline
0.2& $ 187 $ & $143 $& $221$ \\
\hline
0.3 & $202$ & $155 $& $237$ \\
\hline
\hline
 \end{tabular}
 \caption{Comparison of the crossing positions $q_0^{\rm J}$ [$J(q)$], $q_0^{\rm sym}$ [$\Glatt(Q)$], and $q_0^{\rm asym}$ [$L^{\rm asym}(q)$].}
 \label{table_crossing}    
 \end{table}

\section{\label{furcons} Further considerations and Clarifications} 

In this section we comment on a number of subtleties related with some of the concepts introduced,
and provide certain clarifications that we consider necessary. 

({\it i}) It should be evident that, while $\Gamma^{\alpha\mu\nu}$ has been expanded in the
BC basis, mainly in order to make contact with the original BC construction, 
the term $V^{\alpha\mu\nu}$, given in \2eqs{longcoupl}{ABC}, is written in a ``naive'' basis, whose elements,
$n^{\alpha\mu\nu}_i$, are listed in \1eq{thebees}.
From the transformation rules relating the form factors of both bases, given in \2eqs{NtoBC}{BCtoN},
it becomes clear that,  
in general, terms that are ``longitudinally coupled'', in the sense defined in the context of \1eq{longcoupl}, 
when written in the BC basis formed by the tensors given in \2eqs{li}{ti}, 
may have nonvanishing longitudinal and transverse components\footnote{Note that the elements
  $\ell_2^{\alpha\mu\nu}$, $\ell_5^{\alpha\mu\nu}$, and $\ell_8^{\alpha\mu\nu}$ of the BC basis are both
  ``longitudinal'', since they do not vanish when contracted by any of the external momenta,
  and ``longitudinally coupled'', because they satisfy \1eq{eq:transvp}.}.
For instance, to fix the ideas, let us consider the term $v^{\alpha\mu\nu} = q^{\alpha} r^{\mu}p^{\nu}$, which
is one of the elements appearing in $\Gp_{\alpha\mu\nu}(q,r,p)$, multiplied by
$q^{-2}p^{-2}r^{-2}$; in the
BC basis, it may be decomposed into longitudinal and transverse components, as shown in \2eqs{vbc}{vlvt}. 
It is interesting to note the proliferation of terms needed for writing in the BC basis a term as simple as $v^{\alpha\mu\nu}$. 

({\it ii}) The previous exercise indicates that, 
when the entire $\Gp_{\alpha\mu\nu}(q,r,p)$ of \1eq{longcoupl} is cast in the BC basis, namely 
\be
V^{\alpha\mu\nu}(q,r,p)=\sum_{i=1}^{10}{\cal X}_i(q,r,p)\ell_i^{\alpha\mu\nu}(q,r,p)+\sum_{i=1}^{4} {\cal Y}_i(q,r,p)t_i^{\alpha\mu\nu}(q,r,p)\,,
\label{eq:V1}
\ee
it will contain both longitudinal and transverse components.
To appreciate this point in a simplified context,
let us consider the ``abelian'' version of $\Gp_{\alpha\mu\nu}(q,r,p)$, denoted by ${\widehat\Gp}_{\alpha\mu\nu}(q,r,p)$, 
obtained by setting $F=1$ and $H_{\nu\mu} = g_{\nu\mu}$ in \1eq{ABC}, such that 
\begin{align}
{\widehat A}_{\mu\nu}(q,r,p)&=\frac{1}{2}\left\{m^2(r^2)P_{\sigma\mu}(r)\left[g^\sigma_\nu+P^\sigma_\nu(p)\right]
-m^2(p^2)P_{\sigma\nu}(p)\left[g^\sigma_\mu+P^\sigma_\mu(r)\right] \right\}\,,&\nonumber\\
{\widehat B}_{\alpha\nu}(q,r,p)&=\frac{1}{2}\left\{m^2(p^2) P_{\sigma\nu}(p) \left[g^\sigma_\alpha+P^\sigma_\alpha(q)\right]
-m^2(q^2) P_{\sigma\alpha}(q)\left[g^\sigma_\nu+P^\sigma_\nu(p)\right] \right\}\,,&\nonumber\\
{\widehat C}_{\alpha\mu}(q,r,p)&=\frac{1}{2} \left\{m^2(q^2) P_{\sigma\alpha}(q) \left[g_\mu ^{\sigma }+P_\mu^\sigma(r)\right]
-m^2(r^2) P_{\sigma\mu}(r)\left[g_{\alpha}^\sigma+P_\alpha^\sigma(q)\right]\right\} \,.&
\label{abABC}
\end{align}
Then, expanding ${\widehat\Gp}_{\alpha\mu\nu}(q,r,p)$ in the BC basis, using the transformation formulas given in Appendix B, we obtain  
\bea
\widehat{\cal X}_1&=&-\frac{m^2(q^2)}{2 q^2}-\frac{m^2(r^2)}{2 r^2}\,;  \quad 
\widehat{\cal X}_2=\frac{1}{2} \left(\frac{m^2(r^2)}{r^2}-\frac{m^2(q^2)}{q^2}\right)\,;  \quad  \widehat{\cal X}_3=\frac{q^2 m^2(r^2)-r^2 m^2(q^2)}{q^2 r^2 \left(q^2-r^2\right)}\,;  \nonumber\\
 \widehat{\cal X}_4&=&-\frac{m^2(p^2)}{2 p^2}-\frac{m^2(r^2)}{2 r^2}\,; \quad 
\widehat{\cal X}_5=\frac{1}{2} \left[\frac{m^2(p^2)}{p^2}-\frac{m^2(r^2)}{r^2}\right]\,;\quad 
\widehat{\cal X}_6=\frac{1}{p^2-r^2}\left[\frac{m^2(r^2)}{r^2}-\frac{m^2(p^2)}{p^2}\right]\,;\nonumber\\
\widehat{\cal X}_7&=&-\frac{m^2(p^2)}{2 p^2}-\frac{m^2(q^2)}{2 q^2}\,;\quad
\widehat{\cal X}_8=\frac{1}{2} \left[\frac{m^2(q^2)}{q^2}-\frac{m^2(p^2)}{p^2}\right]\,;\quad
\widehat{\cal X}_9=\frac{1}{p^2-q^2}\left[\frac{m^2(q^2)}{q^2}-\frac{m^2(p^2)}{p^2}\right]\,;\nonumber\\
\widehat{\cal X}_{10}&=&0 \,;\nonumber\\
\widehat{\cal Y}_1&=&\frac{q^2 \left[-m^2(p^2)+m^2(q^2)-m^2(r^2)\right]+r^2 \left[m^2(p^2)+m^2(q^2)-m^2(r^2)\right]}{p^2 q^2 r^2 \left(q^2-r^2\right)}\,;\nonumber\\
\widehat{\cal Y}_2&=&\frac{\left(r^2-p^2\right) \left[m^2(q^2)-m^2(r^2)\right]+p^2 \left[m^2(p^2)-2 m^2(r^2)\right]+r^2 m^2(p^2)}{p^2 q^2 r^2 \left(p^2-r^2\right)}\,;\nonumber\\
\widehat{\cal Y}_3&=&-\frac{m^2(r^2) \left(p^2-q^2\right)+\left[2 \left(q^2+q\cdot r\right)+r^2\right] \left[m^2(q^2)-m^2(p^2)\right]}{p^2 q^2 r^2 \left(p^2-q^2\right)}\,;\nonumber\\
\widehat{\cal Y}_4&=&\frac{(q\cdot r) \left[-m^2(p^2)+m^2(q^2)+m^2(r^2)\right]+q^2 m^2(r^2)+r^2 m^2(q^2)}{p^2 q^2 r^2}\,.
\label{vform}
\eea
Evidently, since the $\widehat{\cal Y}_1$ are nonvanishing, 
the transverse part of ${\widehat\Gp}_{\alpha\mu\nu}(q,r,p)$, and therefore that of the entire (abelianized)
vertex $\widehat{\fatg}_{\alpha\mu\nu}(q,r,p)$, contains massless poles. 

The generalization of the above construction to the full ${\Gp}_{\alpha\mu\nu}(q,r,p)$ is straightforward but does not
provide any further conceptual insights;
the resulting expressions for the corresponding ${\cal X}_i$ and ${\cal Y}_i$ are quite lengthy, mainly due
to the complicated ``intertwining'' between the mass terms and the $H_{\nu\mu}$ form factors, $A_1$, $A_3$ and $A_4$,
and will not be reported here.

({\it iii})
The main conclusion that one should draw from the 
above construction is that the expansion into the BC basis of the {\it entire} vertex ${\fatg}_{\alpha\mu\nu}(q,r,p)$,
\ie the sum of {\it both}  ${\Gp}_{\alpha\mu\nu}(q,r,p)$ and ${\Gnp}_{\alpha\mu\nu}(q,r,p)$,
is of no practical usefulness, and may in fact be misleading. 
In particular, let us suppose for a moment that ${\fatg}_{\alpha\mu\nu}(q,r,p)$ (and not just ${\Gnp}_{\alpha\mu\nu}(q,r,p)$
as was done throughout this work) was indeed written in the BC basis. Then, the corresponding form factors, $\mathbb{X}_i$ and
$\mathbb{Y}_i$
would be simply given by 
\bea
\mathbb{X}_i &=& X_i + {\cal X}_i \,,
\label{hypexp1}
\\
\mathbb{Y}_i &=& Y_i + {\cal Y}_i \,;
\label{hypexp2}
\eea
of course, given the intrinsic limitations of the methodology 
employed in this work, we have no access to $Y_i$, but only to ${\cal Y}_i$.
Evidently, due to the form of the  ${\cal X}_i$ and ${\cal Y}_i$, all $\mathbb{X}_i$ and $\mathbb{Y}_i$
would be infested with massless poles; this, in turn, would be clearly reflected in the typical 3D plots
of any {\it individual} $\mathbb{X}_i$ or $\mathbb{Y}_i$. 
However, any such representation would be physically disingenuous, 
because the pole terms from each  $\mathbb{X}_i$ and $\mathbb{Y}_i$, when summed up, would eventually 
organize themselves into a ``longitudinally coupled'' contribution, namely none other than \1eq{longcoupl},
and would cancel from physical amplitudes or lattice observables, such as the $L(q,p,r)$ of \1eq{eq:GammaSym_proj1}. 
Note, in particular, that, under these circumstances, it would be erroneous to consider, 
as part of an ``approximation scheme'', {\it only} the longitudinal part of the full $\fatg^{\alpha\mu\nu}$, 
to be denoted by $\fatg^{\alpha\mu\nu}_{\!\!{\s{\mathbf{L}}}}$, because one would then have\footnote{This quantity was
{\it not} introduced in the previous sections, precisely because of the subtleties associated with its nature.}, 
\be
\label{eq:transvp1}
P_{\alpha\alpha^{\prime}}(q)P_{\mu\mu^{\prime}}(r)P_{\nu\nu^{\prime}}(p) \fatg^{\alpha\mu\nu}_{\!\!{\s{\mathbf{L}}}}(q,r,p) \neq 0 \,,
\ee
and the final (``approximate'') answer would be afflicted by the presence of spurious divergences. 

The way the above problems have been resolved in the present work was simply by not expanding ${\Gp}_{\alpha\mu\nu}(q,r,p)$
in the BC basis, which has been used exclusively for ${\Gnp}_{\alpha\mu\nu}(q,r,p)$, in order for the BC-solution to become directly applicable.
Thus, the approximation employed amounts to setting $Y_i=0$, but keeping the {\it entire} ${\Gp}_{\alpha\mu\nu}(q,r,p)$, or, in the language of the
BC basis, {\it both} ${\cal X}_i$ and  ${\cal Y}_i$ are present; and since \1eq{eq:transvp} remains intact, what one determines and plots
are the $X_i$, which contain no explicit massless poles. 

({\it iv})
The transverse form factors $Y_i$ of  ${\Gnp}_{\alpha\mu\nu}(q,r,p)$, 
whose structures are undetermined by the present gauge-technique-based approach, 
may, in principle, contain divergent contributions, and, in particular, their own poles (simple, or of higher order). 
At present, one may not exclude this possibility,
and further independent studies, based on direct SDE approaches, may shed light on their structure.  
In such a search, the lattice results for $\Glatt(Q)$ provide an interesting constraint.
Specifically, the non-observability of pole divergences at the level of the $\Glatt(Q)$ 
requires that, in the limit $Q^2\to 0$,  
\be
Q^4 Y_1(Q) - 2 Q^2 Y_4(Q) = C \,,
\label{Yconstr}
\ee
where $C$ is some arbitrary finite constant. 
Let us further assume, for the sake of argument, that the pole structure of
$Y_1(Q)$  and $Y_4(Q)$ has the general form 
\be
Y_1(Q) = \sum_{n=1}^{\infty} \frac{a_n}{Q^{2n}}\,, \,\,\, Y_4(Q) = \sum_{n=1}^{\infty} \frac {b_n}{Q^{2n}} \,.
\label{genpolstr}
\ee
Then, \1eq{Yconstr} imposes the following constraints on the coefficients $b_i$ and $c_i$: 
\bea
a_1 &=& \rm{undetermined}\,,
\nonumber\\
a_2 - 2b_1 &=& C\,, 
\nonumber\\
a_n - 2 b_{n-1} &=& 0\,, \,\, n=3,4, ...
\label{abconstr}
\eea
Evidently, the above constraints are trivially satisfied when all $a_n$ and $b_n$ vanish (in which case, $C=0$).

\section{\label{naiveBC} On the ``naive'' implementation of the BC solution}

Finally, having provided a sufficient amount of background material,
we may now revisit the central issue of the BC construction mentioned in the Introduction,
namely the problems with using directly the term 
\be
J_{{\s{\rm{BC}}}} (q) = \frac{\Delta^{-1}(q)}{q^2} \,,
\label{J_BC}
\ee
as ingredient in the BC solution. In particular, let us assume that one were to ignore the presence and
function of the term $V_{\alpha\mu\nu}(q,r,p)$, and suppose that the BC solution of 
\1eq{eq:X_sol} holds at the level of the $\mathbb{X}_i$, namely the full longitudinal form factors. 
As explained in the Introduction, 
due to the finiteness of the gluon propagator, $\Delta^{-1}(0) = m^2(0)$,  the various $\JBC(q)$ contain massless poles, which, through \1eq{eq:X_sol}, will enter into the individual $\mathbb{X}_i$; 
and the combination of all such terms does {\it not} organize itself into a longitudinally coupled contribution. 

\begin{figure}[!h]
\includegraphics[scale=0.42]{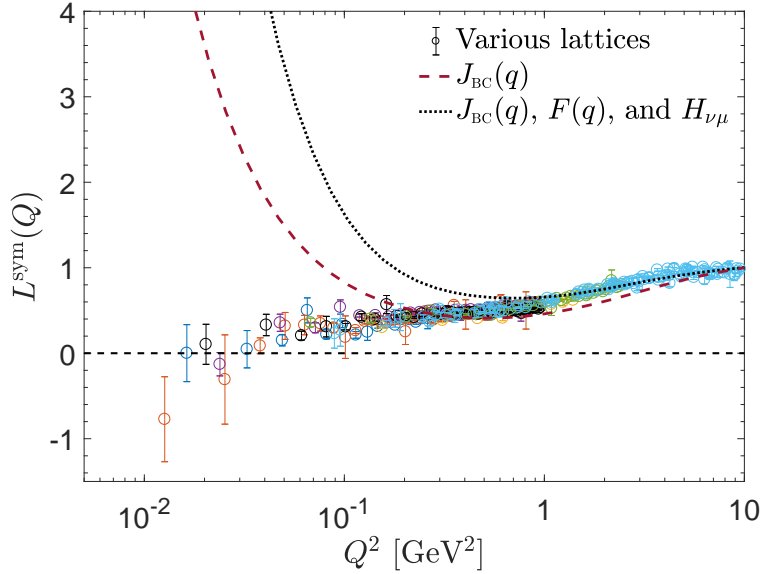}
\caption{ Comparison of the results for $\Glatt_{{\textnormal{\tiny \textsc{BC}}}}(Q)$  given by Eq.~\eqref{eq:GammasymJBC} (red dashed) with the lattice data of~\cite{Athenodorou:2016oyh} (circles). $\Glatt(Q)$ (black dotted curve) shows  the impact of restoring the ghost sector. Evidently, the use of Eq.~\eqref{J_BC} produces a positive infrared divergence, which is incompatible with the lattice results.}
\label{fig:Gamma_symJBC}
\end{figure}

The consequences of this scenario are rather striking.
To appreciate this with one particular example, let us first reduce the algebraic complexity by turning off the ghost sector, and consider
the  ``abelianized'' $\mathbb{X}_i$, given by Eq.~\eqref{eq:X10_mBC}, but now substitute $J(q) \to \JBC(q)$, to obtain (Euclidean space),
$\widehat{\mathbb{X}}_1(q,r,p) = \frac{1}{2}[ J_{{\s{\rm{BC}}}}(r) + J_{{\s{\rm{BC}}}} (q)]$ and     
$\widehat{\mathbb{X}}_3(q,r,p) = \frac{[ J_{{\s{\rm{BC}}}}(q) - J_{{\s{\rm{BC}}}}(r)]}{r^2-q^2}$. 
Then, evaluate the lattice quantity $\Glatt(Q)$, now to be denoted by $\Glatt_{{\textnormal{\tiny \textsc{BC}}}} (Q)$,
by substituting $\widehat{\mathbb{X}}_1$ and $\widehat{\mathbb{X}}_3$ into \1eq{eq:GammaSym_Xi}, and setting 
(temporarily) $\widehat{\mathbb{Y}}_1(Q) = \widehat{\mathbb{Y}}_4(Q) = 0$, to obtain
\bea
\Glatt_{{\textnormal{\tiny \textsc{BC}}}}(Q) &=& \JBC(Q^2) + \frac{Q^2}{2} \left(\frac{d\JBC(Q)}{dQ^2}\right) 
\nonumber\\
&=& \frac{1}{2}\left[\frac{\Delta^{-1}(Q)}{Q^2} + \frac{d\Delta^{-1}(Q)}{dQ^2}\right]\,.
\label{eq:GammasymJBC}
\eea
In the limit $Q^2 \to 0$, the first term is dominant, diverging as a simple pole,
\be
\lim_{Q^2 \to 0} \Glatt_{{\textnormal{\tiny \textsc{BC}}}}(Q) = \frac{1}{2} \frac{m^2(0)}{Q^2} + ...\,,
\label{GammasymJBCQO}
\ee
where the ellipses denote sub dominant terms (remember, in particular, that the derivative term diverges logarithmically).
This pole term, in turn, is clearly visible when contrasting the $\Glatt_{{\textnormal{\tiny \textsc{BC}}}} (Q)$ with the lattice
data, as shown in Fig.~\ref{fig:Gamma_symJBC}; thus, due to the huge discrepancy observed,
the use of \1eq{J_BC}, at least under the assumptions leading to \1eq{GammasymJBCQO}, is
plainly discarded. Note also that the restoration of the ghost sector does not change the situation qualitatively; its inclusion simply
increases the numerical value of the residue of the pole, making the onset of the divergence appear at higher values of $Q^2$,
as seen in Fig.~\ref{fig:Gamma_symJBC}.  

It is clear that the only way to circumvent this discrepancy and still use \1eq{J_BC} is to relax the assumption that
$\widehat{\mathbb{Y}}_1(Q) = \widehat{\mathbb{Y}}_4(Q) = 0$; in fact, one ought to allow these latter form factors to have poles,
which would precisely cancel the corresponding contribution in \1eq{GammasymJBCQO}.
Put in other words, in the absence of a concrete connection between the saturation of the gluon propagator
and the vertex under consideration, the BC construction may not stand on its own, but requires
the inclusion of transverse pieces with a {\it necessarily} non-trivial pole content.
In particular, let us assume that, in complete analogy to \1eq{genpolstr}, 
$\widehat{\mathbb{Y}}_1(Q)$ and $\widehat{\mathbb{Y}}_4(Q)$ are given by 
\be
\widehat{\mathbb{Y}}_1(Q) = \sum_{n=1}^{\infty} \frac{\alpha_n}{Q^{2n}}\,, \,\,\, \widehat{\mathbb{Y}}_4(Q) = \sum_{n=1}^{\infty} \frac {\beta_n}{Q^{2n}}\,.
\label{genpolstr2}
\ee
Then, to accomplish the non-observability of pole contributions in $\Glatt_{{\textnormal{\tiny \textsc{BC}}}} (Q)$, the
constraints of \1eq{abconstr} must hold unchanged ($a_n\to\alpha_n$, $b_n\to \beta_n$),
with the very crucial exception of $n=3$, which must be now modified to
\be
\alpha_3 - 2 \beta_2 = -2 m^2(0) \,;
\label{polecancon}
\ee
evidently, the fulfillment of this last condition requires that at least one of the $\alpha_3$ and $\beta_2$ be nonvanishing. 

Thus, in order 
for the ``naive'' BC construction to be compatible with the lattice results, 
the transverse part of the vertex must possess the {\it minimal} pole structure 
\be
\widehat{\mathbb{Y}}^{\rm min}_1(Q) = \frac{2\beta_2-2m^2(0)}{Q^6} \,,  \,\,\,\,\,\, \widehat{\mathbb{Y}}^{\rm min}_4(Q) = \frac{\beta_2}{Q^4} \,.
\label{Yminpol}
\ee

To make the final connection, 
turn to the expressions for $\widehat{\cal Y}_1$ and $\widehat{\cal Y}_4$ given in \eqref{vform},
pass to Euclidean momenta, and compute them in the symmetric limit; it is fairly straightforward to establish that
\be
\widehat{\cal{Y}}_1(Q) = \frac{m^2(Q)}{Q^6} - \frac{2}{Q^4}\frac{dm^{2}(Q)}{dQ^2}
\,,\,\,\, \widehat{\cal{Y}}_4(Q) =  \frac{3}{2}\frac{m^2(Q)}{Q^4} \,.
\label{calYpoles}
\ee
 
If at this point one were to identify  $\widehat{\cal{Y}}_1(Q)$ and $\widehat{\cal{Y}}_4(Q)$ with $\widehat{\mathbb{Y}}_1(Q)$ and $\widehat{\mathbb{Y}}_4(Q)$,
respectively (which is tantamount to using \1eq{hypexp2} with $Y_i=0$), 
in the limit of $Q^2 \to 0$ one would have
\be
\alpha_3 = m^2(0)\,,\,\,\,\,\,\, \beta_2 = \frac{3}{2}  m^2(0)\,,
\ee
which satisfy precisely the no-pole condition of \1eq{polecancon}.\footnote{Since the first derivative of $m^{2}(Q)$
is finite at the origin~\cite{Aguilar:2011xe,Binosi:2017rwj}, the remaining nonvanishing
coefficient \mbox{$\alpha_2 = - [\frac{2 dm^{2}(Q)}{dQ^2}]_{\s {Q^2=0}}$}
fixes simply the value of $C$ through the second relation of \1eq{abconstr}.}
Thus, the pole structure contained in the $\widehat{\cal{Y}}_1$ and 
$\widehat{\cal{Y}}_4(Q)$ is identical to the {\it minimal} pole structure of \1eq{Yminpol}, 
required for the compatibility with the lattice data. 

We end this discussion with a final comparison between the ``naive'' BC version,  described in this
section, and the one presented in the main part of this work. 
Evidently, the ``naive'' implementation of the BC construction in ``isolation'' 
is physically incomplete, because 
it requires the {\it a-posteriori} inclusion of 
very precise transverse contributions. In particular, in the absence of 
lattice results, one would have no guiding principle on how to construct these terms,
except through the imposition of the 
additional requirement that the combination $P_{\alpha\alpha^{\prime}}(q)P_{\mu\mu^{\prime}}(r)P_{\nu\nu^{\prime}}(p) \fatg^{\alpha\mu\nu}(q,r,p)$ be finite, 
which would lead essentially to the results
of this section. Note, however, that this last requirement alone, although essentially correct,   
establishes no deeper connection with 
the dynamics of the two-point sector of the theory.
Instead, the construction followed in this work
adheres to the theoretical principles that have been spelled out in a series of articles,
being intimately linked
with the intricate dynamics taking place at the level of the gluon propagator,
and, in particular, with the
mass generating mechanism employed. In this way, the results turn out to be
naturally compatible with the lattice, and no {\it a-posteriori} adjustments are
required.

\section{\label{sec:conc} Conclusions}

In the present work we have carried out the nonperturbative derivation of the 
longitudinal part of the three-gluon vertex, $\fatg_{\alpha\mu\nu}(q,r,p)$, from the set of STIs that
it satisfies, given in \1eq{eq:sti_delta}. The procedure followed is a variant of the
well-known BC construction~\cite{Ball:1980ax}, where certain key adjustments have been
implemented in order to account for the fact that the
gluon propagator appearing in the problem is infrared finite.
In particular, in the context of the PT-BFM framework~\cite{Binosi:2009qm},
the origin of the gluonic mass scale, $m^2(q)$, is attributed to the
activation of the Schwinger mechanism by the longitudinally coupled massless poles,
which constitute a purely nonperturbative component of the full vertex $\fatg_{\alpha\mu\nu}(q,r,p)$,
denoted by $V_{\alpha\mu\nu}(q,r,p)$ [see \1eq{longcoupl}].
The inextricable link between $V_{\alpha\mu\nu}(q,r,p)$ and
$m^2(q)$ leads to the partial STI given in \1eq{stiv}~\cite{Aguilar:2011xe,Binosi:2012sj,Ibanez:2012zk,Binosi:2017rwj,Aguilar:2017dco},
while the remainder of the vertex, denoted by $\Gnp_{\alpha\mu\nu}(q,r,p)$, 
satisfies the STI given in \1eq{stig}, which involves the kinetic term $J(q)$, appearing in the decomposition of \1eq{eq:gluon_m_J}.
Given that $V_{\alpha\mu\nu}(q,r,p)$ is completely fixed by the STI of \1eq{longcoupl} and the condition of \1eq{eq:transvp} [see \1eq{ABC}], 
the remaining task boiled down to the application of the BC solution at the
level of the $\Gnp_{\alpha\mu\nu}(q,r,p)$, whose longitudinal form factors, $X_i$, may be thus 
obtained from the STI in \1eq{stig} and its permutations. 

The main ingredient entering in the BC solution is the function $J(q)$, whose form is determined indirectly,
through the subtraction of $m^2(q)$ from $\Delta^{-1}(q)$.
The most prominent feature of $J(q)$ is the zero crossing, whose origin may be traced back
to the ``unprotected'' logarithms contained in it. Note that
the precise form of $J(q)$ is affected by the approximations 
implemented at the level of the dynamical equation that determines $m^2(q)$, and
in particular the value of the coefficient $\gamma$ in the fit of \1eq{eq:gmass}. This fact, in turn, 
introduces minor uncertainties in the results for the form factors, such as the location of 
the corresponding zero crossings displayed by the $X_i$, but does not alter
the main qualitative and quantitative aspects of the answer.

The most prominent feature of the longitudinal form factors is their distinct suppression at
energies below 1 GeV. This property is clearly visible both in the
3D plots of Fig.~\ref{fig:X1_gen_fig}  and Fig.~\ref{fig:X4_gen_fig},
where the size of two form factors becomes inferior to unity (their tree-level value)
for intermediate and infrared momenta. This same trait is also captured
in the 2D plots, corresponding to the two special configurations studied
[left panels of Fig.~\ref{fig:X1_pert} and Fig.~\ref{fig:X1_perta}].
It is important to emphasize that, in addition to lattice simulations,
this suppression has also been observed in the studies carried out using different approaches, such as 
the direct SDE-based derivations of~\cite{Blum:2014gna,Huber:2018ned},
shown in Fig.~\ref{fig:comprevious}.

\begin{figure}[!h]
\begin{minipage}[b]{0.45\linewidth}
\centering
\includegraphics[scale=0.4]{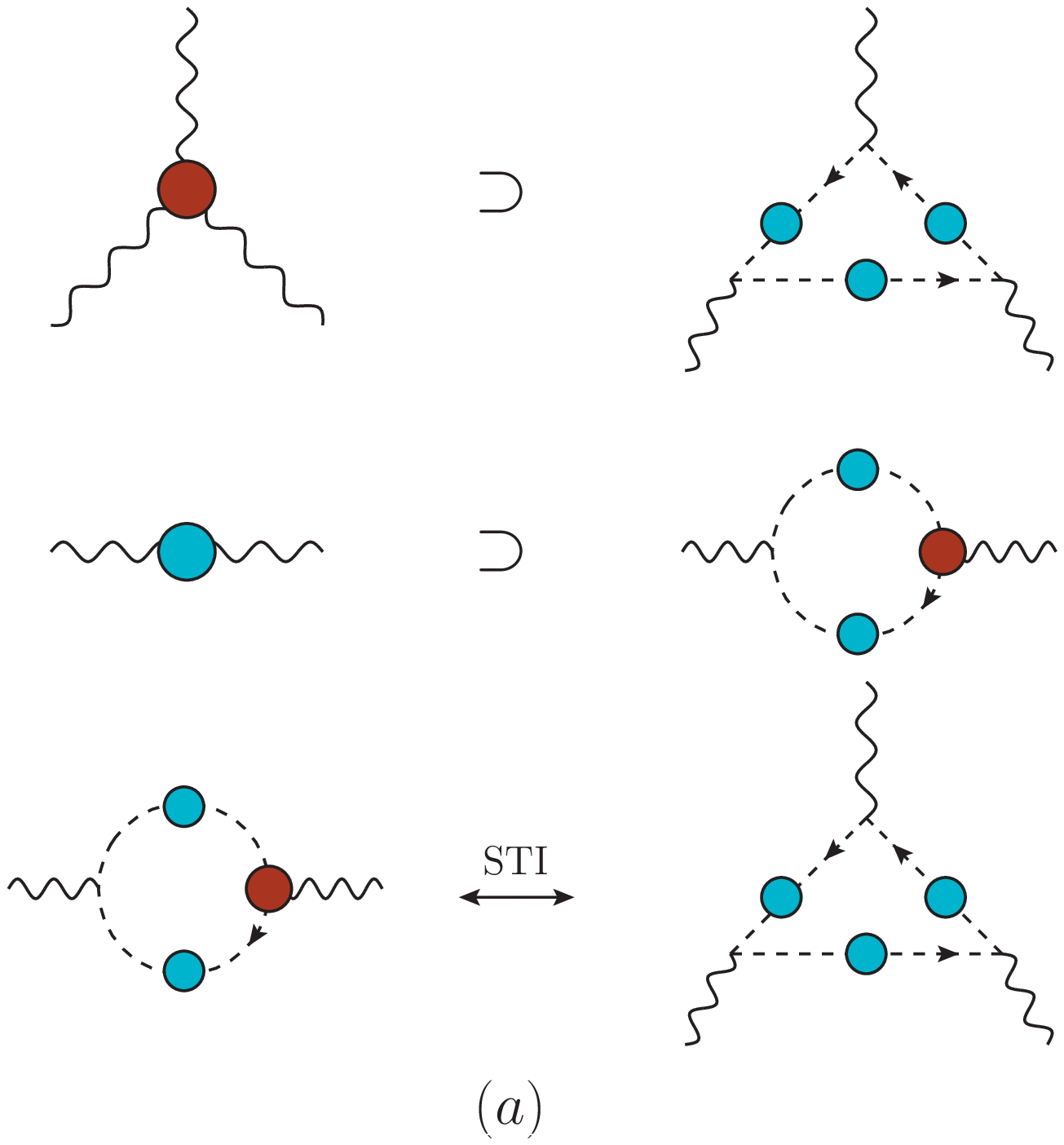}
\end{minipage}
\hspace{0.25cm}
\begin{minipage}[b]{0.45\linewidth}
\includegraphics[scale=0.32]{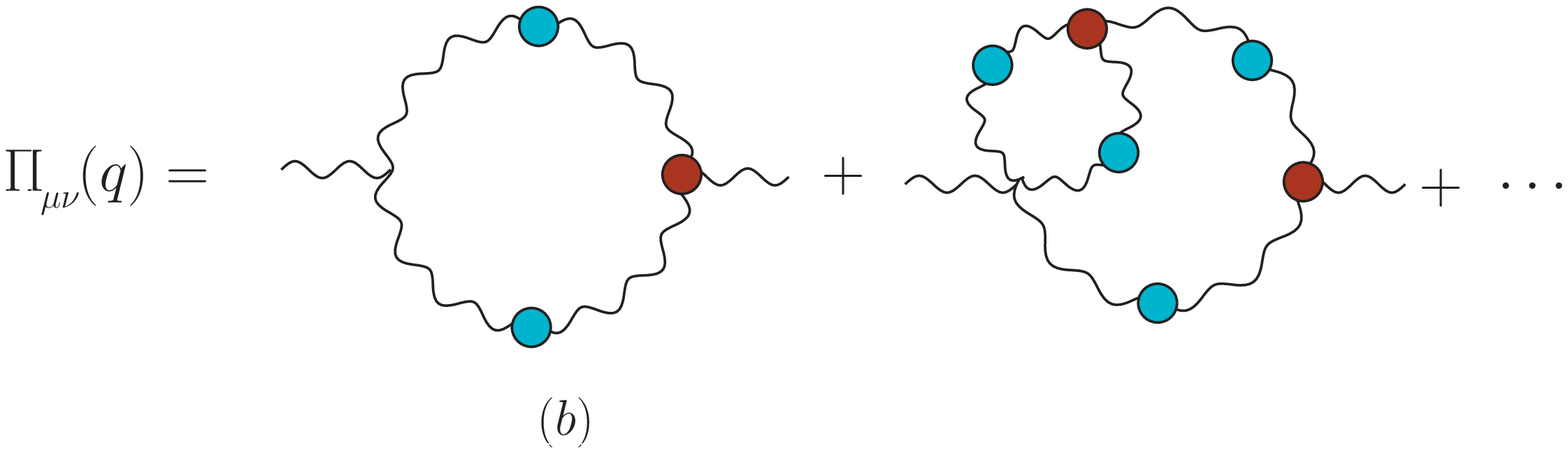}
\begin{flushleft}
\includegraphics[scale=0.4]{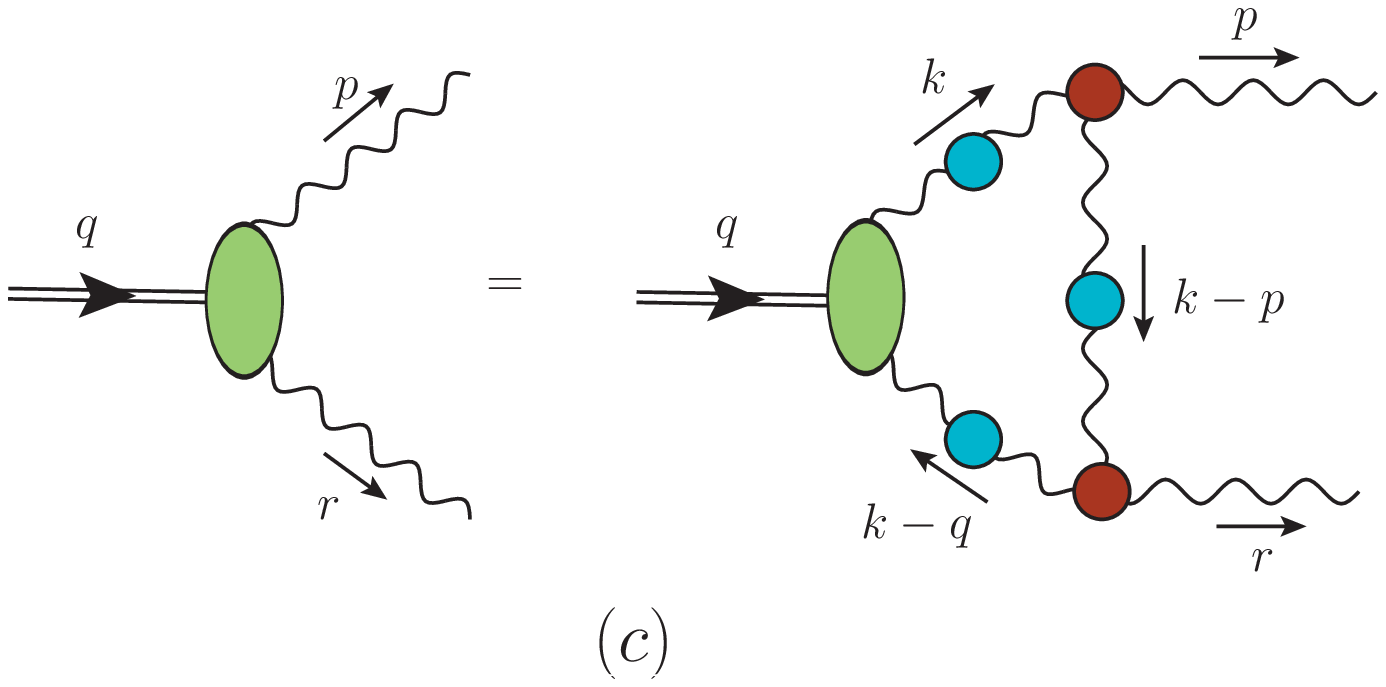}
\end{flushleft}
\end{minipage}
\caption{Panel $(a)$: 
    The SDE diagram of the three gluon vertex responsible for the suppression (first row),  
the one in the gauge-technique approach (second row), and their schematic connection
implemented by the STI (third row). Panel $(b)$:  The gluon self-energy
contributions containing the three gluon vertex. Panel $(c)$:
The  homogeneous BSE which describes the 
formation of the (colored) massless excitations contained in $V_{\alpha\mu\nu}$, or one of the contributions entering in the glueball BSEs.}
\label{fig:epilogue}
\end{figure}

It is interesting to comment on the origin of the
suppression (and the zero-crossing) in the gauge-technique (BC solution) and the SDE analyses.
Evidently, in both approaches the resulting suppression is the outcome of the ``competition''
between the infrared finite contributions originating from diagrams containing
``massive'' gluons and the infrared divergent logarithms stemming from
diagrams containing massless ghosts. As the momenta become smaller, the
``unprotected'' logarithms take over, causing the overall suppression, which culminates
with a negative logarithmic divergence at the origin.  
In the case of the SDE analysis, where the corresponding integral equation for $\Gnp_{\alpha\mu\nu}(q,r,p)$
is considered directly, the diagram responsible for the suppression is the triangle ghost graph
[Fig.~\ref{fig:epilogue}, first row of panel ($a$)].
Instead, in the case of the gauge-technique, where the form factors of  $\Gnp_{\alpha\mu\nu}(q,r,p)$
are built out of the quantities appearing in the corresponding STIs, 
the ingredient causing the suppression and the zero crossing is the
ghost loop that appears in the diagrammatic expansion of the gluon propagator
[Fig.~\ref{fig:epilogue}, second row of panel ($a$)]. This particular graph 
furnishes unprotected logarithms, which  
eventually find their way into the structure of the function $J(q)$. Evidently, the STIs
relate these two graphs, as shown schematically in Fig.~\ref{fig:epilogue}, third row of panel ($a$). 

The 3D results obtained for the $X_i$ may be employed in a variety of situations 
where the three-gluon vertex is expected to play a significant role, and especially in
circumstances where integrations over the entire range of momenta are required.
In what follows we will mention a few notable cases that belong to this general category.

The three-gluon vertex is instrumental 
for the SDE that governs the momentum evolution of the gluon propagator, entering in the
diagrams shown in the panel ($b$) of Fig.~\ref{fig:epilogue}.
After the implementation of the decompositions given in \2eqs{eq:gluon_m_J}{eq:Gnp},
the original SDE furnishes 
two integral equations~\cite{Binosi:2012sj,Aguilar:2014tka}, which determine the quantities $J(q)$ and $m^2(q)$.
The use of the 3D data obtained here for the  
$X_i$, instead of approximate Ans\"atze,
is expected to provide a tighter control on the behavior of these two quantities.

As has been shown in~\cite{Aguilar:2011xe,Ibanez:2012zk}, the  
formation of the (colored) massless excitations contained in $V_{\alpha\mu\nu}(q,r,p)$ 
hinges on the nonvanishing of the
vertex function ${\cal B}_{\mu\nu}$, introduced in Fig.~\ref{fig:pole}.
This possibility, in turn, is determined by the homogeneous BSE
shown in  panel ($c$) of Fig.~\ref{fig:epilogue}, where the ``green ellipse'' represents ${\cal B}_{\mu\nu}$.
Evidently, the kernel of this BSE, and hence the type of solutions obtained, 
depend crucially on the details of the product $\fatg\fatg$;
in the early treatments cited above, this product was simply approximated
by its tree-level value, \ie $\fatg \fatg \to \Gnp^{(0)} \Gnp^{(0)}$,
and nontrivial solutions were found that
corroborate the proposed mass generating mechanism. In fact, the solutions obtained are
intimately related with the first derivative of the running gluon mass, $dm^2 (q^2)/dq^2$~\cite{Aguilar:2011xe}, from which
$m^2 (q^2)$ may be reconstructed~\cite{Binosi:2017rwj}.
The detailed knowledge of the $X_i$ allows for a more sophisticated treatment of this problem, 
achieving a higher degree of self-consistency between all the ingredients involved.

The treatment of systems of BSEs  is of central importance in the studies dedicated to the formation of glueballs.
The particular BSE shown in  panel ($c$) of Fig.~\ref{fig:epilogue} is
present in all such analyses\footnote{The ``green ellipse'' in Fig.~\ref{fig:epilogue}
  represents now the corresponding glueball amplitudes. Note also that the color structure of the
problem is different than that of the massless colored excitations mentioned above.}.
Previous studies indicate that in order to obtain masses compatible with
lattice simulations, the total integrated strength of the kernel must undergo a
considerable suppression~\cite{Meyers:2012ka,Fukamachi:2016wxf}. A need for an analogous attenuation has been
also observed in the recent study of hybrid mesons~\cite{Xu:2018cor}; the 
required suppression has been implemented by resorting to 
a simplified Ansatz for the form factors associated with the tensors 
comprising $\Gamma^{(0)}_{\alpha\mu\nu}(q,r,p)$. Evidently, the results obtained here
offer the possibility of refining future studies in this direction, 
furnishing ingredients that, despite the approximations implemented
in deriving them, can trace their origins to the fundamental underlying theory.

An important limitation of the method employed in this work is that
the structure of the transverse form factors $Y_i$ of the vertex
$\Gnp_{\alpha\mu\nu}(q,r,p)$ [see \2eqs{eq:3g_tr_structure}{hypexp2}]
remains completely undetermined. Of course, this particular
drawback is typical to all gauge-technique based approaches, even though, in some cases,
such as the electron-photon or the quark-gluon vertices, 
partial information on these form factors may be extracted from the
so-called ``transverse'' Ward identities~\cite{Takahashi:1985yz,Kondo:1996xn,He:2006ce,Qin:2013mta,Aguilar:2014lha,Albino:2018ncl}.
Note that the $Y_i$ may be numerically relevant in some of the problems 
mentioned earlier. Moreover, their inclusion is important
in situations where $\Gnp_{\alpha\mu\nu}(q,r,p)$ forms part of an integral equation that
must be multiplicatively renormalized; a particularly relevant 
example of such a case is again the SDE appearing in the panel $(b)$ of Fig.~\ref{fig:epilogue}.
A systematic way for obtaining approximate expressions for these components could be developed at the level of  
the SDE satisfied by $\Gnp_{\alpha\mu\nu}(q,r,p)$,
where the tensorial structures associated
with the $Y_i$ will have to be appropriately projected out of the corresponding integral equation.
Calculations in that direction are already underway, and we hope to report on the 
results in the near future.

\appendix

\section{\label{app:pert} One-loop results}
  
In this Appendix we present the one-loop results for form factors in the ({\it i}) totally symmetric, ({\it ii}) asymmetric, and 
 ({\it iii}) the general orthogonal configurations. All of them were obtained by the direct evaluation of the one-loop diagrams contributing to
the three-gluon vertex. 
The relevant expressions are obtained from~\cite{Davydychev:1996pb} (Landau gauge),
and are renormalized in the Taylor scheme; this means that the corresponding $A_i$ are  
not renormalized at $Q^2 = \mu^2$, but rather at the ``soft-ghost'' configuration, as discussed at the end of Sec.~\ref{sec:BCS}. 
  
In particular, $Z_3$ will be obtained directly from \1eq{eq:renorm_sti}, by setting
\bea
Z_A &=& 1 + \frac{ C_\mathrm{A}\alpha_s }{ 144\pi }\left[ 78\left\lbrace \frac{2}{\epsilon } - \gamma_{\s E} + \ln (4 \pi )\right\rbrace + 97 \right] \,, \nonumber\\
Z_c &=& 1 + \frac{ C_\mathrm{A}\alpha_s }{16 \pi } \left[ 3\left\lbrace \frac{2}{\epsilon } - \gamma_{\s E} + \ln (4\pi )\right\rbrace + 4 \right] \,, \nonumber\\
Z_1 &=& 1\,, 
\label{eq:renorm_zs}
\eea
where $\gamma_{\s E}$ denotes the Euler-Mascheroni constant, and 
$\epsilon = 4 - d$, with $d$ being the dimension of spacetime in dimensional regularization.

Substituting the results of Eqs.~\eqref{eq:renorm_zs} into the STI for the renormalization constants given by~\eqref{eq:renorm_sti}, one finds 
\be
Z_3 = 1 + \frac{ C_\mathrm{A}\alpha_s}{144 \pi }\left[ 51\left\lbrace \frac{2}{\epsilon } - \gamma_{\s E} + \ln (4 \pi )\right\rbrace + 61\right] \,. 
\label{eq:Z3}
\ee
Then, one may obtain ultraviolet
finite (cutoff-independent) one-loop results for the three-gluon vertex by performing the renormalization as described above. In what follows
we will implement this procedure to obtain the corresponding results for the 
symmetric and asymmetric configurations.

\begin{enumerate}

\item{\emph{Symmetric configuration:}}  The kinematics of this configuration
is defined in Eq.~\eqref{defsym}. Then, at one-loop we have
\begin{align}
&X_1^{(1)}(Q) = 1 + \frac{ C_\mathrm{A}\alpha_s }{48 \pi } \left[ 17\ln \left(\frac{Q ^2}{\mu^2}\right) + 3 - 3 {\rm I} \right] \,, \quad 
X_2^{(1)}(Q) = 0 \,,  \label{eq:3g_pert} \\
&X_3^{(1)}(Q) = -\frac{ C_\mathrm{A}\alpha_s }{48 \pi  Q^2} \left( 38 - 7 {\rm I} \right)\,, \qquad\qquad \qquad \;  \quad  
X_{10}^{(1)}(Q) = 0 \,, \nonumber\\
&Y_1^{(1)}(Q) = -\frac{ C_\mathrm{A}\alpha_s }{432 \pi Q^4} \left( 587 - 193 {\rm I} \right)\,, \qquad \qquad \;\;  \quad 
Y_4^{(1)}(Q) = -\frac{ C_\mathrm{A}\alpha_s }{864 \pi  Q^2} \left(  365 + 179 {\rm I} \right)\!, \nonumber 
\end{align}
with ${\rm I}$ defined as~\cite{Celmaster:1979km}
\be 
{\rm I} = \frac{1}{3}\left[ \psi_1 \left(\frac{1}{3}\right) - \psi_1 \left(\frac{2}{3}\right) \right] = 2.34391 \,,
\label{eq:Icel}
\ee
where
$\psi_1(z)$ is the trigamma function, related to $\Gamma(z)$ by
\be
\psi_1(z) = \frac{d^2}{dz^2}\ln [\Gamma(z)] \,.
\label{trigamma}
\ee

The form factors $X_2$ and $X_{10}$ vanish in the symmetric configuration since they are antisymmetric under the exchange of at least two arguments [see Eq.~\eqref{eq:Xi_more_bose}].

\item{\emph{Asymmetric configuration:}}

In the asymmetric configuration, defined in Eq.~\eqref{defasym}, the tensor structure of the three-gluon vertex reduces to~\cite{Davydychev:1996pb}
\bea
\Gamma_{\alpha\mu\nu}(q, -q, 0) &=& 2 g_{\alpha\mu} q_\nu [ X_1(q,-q,0) - q^2 X_3(q,-q,0)] 
- 2 q_\alpha q_\mu q_\nu X_3(q,-q,0)  \notag\\ 
& -& ( q_\alpha g_{\mu\nu} + q_\mu g_{\alpha\nu} ) [ X_1(0, q, -q) - X_2(0, q, -q) ] \,. 
\label{eq:asym_tensors}
\eea
Then, the corresponding form factors at one-loop read
\bea
&&X_1^{(1)}(q,-q,0) = 1 + \frac{ 17 C_\mathrm{A} \alpha_s }{48 \pi } \ln \left( \frac{q ^2}{\mu^2} \right) \,, \nonumber\\
&&X_3^{(1)}(q,-q,0) = -\frac{ 37 C_\mathrm{A} \alpha_s }{96 \pi  q^2} \,, \nonumber\\
&&X_1^{(1)}(0, q, -q) - X_2^{(1)}(0, q, -q) = X_1^{(1)}(q,-q,0) \,. \label{eq:X1_and_X3_asym}
\eea
Notice that we cannot disentangle $X_1^{(1)}(0, q, -q)$ and $X_2^{(1)}(0, q, -q)$.

One should also note that, due to our choice of the Taylor renormalization prescription, $X_1$ reduces to its tree-level value
for $q = \mu$ in the asymmetric configuration,  $X_1^{(1)}(\mu,-\mu,0) = 1$, instead of satisfying this condition at the symmetric point.

\item{\emph{General orthogonal configuration:}}

In this configuration,  the momenta $q^2$ and $r^2$ are independent,
but the angle $\theta$ is fixed at $\theta = \pi/2$; therefore, one has $p^2 = q^2 + r^2$.
In this case we have determined only $X_1$, which is given by 
\begin{align}
\hspace{-0.8cm}
X_1(q^2,r^2,\pi/2) =& 1 + \frac{ C_\mathrm{A} \alpha_s }{ 768 \pi q^2 r^2 } \bigg\lbrace 2 \left( 9 q^4 + 128 q^2 r^2 + 3 r^4 \right) \ln \left(\frac{q^2}{\mu ^2}\right) - 12  ( q^4 - 8 q^2 r^2 + r^4 )   \nonumber \\
& \hspace{-1.5cm} + 2  \left( 3 q^4 + 128 q^2 r^2 + 9 r^4 \right) \ln \left(\frac{r^2}{\mu ^2}\right) 
- 24  \left( q^4 + 10 q^2 r^2 + r^4 \right) \ln \left(\frac{q^2 + r^2}{\mu ^2}\right) \nonumber \\
& \hspace{-1.5cm} - 3 i \frac{( q^2 + r^2 )}{qr}\left( q^2 - r^2 \right)^2 \bigg[ \text{Li}_2 \left( - z \right) - \text{Li}_2 \left(z \right)
 + \text{Li}_2 \left( z^{-1} \right) - \text{Li}_2 \left( - z^{-1} \right) \bigg] \bigg\rbrace \,, \label{eq:X1_ortho_pert}
\end{align}
with  $q=|q|$, $r=|r|$, \mbox{$z =(q - i r)/( q + i r)$},
and 
\be 
\mathrm{Li}_2(z) = - \int_0^z \frac{\ln(1 - t)}{t} dt \,,
\ee
is the dilogarithm (or Spence function). Note that the above  expression is symmetric under $q \leftrightarrow r$.

\item{\emph{Orthogonal symmetric configuration:}}

This is a particular limit of  Eq.~\eqref{eq:X1_ortho_pert}
where  $q^2 = r^2$. We obtain 
\be 
X_1(q^2,q^2,\pi/2) = 1 + \frac{ C_\mathrm{A} \alpha_s }{ 96 \pi }\left[ 34 \ln \left( \frac{q^2}{\mu^2} \right) - 36 \ln(2) + 9 \right] \,.
\label{ortho-sym}
\ee

\end{enumerate}

\newpage

\section{\label{changebasis} The BC and naive bases}

Let us consider an arbitrary tensor with three Lorentz indices ($\alpha$, $\mu$, $\nu$) and three momenta $(q,r,p)$, to be denoted
by $S^{\alpha\mu\nu}(q,r,p)$.
We expand $S^{\alpha\mu\nu}(q,r,p)$ in two different bases, the ``naive'' and the BC basis,
\begin{align}
S^{\alpha\mu\nu}(q,r,p)&=\sum_{i=1}^{14} N_i(q,r,p)\,n_i^{\alpha\mu\nu}\,,&\nonumber\\
&=\sum_{i=1}^{10} L_i(q,r,p)\,\ell_i^{\alpha\mu\nu}+\sum_{i=1}^{4} T_i(q,r,p)\,t_i^{\alpha\mu\nu}\,,&
\label{eq:V}
\end{align}
where the BC elements $\ell_i^{\alpha\mu\nu}$ and $t_i^{\alpha\mu\nu}$ are given in \2eqs{li}{ti}, 
and we define the elements of the naive basis to be
\begin{align}
n_1^{\alpha\mu\nu} &=  q^{\alpha} g^{\mu\nu};&
n_2^{\alpha\mu\nu} &=  q^{\alpha} q^{\mu} q^{\nu};&
n_3^{\alpha\mu\nu} &=  q^{\alpha} q^{\mu} r^{\nu};&
n_4^{\alpha\mu\nu} &= q^{\alpha} r^{\mu} q^{\nu};&
n_5^{\alpha\mu\nu} &=  q^{\alpha} r^{\mu} r^{\nu},
\nonumber\\
n_6^{\alpha\mu\nu} &=  r^{\alpha} g^{\mu\nu};&
n_7^{\alpha\mu\nu} &=  r^{\alpha} q^{\mu} q^{\nu};&
n_8^{\alpha\mu\nu} &= r^{\alpha} q^{\mu} r^{\nu};&
n_9^{\alpha\mu\nu} &= r^{\alpha} r^{\mu} q^{\nu};&
n_{10}^{\alpha\mu\nu} &= r^{\alpha} r^{\mu} r^{\nu},
\nonumber\\
n_{11}^{\alpha\mu\nu} &= q^{\mu} g^{\nu\alpha};&
n_{12}^{\alpha\mu\nu} &= q^{\nu} g^{\mu\alpha};&
n_{13}^{\alpha\mu\nu} &= r^{\mu} g^{\nu\alpha};&
n_{14}^{\alpha\mu\nu} &= r^{\nu} g^{\mu\alpha} . 
\label{thebees}	
\end{align}
The form factors $N_i$ can be written in terms of the form factors $L_i$ and $T_i$ as 
\begin{align}
N_1&=L_4-L_5-(p\cd r)L_6+(p\cd r) (q\cd r)T_2 -(q\cd r)T_4\,,&\nonumber\\
N_2&=2 L_9+r^2 T_3\,,&\nonumber\\
N_3&=-L_6-L_{10}+(q\cd r)T_2 + T_4\,,&\nonumber\\
N_4&=L_9-(q\cd r)T_3\,,&\nonumber\\
N_5&=-L_6+(q\cd r)T_2\,,&\nonumber\\
N_6&=2 L_4-2 (p\cd r)L_6-q^2 (p\cd r)T_2 + q^2 T_4\,,&\nonumber\\
N_7&=L_3+2 L_9-L_{10}-(p\cd r)T_1 +r^2 T_3-T_4\,,&\nonumber\\
N_8&=-L_3-2 L_6-L_{10}+(p\cd q)T_1-q^2 T_2+T_4\,,&\nonumber\\
N_9&=L_9-L_{10}-(q\cd r)T_3-T_4\,,&\nonumber\\
N_{10}&=-2L_6-q^2 T_2\,,&\nonumber\\
N_{11}&=-2L_7+2 (p\cd q)L_9+r^2(p\cd q) T_3-r^2 T_4\,,&\nonumber\\
N_{12}&=L_1+L_2-(q\cd r)L_3+(p\cd r) (q\cd r)T_1-(p\cd r)T_4\,,&\nonumber\\
N_{13}&=-L_7-L_8+(p\cd q)L_9-(p\cd q) (q\cd r)T_3+(q\cd r)T_4\,,&\nonumber\\
N_{14}&=-L_1+L_2+(q\cd r)L_3-(p\cd q) (q\cd r)T_1+(p\cd q)T_4\,,&
\label{NtoBC}
\end{align}
where we have omitted the momenta dependence of the form factors.

On the other hand, the change of basis can be inverted in order to obtain $L_i$ and $T_i$ 
in terms of $N_i$,
\begin{align}
L_1&=\frac{1}{4} \left\{2 [(q\cd r) (N_{10}-N_2+N_3+N_4-N_5+N_7-N_8-N_9)+N_{12}-N_{14}]\right.&\nonumber\\
&\left.+(p\cd q)(N_3+N_4-N_5-N_9)+(p\cd r) (N_3+N_4-N_5-N_9)\right\}\,,&\nonumber\\
L_2&=\frac{1}{4} \left\{2 [(q\cd r) (-N_{10}-N_2-N_3+N_4+N_5+N_7+N_8-N_9)+N_{12}+N_{14}]\right.&\nonumber\\
&\left.+(p\cd q)(-N_3-N_4+N_5+N_9)+(p\cd r) (N_3+N_4-N_5-N_9)\right\}\,,&\nonumber\\
L_3&=\frac{(p\cd r) (-N_{10}-N_3+N_5+N_8)+(p\cd q) (-N_2+N_4+N_7-N_9)}{(p\cd q)-(p\cd r)}\,,&\nonumber\\
L_4&=\frac{1}{4} \left\{-2(p\cd r)N_{10} +q^2 (-N_3-N_4+N_5+N_9)+2N_6\right\}\,,&\nonumber\\
L_5&=\frac{1}{4} \left\{2 [-2 N_1+(q\cd r) (-N_3-N_4+N_5+N_9)+N_6]-2(p\cd r) (N_{10}-2N_5)\right.&\nonumber\\
&\left. +q^2(-N_3-N_4+N_5+N_9)\right\}\,,&\nonumber\\
L_6&=-\frac{(q\cd r)N_{10} +q^2N_5 }{q^2+2 (q\cd r)}\,,&\nonumber\\
L_7&=\frac{1}{4} \left\{-2 N_{11}+2(p\cd q) N_2+r^2 (-N_3-N_4+N_5+N_9)\right\}\,,&\nonumber\\
L_8&=\frac{1}{4} \left\{2 N_{11}-4 N_{13}-2(p\cd q) (N_2-2N_4) +2 (q\cd r) (N_3+N_4-N_5-N_9)\right.&\nonumber\\
&\left. +r^2(N_3 +N_4-N_5-N_9)\right\}\,,&\nonumber\\
L_9&=\frac{(q\cd r)N_2+r^2N_4}{2 (q\cd r)+r^2}\,,&\nonumber\\
L_{10}&=\frac{1}{2} (-N_3+N_4+N_5-N_9)\,,&\nonumber\\
T_1&=\frac{-N_2-N_3+N_4+N_5+N_7+N_8-N_9-N_{10}}{(p\cd q)-(p\cd r)}\,,&\nonumber\\
T_2&=\frac{2 N_5-N_{10}}{q^2+2 (q\cd r)}\,,&\nonumber\\
T_3&=\frac{N_2-2N_4}{2 (q\cd r)+r^2}\,,&\nonumber\\
T_4&=\frac{1}{2} (N_3+N_4-N_5-N_9)\,.&
\label{BCtoN}
\end{align}

As a concrete example, consider the vector $v^{\alpha\mu\nu}\equiv q^\alpha r^\mu p^\nu$, which can be written in the naive basis as
\be
v^{\alpha\mu\nu}=-n_4^{\alpha\mu\nu} - n_5^{\alpha\mu\nu}\,,
\label{vni}
\ee
\textit{i.e.}, $N_4=N_5=-1$, while all the other form factors vanish.
Using the transformation rules of \1eq{BCtoN} (and $q+p+r=0$), 
we can write this vector in the BC basis as
\be
v^{\alpha\mu\nu} = v^{\alpha\mu\nu}_{{\s{\mathbf{L}}}} +  v^{\alpha\mu\nu}_{{\s{\mathbf{T}}}}\,,
\label{vbc}
\ee
with
\bea
v^{\alpha\mu\nu}_{{\s{\mathbf{L}}}} &=& -(q\cd r)\ell_2^{\alpha\mu\nu}+
\frac{p^2}{p^2+2(p\cd q)}\,\ell_3^{\alpha\mu\nu}
-(p\cd r)\ell_5^{\alpha\mu\nu}+\frac{q^2}{q^2+2(q\cd r)}\,\ell_6^{\alpha\mu\nu}
-(p\cd q)\ell_8^{\alpha\mu\nu} 
\nonumber\\
&+& \frac{r^2}{r^2+2(p\cd r)}\,\ell_9^{\alpha\mu\nu}
-\ell_{10}^{\alpha\mu\nu} \,,
\nonumber\\
v^{\alpha\mu\nu}_{{\s{\mathbf{T}}}} &=& -\frac{2}{p^2+2(p\cd q)}\,t_1^{\alpha\mu\nu}
-\frac{2}{q^2+2(q\cd r)}\,t_2^{\alpha\mu\nu} -\frac{2}{r^2+2(p\cd r)}\,t_3^{\alpha\mu\nu}\,.
\label{vlvt}
\eea

\acknowledgments 

The research of J.~P. is supported by the Spanish MEyC under grants FPA2017-84543-P and SEV-2014-0398, and Generalitat Valenciana  
under grant Prometeo~II/2014/066.
The work of  A.~C.~A and M.~N.~F. are supported by the Brazilian National Council for Scientific and Technological Development (CNPq) under the grants 305815/2015 and  142226/2016-5, respectively. A.~C.~A and C.~T.~F. also acknowledge the financial support
from  S\~{a}o Paulo Research Foundation (FAPESP) through the projects  2017/07595-0, 
2017/05685-2, 2016/11894-0, and 2018/09684-3.  This study was financed in part by the Coordena\c{c}\~{a}o de Aperfei\c{c}oamento de Pessoal de N\'{\i}vel Superior - Brasil (CAPES) - Finance Code 001 (M.~N.~F.). This research was performed using the Feynman Cluster of the
John David Rogers Computation Center (CCJDR) in the Institute of Physics ``Gleb
Wataghin", University of Campinas.


\end{document}